%% file: EPJC_V2.tex
\documentclass[11pt]{article}
\usepackage[DIV13]{typearea}
\usepackage{indentfirst}
\usepackage{amsmath}
\usepackage{amsfonts}
\usepackage{mathrsfs}
\usepackage{amssymb}
\usepackage{mathtools}
\usepackage{longtable}
\usepackage{epsfig}
\usepackage{graphicx}
\usepackage{subcaption}
\usepackage{slashed}
\usepackage{multicol}
\usepackage[usenames,dvipsnames]{color}
\usepackage{cite}
\usepackage{bbold}
\RequirePackage[colorlinks=true,urlcolor=blue,anchorcolor=blue,citecolor=blue,filecolor=blue,
               linkcolor=blue,menucolor=blue,linktocpage=true,pdfproducer=medialab]{hyperref}
 \usepackage{cancel}
 \usepackage{bbm}
 \usepackage{lscape}
\usepackage{feynmp}
\usepackage{multirow}
\usepackage{pdfpages}
\usepackage{fancyhdr}
\usepackage{array}
\usepackage{units}
\usepackage[normalem]{ulem}
\usepackage{booktabs}
\usepackage[left=.9in, right=.9in]{geometry}
\usepackage[titletoc]{appendix}
\usepackage{hhline}
\usepackage[T1]{fontenc}
\usepackage{amssymb}
\usepackage[utf8]{inputenc}
\newcommand{\Bsigma}{\mbox{$\hspace{0.12em}\shortmid\hspace{-0.62em}\sigma$}}
\usepackage{framed,feynmp,graphicx}
\usepackage{catchfile}
\newcommand{\getenv}[2][]{%
  \CatchFileEdef{\temp}{"|kpsewhich --var-value #2"}{}%
  \if\relax\detokenize{#1}\relax\temp\else\let#1\temp\fi
}
\getenv[\USER]{USER}

\allowdisplaybreaks
\renewcommand{\a}{\alpha}
\renewcommand{\b}{\beta}

\newcommand{\g}{\gamma}
\newcommand{\m}{\mu}
\renewcommand{\r}{\rho}
\newcommand{\s}{\sigma}
\newcommand{\e}{\varepsilon}
\newcommand{\f}{\Phi}

\renewcommand{\m}{\mu}
\newcommand{\n}{\nu}
\newcommand{\A}{\mathcal{A}}
\newcommand{\B}{\mathcal{B}}
\newcommand{\D}{\mathcal{D}}
\newcommand{\F}{\mathcal{F}}
\renewcommand{\L}{\mathcal{L}}
\renewcommand{\O}{\mathbf{O}}

\newcommand{\T}{\mathbf{T}}
\newcommand{\U}{\mathbf{U}}
\newcommand{\V}{\mathbf{V}}

\newcommand{\at}{\tilde{a}}
\newcommand{\bt}{\tilde{b}}
\newcommand{\fda}{\f^\dagger}

\newcommand{\BBu}{B^{\mu\nu}}
\newcommand{\BBd}{B_{\mu\nu}}

\newcommand{\WWd}{W_{\mu\nu}}

\newcommand{\st}{s_\theta}
\newcommand{\ct}{c_\theta}

\newcommand{\sdt}{s_{2\theta}}
\newcommand{\Bt}{{\tilde{B}}}
\newcommand{\Wt}{{\tilde{W}}}
\newcommand{\ETmiss}{\slashed{E}_T}
\renewcommand{\max}{\mathrm{max}}
\DeclareMathOperator{\Br}{Br}
\newcommand{\cl}{\%\,\,  \text{C.L.}}
\newcommand{\tr}{{\rm Tr}}
\renewcommand{\to}{\rightarrow}

\newcommand{\diag}{\text{diag}}
\newcommand{\de}{\partial}
\newcommand{\derp}{\partial}
\newcommand{\nn}{\nonumber}

\newcommand{\hc}{\mathrm{h.c.}}
\newcommand{\olra}{\overleftrightarrow}
\newcommand{\beq}{\begin{equation}}
\newcommand{\eeq}{\end{equation}}
\newcommand{\bea}{\begin{eqnarray}}
\newcommand{\eea}{\end{eqnarray}}
\renewcommand{\[}{\begin{equation}}
\renewcommand{\]}{\end{equation}}
\renewcommand{\arraystretch}{1.5}
\newcommand{\tpdf}{\texorpdfstring}
\definecolor{orange}{rgb}{1,0.5,0}

\newcommand{\blue}[1]{\color{blue} #1 \color{black}}

\newcommand{\LL}{\mathscr{L}}
\newcommand{\TL}{\mathbf{T}}
\newcommand{\VL}{\mathbf{V}}

\newcommand{\DLR}{\mathbf{D}}

\newcommand{\unity}{\mathbb 1}

\newcommand{\nr}{\refstepcounter{diagram}(FR.\arabic{diagram})}
\newcounter{diagram}

\newcommand{\email}[1]{\href{mailto:#1}{\tt #1}}

\def\cY{{\bf Y}}

\def\cF{{\cal F}}
\def\cY{{\bf Y}}

\newcolumntype{C}[1]{>{\centering\let\newline\\\arraybackslash\hspace{0pt}}m{#1}}

\begin{document}
\vspace*{-1cm}
\phantom{hep-ph/***} 
{\flushleft
{\blue{IFT-UAM/CSIC-16-141}}
\hfill{\blue{KCL-PH-TH/2016-72}}
\vskip 0.2cm
{\blue{FTUAM-16-49}}
\hfill{\blue{CP3-17-04}}
}
\vskip 1.5cm
\begin{center}
{\LARGE ALPs Effective Field Theory and Collider Signatures\\[3mm] }
\vskip .3cm
\end{center}
\vskip 0.5  cm
\begin{center}
{\large I.~Brivio}~$^{a,b)}$,
{\large M.B.~Gavela}~$^{a)}$,
{\large L.~Merlo}~$^{a)}$,
{\large K.~Mimasu}~$^{c,d)}$, \\
{\large J.M.~No}~$^{c,e)}$,
{\large R.~del~Rey}~$^{a)}$,
{\large V.~Sanz}~$^{c)}$
\\
\vskip .7cm
{\footnotesize
$^{a)}$~
Departamento de F\'isica Te\'orica and Instituto de F\'{\i}sica Te\'orica, IFT-UAM/CSIC,\\
Universidad Aut\'onoma de Madrid, Cantoblanco, 28049, Madrid, Spain\\
\vskip .1cm
$^{b)}$~
Niels Bohr International Academy, University of Copenhagen, DK-2100 Copenhagen, Denmark\\
\vskip .1cm
$^{c)}$~
Department of Physics and Astronomy, University of Sussex, Brighton BN1 9QH, UK\\
\vskip .1cm
$^{d)}$~
Centre for Cosmology, Particle Physics and Phenomenology (CP3), Universit\'e catholique de Louvain, \\
Chemin du Cyclotron, 2, B-1348 Louvain-la-Neuve, Belgium \\
\vskip .1cm
$^{e)}$~
Department of Physics, King's College London, Strand, WC2R 2LS London, UK

\vskip .5cm
\begin{minipage}[l]{.9\textwidth}
\begin{center} 
\textit{E-mail:} 
\email{ilaria.brivio@nbi.ku.dk},
\email{belen.gavela@uam.es},
\email{luca.merlo@uam.es},
\email{ken.mimasu@uclouvain.be},
\email{jose$\_$miguel.no@kcl.ac.uk},
\email{rocio.rey@uam.es,}
\email{v.sanz@sussex.ac.uk}
\end{center}
\end{minipage}
}
\end{center}

\begin{abstract}
We study the leading effective interactions between the Standard Model fields and a generic singlet CP-odd (pseudo)Goldstone boson. Two possible frameworks for electroweak symmetry breaking are considered: linear and non-linear. For the latter case, the basis of leading effective operators is determined and compared with that for the linear expansion. Associated phenomenological signals at colliders are explored for both scenarios, deriving new bounds and analyzing future prospects, including LHC and High Luminosity LHC sensitivities. Mono-$Z$, mono-$W$,  $W$-photon plus missing energy and on-shell top final states are most promising signals expected in both frameworks. In addition, non-standard Higgs decays and mono-Higgs signatures are especially prominent and expected to be dominant in non-linear realizations.

\end{abstract}

\vskip 1cm

\pagebreak
\tableofcontents
\pagebreak

%
%%%%%%%%%%%%%%%%%%%%%%%       1.       %%%%%%%%%%%%%%%%%%%%%
%
\section{Introduction}
\label{Sect:Intro}

The Higgs discovery  has set spin zero particles in the spotlight of searches for beyond the Standard Model (BSM) physics.
This may have been the first incursion into new territory: 
 scalar and pseudoscalar particles --elementary or not-- as heralds of  new physics.

Extra spin zero particles are in fact proposed by candidate solutions to major and pressing problems in particle physics. For instance, an option to explain 
the nature of dark matter (DM) 
 is a new scalar particle within a $Z_2$ invariant setup.  A different and outstanding example is the strong CP problem of QCD, for which the paradigmatic solution 
 relies on an anomalous global  $U(1)$ symmetry which is spontaneously broken; the associated (pseudo)Nambu-Goldstone boson, the axion, is in addition an optimal 
 candidate to explain the DM of the universe. The original 
formulation, the so-called PQWW axion~\cite{Peccei:1977hh,Weinberg:1977ma,Wilczek:1977pj}, is now disfavoured by data, while other popular constructions that 
deal with the so-called invisible axion, such as the DFSZ~\cite{Zhitnitsky:1980tq,Dine:1981rt} and the KSVZ~\cite{Kim:1979if,Shifman:1979if} models, are still 
viable solutions to the strong CP problem.  The magnitude of the couplings of axions to ordinary matter is inversely proportional to the scale of $U(1)$ spontaneous 
symmetry breaking, which is much higher than the electroweak scale in the latter, ``invisible'', constructions. Lower axion scales are considered though in other 
implementations of the Peccei-Quinn solution to the strong 
CP problem~\cite{Dimopoulos:1979qi, Dimopoulos:1979pp, Tye:1981zy, Holdom:1982ex, Kim:1984pt, Kaplan:1985dv, Randall:1992ut, Rubakov:1997vp, Berezhiani:2000gh, Gianfagna:2004je, Hook:2014cda, Fukuda:2015ana, Albaid:2015axa, Chiang:2016eav, Wilczek:2016gzx}.

Many other extensions of the Standard Model of Particle Physics (SM) feature one or several spontaneously broken global U(1) symmetries, 
thus predicting the existence of massless Nambu-Goldstone excitations whose couplings need not abide by the same stringent constraints of the 
original QCD axion: axion-like particles (ALPs).  ALPs, if they get a small mass due to non-perturbative effects or other explicit symmetry breaking mechanism, are 
also good DM candidates and/or may  affect the thermal evolution of the universe.  The impact of ALPs, at both high- and low-energies, 
depends on their nature and on the type and strength of their couplings.  In practice, the relevant  generic characteristic of Nambu-Goldstone bosons is that they 
only enjoy derivative couplings, because of the underlying shift symmetry.

The ultimate nature of the Higgs particle itself is still at stake. Is this scalar elementary or composite? Should we accept the uncomfortable fine-tuning associated 
to the electroweak hierarchy problem as a feature of Nature or is there some dynamic explanation for it? Much of the effort in this direction is based on the search 
for symmetries which would justify a low Higgs mass. Two major frameworks are being considered: either linear realizations of electroweak symmetry breaking (EWSB), 
typical of weakly coupled new physics, as for instance in many supersymmetric models, or non-linear ones such as those 
in the so-called ``composite Higgs models'' and other constructions involving new strongly 
interacting physics. The model-independent way of formulating the ultimate exploration of the Higgs nature in low-energy data is provided by the use of effective 
Lagrangians: a ``linear'' expansion~\cite{Buchmuller:1985jz,Grzadkowski:2010es} (often called SMEFT) in terms of towers of gauge invariant operators built 
out of SM fields and ordered by their mass dimension, is used when assuming linear realizations of EWSB, 
while ``non-linear'' expansions~\cite{Feruglio:1992wf,Alonso:2012jc,Azatov:2012bz,Alonso:2012px,Alonso:2012pz,Buchalla:2013rka,Brivio:2013pma} --sometimes called 
``chiral'' or HEFT-- are  the optimal instrument to treat regimes which are not necessarily weakly interacting. The non-linear formulation has the disadvantage of 
depending on a larger 
 number of free parameters, while it has the advantage of being more general; in particular, it reduces to the SM Lagrangian 
 in a particular limit.\footnote{The exception is the case in which the scalar manifold of the SM does not have a fixed point, as it would not admit then a linear 
 representation for the Higgs, see Ref.~\cite{Alonso:2016oah}.} The non-linear expansion does not presuppose that the Higgs particle at low energies belongs to an electroweak doublet, 
 a crucial question to be explored in the years to come.

This paper explores  the physics of an extra singlet scalar which is a CP-odd (pseudo)Nambu-Goldstone boson. We will  
 formulate in all generality its leading CP-invariant effective couplings to SM fields, which must be purely derivative couplings when its mass is neglected. 
 This first --theoretical-- part   is general by nature and   
 holds for ALP scales larger than the electroweak one (in the EWSB non-linear case also larger than its implicit BSM scale). While the dominant ALP interactions in the linear --SMEFT-- expansion have been formulated long ago~\cite{Georgi:1986df}, the analogous analysis for the non-linear regime is missing  and will be developed here. We will first 
 concentrate on determining a complete basis of CP-even bosonic operators containing one ALP insertion;  nevertheless, the  fermionic operators are also 
 derived in this paper, building a complete and non-redundant chiral set. The relation and differences between the dominant operators in both 
 expansions --linear and chiral-- will be 
 subsequently discussed. It is interesting to note that all results to be obtained below apply as well to a different case: the complete basis of 
 CP-odd derivative couplings of an hypothetical CP-even scalar (see also Ref.~\cite{Gavela:2014vra} for a generic CP-even scalar).

Up to now, most phenomenological ALPs analyses concentrated on their couplings to  photons, 
gluons and/or quarks, as they dominate at low energies and determine 
astrophysical and cosmological constraints for very light ALPs. Nevertheless, ALPs may well show up first 
at colliders~\cite{Mimasu:2014nea,Jaeckel:2015jla,Alves:2016koo} or in rare mesonic decays~\cite{Dolan:2014ska,Izaguirre:2016dfi}, and the $SU(2)_L\times U(1)_Y$ invariant 
formulation of their interactions developed here provides new beautiful channels involving the electroweak gauge 
bosons  and the Higgs particle. In the second --phenomenological-- part of this work, the foreseen impact of 
those couplings at colliders and in particular at LHC will be analyzed  for the first time, identifying the new signals and performing 
a detailed analysis of experimental bounds and prospects. \footnote{ The set of Feynman rules stemming from the bosonic ALP effective Lagrangian can be found in App.~\ref{App:feynman_rules}.}  
 Unlike  for the theoretical results, this search for (pseudo)Nambu-Goldstone bosons at colliders  implicitly assumes an overall ALP scale not 
 far from the electroweak one, e.g. $\mathcal{O}$(TeV), for observability.
 
 The structure of the paper can be easily inferred from the Table of Contents.

%%%
%%%%%%%%%%%%%%%%%%%%%%%%%% 2. 
%%%
\section{ The ALP linear Lagrangian}
\label{Sect:2pointF_linear}
In linear realizations of EWSB with only SM fields at low-energies,  the leading order (LO) effective Lagrangian  is simply the SM one, 
\begin{equation}\label{LSM_lin}
\L_\text{SM} \supset D_\mu\Phi^\dag D^\mu\Phi+\sum_\psi i\bar{\psi}\slashed{D}\psi-\left(\bar{Q}_L\,\cY_D\,\Phi d_R+\bar{Q}_L\,\cY_U\,\tilde\Phi u_R+\bar{L}_L\,\cY_E\,\Phi e_R+\hc\right)\,,
\end{equation}
where $\tilde{\Phi}=i\s^2\Phi^*$\, and 
  $\cY_D$, $\cY_U$ and $\cY_E$ are $3\times 3$ matrices in flavour space which encode the Yukawa couplings for down quarks, up quarks 
  and charged leptons, respectively.  Consider now an additional particle, singlet under the SM charges, which is a (pseudo)Nambu-Goldstone boson of a 
  spontaneously broken symmetry at energies higher than the electroweak scale $v$ (set by the $W$ mass).  Neglecting its mass, its couplings would be pure derivative ones 
  because of the underlying shift symmetry. Denoting by $f_a$ the scale associated to the physics of this ALP particle $a$,  insertions of the 
  latter in effective operators will be weighted down by powers of $a/f_a$. Focusing on interactions involving only one ALP, 
  the next-to-leading-order (NLO) effective linear ALP Lagrangian has been determined long ago~\cite{Georgi:1986df}. 

In this paper we mostly focus on the bosonic operators involving $a$, determining a complete and non-redundant set. For linear EWSB realizations 
  the most general  linear bosonic Lagrangian, including only the  NLO corrections involving $a$, is given by 
\beq
\LL_\text{eff}^\text{linear}=\LL^{\text{LO}}+\, \delta\LL_a^\text{bosonic}\,, 
\label{Lbosonic-lin}
\eeq
where now the leading order Lagrangian is the SM one plus the ALP kinetic term,
\beq
\LL^{\text{LO}}=\LL_\text{SM}\,+\frac{1}{2}(\partial_\mu a)(\partial^\mu a)\,,
\label{Lbosonic-lin}
\eeq
while the NLO bosonic corrections are given by
\beq
\delta\LL_a^\text{bosonic}\,=\,c_{\tilde{W}}\A_{\tilde{W}}+c_{\tilde{B}}\A_{\tilde{B}}+c_{\tilde{G}}\A_{\tilde{G}}+c_{a\Phi}\O_{a\Phi}\,,
\label{deltaLbosonic-lin-initial}
\eeq
with
\begin{eqnarray}
\A_{\tilde{B}} &=&-\BBd\tilde{B}^{\mu\nu}\dfrac{a}{f_a}\,,\label{ABtilde}\\
\A_{\tilde{W}} &=&-\WWd^a\tilde{W}^{a\mu\nu}\dfrac{a}{f_a}\,,\label{AWtilde}\\
\A_{\tilde{G}} &=&-G^a_{\mu\nu}\tilde{G}^{a\mu\nu}\dfrac{a}{f_a}\,,\label{AGtilde}\\
\O_{a\Phi}&=& i (\Phi^\dag\overleftrightarrow{D}_\mu\Phi)\frac{\de^\mu a}{f_a}\,,\label{OaPhi}
\end{eqnarray}
and $\tilde{X}^{\mu\nu}\equiv \frac12 \epsilon^{\mu\nu\rho\sigma}X_{\rho\sigma}$.  The action of the shift symmetry on the first three operators, $a\to a+\alpha$, with $\alpha$ constant, yields 
\beq\label{aXXtilde_shift}
\tr[X_{\mu\nu}\tilde{X}^{\mu\nu}]\dfrac{a}{f_a}\equiv
\derp_\mu K_X^\mu\dfrac{a}{f_a}\to
\derp_\mu K_X^\mu\dfrac{a+\alpha}{f_a}=
-K_X^\mu\partial_\mu \dfrac{a}{f_a}+\dfrac{\alpha}{f_a}\partial_\mu K_X^\mu\,, 
\eeq
 and thus the corresponding associated current is anomalous as $\delta \LL = \dfrac{\alpha}{f_a}\partial_\mu K_X^\mu\,$. 
 Even if this correction is a total derivative, in the case of $\A_{\tilde G}$ the existence of instantonic configurations in the QCD Lagrangian 
implies that the action is modified because the integral of $\de_\mu K^\mu_G$ does not vanish (although a discrete version of the shift symmetry is preserved); it is nevertheless often added to the Lagrangian given its relevance for the case of the true QCD axion and the solution of the strong CP problem. \footnote{In any case, $\A_{\tilde G}$ will play no role whatsoever in the analysis to be performed in this work.}

After electroweak symmetry breaking, $\O_{a\Phi}$ induces a  two-point function contribution, tantamount to $a$ acting as an additional contribution to the longitudinal component of the electroweak gauge fields. An easy way of determining its  impact on observables is to trade it for a fermionic vertex~\cite{Georgi:1986df}, either chirality conserving or chirality flipping, or a combination of them.  
 For instance, the Higgs field redefinition
\beq
\Phi\to e^{ic_\Phi\, a/f_a}\Phi
\label{Oaphiequiv}
\eeq
 applied to  the bosonic Lagrangian Eq.~\eqref{Lbosonic-lin}, induces a correction stemming from the Higgs kinetic energy term (see Eq.\eqref{LSM_lin})
which cancels exactly $\O_{a\Phi}$ up to $\mathcal{O}(a/f_a)$, while the Yukawa terms in that equation induce a new Yukawa-axion coupling for which $\O_{a\Phi}$ can be entirely traded (see App.~\ref{App:field_redef} for details and  a general discussion of possible field redefinitions). 
 The overall effect  
 is thus the  
 replacement in Eq.~(\ref{deltaLbosonic-lin-initial})   
 \begin{equation}
 \O_{a\Phi} \longrightarrow \O^\psi_{a\Phi}\,,
  \end{equation}
where
\beq
\label{axion-Yukawa}
\O^\psi_{a\Phi}\equiv i\left(\bar{Q}_L\,\cY_U\,\tilde\Phi u_R-\bar{Q}_L\,\cY_D\,\Phi d_R-\bar{L}_L\,\cY_E\,\Phi e_R\right)\frac{a}{f_a} + \hc\,,
\eeq
which exhibits a relative minus sign between the Yukawa-ALP type of interaction for up and down fermions. This coupling can then be written in a more compact way as \beq
\label{Axion-Yukawa-compact}
\O^\psi_{a\Phi}= i\,\frac{a}{f_a}\sum_{\psi=Q,\,L} \left(\bar\psi_{L}{\bf \mathcal{\bf Y}_\psi}{\bf \Phi}\sigma_3 \psi_{R}\right)+ \hc\,,
\eeq
where $Q_R\equiv \{u_R,d_R \} \, \, \left( L_R\equiv \{0,e_R \} \right) $ --with $\sigma_3$ acting on weak isospin space-- and where the block matrices $\mathcal{{\bf Y}}_\psi$ and ${\bf \Phi}$ are defined by
\begin{equation}
\mathcal{{\bf Y}}_Q \equiv \rm{diag}\left( \cY_U\,, \cY_{D}\right) \,,\qquad \qquad 
\mathcal{{\bf Y}}_L \equiv \rm{diag}\left( 0\,, \cY_{E}\right) \,,\qquad \qquad  \bf{\Phi}=\rm{diag}(\tilde\Phi\,,\Phi)\,.
\label{Yukawa_matrices}
\end{equation}

Alternatively, using the equations of motion of 
 $\LL^{\text{LO}}$, $\O_{a\Phi}$ could be entirely traded by a flavour-blind and chirality-conserving fermionic operator\,,
\beq
\O_{a\Phi}\longrightarrow-\frac{1}{2}\,\frac{\de_\mu a}{f_a} \sum_{ \psi=Q,\,L} \left(\bar\psi\gamma_\mu \gamma_5 \sigma_3 \psi\right)+ \hc\,,
\label{Axion-fermion-from-bosonic}
\eeq
where again terms with more than one axion insertion have been neglected. In this work we choose to use  the chirality-flipping version of the fermionic couplings,
 though. In summary,  the expression for $\delta\LL_a^\text{bosonic}$ to be used below reads
 \beq
\delta\LL_a^\text{bosonic}\,=\,c_{\tilde{W}}\A_{\tilde{W}}+c_{\tilde{B}}\A_{\tilde{B}}+c_{\tilde{G}}\A_{\tilde{G}}+c_{a\Phi}\O^\psi_{a\Phi}\,,
\label{deltaLbosonic-lin}
\eeq
with $\O^\psi_{a\Phi}$ as defined in Eq.~(\ref{Axion-Yukawa-compact}).

For completion, it is worth mentioning that when the complete NLO Lagrangian is considered in the linear case, additional fermionic operators are present. In fact the most general NLO ALP Lagrangian is given by ~\cite{Georgi:1986df,Choi:1986zw, Salvio:2013iaa}
 \beq
 \delta\LL_a^\text{total}\,=\,c_{\tilde{W}}\A_{\tilde{W}}+c_{\tilde{B}}\A_{\tilde{B}}+c_{\tilde{G}}\A_{\tilde{G}}+ \frac{\de_\mu a}{f_a}\sum_{\substack{\psi=Q_L,\,Q_R, \\\,L_L,\,L_R}} \bar\psi \gamma_\mu X_\psi \psi \,\,,
\label{general-NLOLag-lin}
\eeq
where  $X_\Psi$ are  $3\times3$ hermitian matrices in flavour space. The chirality-conserving operator in the last term of this equation could alternatively be traded using the equations of motion (EOM) by a chirality-flipping coupling:
\[
 \frac{\de_\mu a}{f_a}\sum_{\substack{\psi=Q_L,\,Q_R, \\\,L_L,\,L_R}} \bar\psi \gamma_\mu X_\psi \psi \, \longrightarrow  \frac{ia}{f_a}
 \sum_{\psi= Q, L} \bar{\psi}_L  {\bf{\Phi}} \left( X_{\psi_L} {\bf Y}_\psi - \mathbf{Y}_\psi X_{\psi_R} \right) \psi_R \, + \hc \,.
 \label{general-fermionic-Yukawa}
\]
In this equation, the  products $X_{\psi_L} {\bf Y}_\psi$ and $\mathbf{Y}_\psi X_{\psi_R}$ are completely generic matrices and in consequence,  in the complete linear basis, operators of the type $a\,\bar{\psi}_L {\bf{\Phi}} \psi_R  $ are not Yukawa suppressed. Note as well that it would be redundant to consider simultaneously a bosonic coupling such as $\O_{a\Phi}$ in Eq.~(\ref{OaPhi}) and the general fermionic couplings in Eq.~(\ref{general-NLOLag-lin}) or (\ref{general-fermionic-Yukawa}), as the effects of the former are already included in the flavour blind components of the $X_\psi$ matrices, see Eq.~(\ref{Axion-fermion-from-bosonic}). 
In this paper, we concentrate on the thorough exploration of observables induced by the purely bosonic ALP  couplings as expressed in Eq.~(\ref{deltaLbosonic-lin}), for the case of linear EWSB realizations. 

\subsection{ Previous phenomenological bounds}
The experimental bounds on the couplings of axions -- and in general ALPs --  to gluons, photons and fermions have been abundantly 
considered in the linear EWSB scenario (see e.g.~\cite{Agashe:2014kda,Bjorken:1984mt,Vinyoles:2015aba,Raffelt:2006cw,Friedland:2012hj,Ayala:2014pea,Khachatryan:2014rra,Aad:2015zva,Krnjaic:2015mbs,Clarke:2013aya,Aprile:2014eoa,Viaux:2013lha}), including as well their impact at colliders for the case $f_a\sim\mathcal{O}(\text{TeV})$~\cite{Mimasu:2014nea, Jaeckel:2015jla}. Additionally, constraints on the linear coupling of the ALP to $W^\pm$ gauge bosons have recently been obtained in Ref.~\cite{Dolan:2014ska,Izaguirre:2016dfi}.

{\bf Coupling to photons: } Both $\A_{\tilde{B}}$ and $\A_{\tilde{W}}$ --Eqs.~(\ref{ABtilde})-(\ref{AWtilde})-- contribute to the interaction of the ALP with two photons,
 \beq
 \delta\LL_a^{\text{bosonic}} \supset  - \frac{1}{4}g_{a\gamma\gamma} \,a\,F_{\m\n}\tilde{F}^{\m\n}\,,
 \label{gagammagamma-def}
 \eeq
 where $F_{\m\n}$ denotes the electromagnetic field strength, and the dimensionful coupling $g_{a\g\g}$ is given by
\begin{equation}\label{def_agg}
 g_{a\g\g} =\frac{4}{f_a}(c_{\tilde{B}}\ct^2+c_{\tilde{W}}\st^2)\,,
\end{equation} 
where $\ct$ ($\st$) denotes the cosinus (sinus) of the Weinberg angle. Bounds on $g_{a\g\g}$ can be inferred as a function of the 
ALP mass $m_a$ from various astrophysical constraints and low energy data, which rely only on an hypothetical ALP-photon coupling and not on fermion-ALP interactions,  
 as discussed e.g. in Ref.~\cite{Mimasu:2014nea}. They 
  enforce the combination $|c_\theta^2c_\Bt + s_\theta^2 c_\Wt|/f_a$ to cancel to one part 
in $10^{3}$ ($10^{8}$) for $m_a=\unit[1]{MeV\, (keV)}$. 
 Indeed, for $m_a \simeq \unit[1]{MeV}$ the best present constraint is set by  
 Beam Dump experiments,  $g_{a\gamma\gamma} \lesssim \unit[10^{-5}]{GeV^{-1}}$\cite{Mimasu:2014nea,Bjorken:1984mt}~\footnote{
 There are in fact stronger bounds on g for $m_a = \unit[1]{MeV}$ from SN1987a measurements, which rule out the range $\unit[10^{-9}]{ GeV^{-1}} < g_{a\gamma\gamma} < \unit[10^{-6}]{GeV^{-1}}$ (see e.g. Ref.~\cite{Mimasu:2014nea}). The one given here is describing a window that remained untested for $m_a \sim \unit[1]{ MeV}$ and $g_{a\gamma\gamma} \sim \unit[10^{-5}]{GeV^{-1}}$ . This window may be excluded through the model-dependent constraints in Ref.~\cite{Millea:2015qra}, in which case the bound for 1 MeV axions would be $g_{a\gamma\gamma} \ll \unit[10^{-12}]{ GeV^{-1}}$}, that is  
\begin{equation}\label{agg_constraint}
|c_{\tilde{B}}\ct^2+c_{\tilde{W}}\st^2| \lesssim 0.0025 \,\left(\frac{f_a}{\unit[1]{TeV}} \right)\quad (90\cl)   \quad \text{for}\quad m_a \leq \unit[1]{MeV}\,.
\end{equation}
For substantially lower masses astrophysical constraints may apply, e.g. for  $m_a=\unit[1]{keV}$  the  combination of helioseismology, solar neutrino  data
observations~\cite{Vinyoles:2015aba} and Horizontal Branch stars  data~\cite{Raffelt:2006cw,Friedland:2012hj,Ayala:2014pea} results in $g_{a\g\g}\lesssim \unit[10^{-10}]{GeV^{-1}}$, that is, 
\begin{equation}\label{agg_constraint_kev}
|c_{\tilde{B}}\ct^2+c_{\tilde{W}}\st^2| \lesssim 2.5 \cdot 10^{-8} \,\left(\frac{f_a}{\unit[1]{TeV}} \right) \quad \text{for}\quad  m_a\leq\unit[1]{keV}\,.
\end{equation}

These strong constraints  on  
  $g_{a\gamma \gamma}$ could suggest that each of the two coefficients involved, $c_{\tilde{B}}$ and $c_{\tilde{W}}$, may be individually subject to bounds of the same order of magnitude. Nevertheless,  often symmetry reasons force a given theory to produce couplings to photons much suppressed with respect to $Z$ couplings. 
In any case, from the point of view of effective theory they are two independent degrees of freedom: the combination orthogonal to that in  Eq.~(\ref{def_agg}) should 
be probed and bounded independently. In practice, in most of the phenomenological analysis to be developed in this work the constraint 
\beq
 c_\Bt = -t_\theta^2 c_\Wt
 \label{cWB}
 \eeq
  will be systematically enforced.

\vspace{0.3cm}
{\bf Coupling to gluons: } In turn, the effective ALP-gluon $g_{agg}$ coupling is analogously defined by
\beq
 \delta\LL_a^{\text{bosonic}} \supset  - \frac{1}{4}g_{agg} \,a\,G_{\m\n}^a\tilde{G}^{a\m\n}\,,
  \label{gaglueglue-def}
 \eeq
  where $G_{\m\n}$ denotes the QCD field strength. It receives contributions from the NLO effective operator
   $\A_{\tilde{G}}$ in Eq.~(\ref{AGtilde}), where  
   \beq
g_{agg}=\frac{4}{f_a}\,c_{\tilde{G}} \,
 \eeq
    can  be directly constrained at energies 
above the QCD scale $\Lambda_{QCD}$ via axion-pion mixing effects, and also via mono-jet searches at hadron colliders.

Bounds on $\mathrm{Br}(K^+ \to \pi^+ + \mathrm{nothing})$~\cite{Adler:2004hp} can be used to constrain the process $K^+ \to \pi^+ \,\pi^0 \,(\pi^0 \to a)$, 
where the pion-axion mixing arises through the anomalous coupling of mesons and of the axion to gluons~\cite{Choi:1986zw, Carena:1988kr}. These bounds 
have been used to constrain $f_a$ in contexts 
where the coupling of the ALP to gluons is only present due to the 
anomaly, \textit{i.e.} where $\L \supset \frac{\alpha_s}{8\pi}\frac{a}{f_a}G \tilde{G}$ (see, for example, Ref.~\cite{Fukuda:2015ana}). They can 
be reinterpreted in terms of the generic ALP-gluon coupling, Eq.\eqref{gaglueglue-def}, yielding
\[
g_{agg} \lesssim \unit[1.1\cdot10^{-5}]{GeV^{-1}} \quad (90\cl) \quad \text{for}\quad m_a \lesssim \unit[60]{MeV}\,.
\]

Slightly higher ALP masses  have been considered at colliders, assuming only the coupling in Eq.~(\ref{gaglueglue-def}). Limits of order
\[\label{a-f-5}
g_{agg} \lesssim \unit[10^{-4}]{GeV^{-1}}\quad (95\cl) 
\quad \text{for}\quad m_a \lesssim \unit[0.1]{GeV}\,, 
\]
were obtained \cite{Mimasu:2014nea} by recasting 8 TeV LHC 
analyses~\cite{Khachatryan:2014rra,Aad:2015zva}.
\vspace{0.3cm}

{\bf Coupling to fermions: } Interesting bounds on ALP-fermion interactions can be obtained from several set of experimental data. 
For instance, considering those stemming from the purely bosonic operator $\O_{a\Phi}$ --see Eq.~(\ref{Axion-Yukawa-compact})-- or, in other words, 
the flavour-diagonal couplings in the  last operator in Eq.~\eqref{general-NLOLag-lin} as expressed in Eq.~\eqref{general-fermionic-Yukawa} with $X_{L,R}^{ij}=X_{L,R}^{ii}\delta^{ij}$ and 
\beq
g_{a\psi}=X_{L}-X_{R}\, ,
\eeq
their contribution to the effective Lagrangian in the fermion mass basis reads
\[ 
\delta\LL_a^{\text{bosonic}} \supset   \frac{ia}{f_a}  \sum_{\psi=Q,\,L} g_{a\psi} m^{\text{diag}}_{\psi}\, \bar{\psi} \gamma_5 \psi \,
\]
where $m^{\text{diag}}_\psi$ is the fermion mass matrix resulting from diagonalizing the product $v \mathbf{Y}_\psi/\sqrt{2}$.  
The severity of the constraints on $g_{a\psi}$ depends on the ALP mass range considered. 
  The least constrained is the high-mass region, tested through rare meson decays and in DM direct detection  searches (the latter being very model-dependent)~\cite{Krnjaic:2015mbs}.  The former provide bounds on ALP-fermion couplings below $10$ GeV and in particular Beam Dump experiments (CHARM) constraints read~\cite{Bergsma:1985qz,Dolan:2014ska}:
\[\label{a-f-2}
g_{a\psi}/f_a < \unit[\left(3.4\cdot 10^{-8} - 2.9\cdot10^{-6}\right)]{GeV^{-1}} \quad(90\cl)  \quad \text{for}\quad \unit[1]{MeV}\lesssim m_a \lesssim \unit[3]{GeV}\,.
\]
Lighter ALPs have been tested in axion searches in Xenon100~\cite{Aprile:2014eoa} through the axio-electric effect in liquid xenon (analogue of the photo-electric process with the absorption of an axion instead of a photon), bounding ALP couplings to electrons: 
\[\label{a-f-3}
g_{ae}/f_a<\unit[1.5\cdot 10^{-8}]{GeV^{-1}}\quad\quad(90\cl)\quad \text{for}\quad  m_a<\unit[1]{keV}\,.
\]
Finally, the strongest bounds apply to very low ALP masses. They are inferred from high-precision photometry of the red giant branch of the color-magnitude diagram for globular clusters~\cite{Viaux:2013lha}. Measurements of axionic recombination and de-excitation, Compton scattering and axion-bremsstrahlung set very strong bounds again on the coupling to electrons:
\[\label{a-f-4}
g_{ae}/f_a < \unit[8.6\cdot 10^{-10}]{GeV^{-1}}\quad (95\cl) \quad \text{for}\quad m_a \lesssim \unit[]{eV}\,.
\]

The above set of fermionic bounds could suggest to infer new limits on the coefficient of the linear bosonic operator  $\O_{a\Phi}$  of the bosonic linear ALP basis, Eq.~(\ref{OaPhi}), if considered by itself, via the equivalence discussed in Eqs.~(\ref{Oaphiequiv})-(\ref{Axion-Yukawa-compact}). This bound would depend on the ALP mass, and would be conservatively summarized in 
\beq \label{boundaphifermions}
|c_{a\Phi}|/f_a\,<\,\unit[\left(3.4\cdot 10^{-8} - 2.9\cdot10^{-6}\right)]{GeV^{-1}} \quad(90\cl)   \quad \text{for}\quad m_a \lesssim \unit[3]{GeV}\,,
\eeq  
except for ALPs with masses in the $\unit[1]{keV}-\unit[1]{MeV}$ range, where the bounds from rare meson decay and DM searches are much weaker. Nevertheless, more than one effective operator can contribute to the rare processes under discussion and, in consequence, strictly speaking a bound can only be set on the corresponding combination of operators, see further below, in the same spirit that the bounds on $a\gamma\gamma$ decay do not nullify simultaneously the two couplings in the set $\{a_{\tilde W}, a_{\tilde B}\}$, but only a combination of them, see Eq.~(\ref{cWB}). For the time being, the value of $c_{a\Phi}$ will be thus left free for further exploration below.

\vspace{0.3cm}

Axion-like particles are also appealing DM candidates and further bounds apply if such hypothesis is considered. 
 Indeed,  heavy ALPs (in the GeV-TeV mass range) have largely been searched for at colliders as weakly interacting massive particles (WIMPs).   However, the phenomenological analysis in this work will focus on a low-mass region with ALP masses below the MeV range;  
  DM candidates in this range are known as weakly interacting slim particles (WISPs) and could be produced non-thermally through the misalignment mechanism~\cite{Preskill:1982cy, Abbott:1982af, Dine:1982ah, Nelson:2011sf, Masso:2004cv}. ALP DM candidates capable of generating the correct relic abundance call for a large enough initial field value. Because of their (pseudo)Nambu-Goldstone  nature,  these ALPs are the phase of a complex field and thus have field values limited to $-\pi f_a< a(x) < \pi f_a$, implying  that standard ALP CDM (cold DM) producing the correct relic density would require large ALP scales~\cite{Arias:2012az}: $f_a \gtrsim \unit[3.2\cdot 10^{10}]{GeV} \left(m_0 / \unit[]{eV}\right)^{1/4}$ (smaller scale values cannot explain the totality of the relic abundance).  In what follows, ALPs will not be required to necessarily account for the DM of the universe. 

\vspace{0.3cm}

{\bf Coupling to massive vector bosons: } In contrast to the present constraints  discussed above, the couplings of ALPs  to the heavy SM bosons 
have been largely disregarded although they appear at  NLO of the linear expansion, that is, at the same order as the pure photonic, gluonic and fermionic ALP couplings.

The associated signals stemming from the linear $\delta\LL_a^\text{bosonic}$ in Eq.~(\ref{deltaLbosonic-lin}) 
are illustrated  in the column on the right hand side of the Feynman rules detailed in App.~\ref{App:feynman_rules}; they   
include in particular interaction vertices of the ALP with electroweak gauge bosons such as:  $a\gamma Z$, $aZZ$,  $aW^+W^-$, $a\gamma W^+W^-$ and $aZW^+W^-$. Besides the collider signatures that will be presented in the phenomenological sections of this paper, rare  decays provide an additional handle on the ALP couplings to massive vector bosons.

Consider the ALP-$W^+W^-$ interaction defined by
\beq
 \delta\LL_a^{\text{bosonic}} \supset  - \frac{1}{4}g_{a WW} \,a\,W_{\m\n}\tilde{W}^{\m\n}\,,
 \label{gaWW-def}
 \eeq
   which  may induce flavour changing rare meson decays via $W$ exchange at one loop, and an ALP radiated from the $W$ boson.
 $\A_\Wt$ contributes to such processes, with  $g_{a W W}=4 c_\Wt / f_a$. Upon considering the action of $\A_\Wt$  by itself, the same coupling 
 induces the subsequent ALP decay into two photons. NA48/2, NA62 and Beam Dump experiments  have been analysed in this 
 context in Ref.~\cite{Izaguirre:2016dfi},  which extends to higher ALP masses the bounds in Eqs.~\eqref{agg_constraint} and \eqref{agg_constraint_kev} of 
 Ref.~\cite{Mimasu:2014nea}, indicating a constraint~\footnote{We are indebted to Brian Shuve for stressing the impact of present rare decay bounds for light axions.}
\beq
 f_a/c_{\tilde W} \gtrsim \unit[4-8000]{TeV}\,,\quad \text{for}\quad m_a < \unit[500]{MeV}\,. \label{rarebound}
\eeq

Other limits have been obtained from the bounds on rare meson decays  into invisible products, $B \to K + a$ and $K \to \pi + a$ with $a\to \text{inv.}$. This is 
nevertheless at the price of assuming, in addition to $\A_\Wt$, the existence of some supplementary ALP decay channel into invisible sectors that furthermore is required 
to be largely dominant~\cite{Izaguirre:2016dfi}.

The bounds just discussed are precious and in particular the approach of having started considering just one operator at a time is a valid one. Nevertheless, 
with the discussed level of accuracy for $c_{\tilde W}$ when considered just by itself, it may be pertinent to take into account 
the possible competing action of other specific ALP-SM couplings in the EFT, for instance those  where the ALP would not be attached to the $W$ boson but to the intermediate fermion in the loop. These stem from the fermionic couplings  --in particular the top quark coupling-- induced by the bosonic operator $\O_{a\Phi}$ in Eq.~(\ref{deltaLbosonic-lin}), or from other ALP-fermion interactions such as the generic ones in Eq.~(\ref{general-NLOLag-lin}) for the linear case. Indeed,  alike to the analysis that lead to Eq.~(\ref{cWB}), from the point of view of the effective field theory in the linear EWSB framework, only combinations of the couplings in the set
\beq
\{c_{\tilde W}, c_{a \Phi},c_{\psi_i}\}
\eeq 
can  be strictly bound by such data,  where $c_{\psi_i}$ refers to the coefficients of the fermionic couplings in the complete NLO 
linear ALP Lagrangian Eq.~(\ref{general-NLOLag-lin}) which are not tantamount to $c_{a \Phi}$ via EOM. 

In this paper we will explore the complementary information that the LHC can provide in various tree-level channels, 
e.g. mono-$W$, which are insensitive to the presence of the operator coefficient $c_{a\Phi}$ but share with the rare-decay analyses 
the dependence on the linear operator coefficient $c_{\tilde W}$.
This complementarity is also manifest as the LHC has access to a larger kinematic range. Hence the breakdown of the ALP Effective Theory, 
and possible discovery of new physics, may be possible at the LHC but be hidden in physics at B-factories. For these reasons, 
in the phenomenological sections we will obtain LHC bounds on operators involved in tree-level ALP-$W$ couplings (among others) and 
without the prejudice from rare-decays. The combined impact at LHC of $c_{\tilde W}$ and $c_{a\Phi}$ plus general ALP-fermion couplings, 
as well as the impact of non-linear operators on rare decays is a subject for future work.

\vspace{0.5cm}
\section{The Bosonic Chiral ALP Lagrangian} 
\label{Sect:basis}
This section explores the leading effective couplings between an ALP  and the SM fields, in the general framework of a non-linear (often referred to as chiral or HEFT) 
realization of EWSB.   The complete set of LO and NLO bosonic CP-even couplings involving one ALP  will be determined (again, they could also be read as the complete bosonic set of derivative CP-odd couplings involving a CP-even singlet scalar). It will be assumed that the 
characteristic  scale $f_a$  associated to the Nambu-Goldstone boson origin of the ALP  is at least of the same order of magnitude or larger than the cut-off of the 
BSM electroweak theory $\Lambda$. The ALP scale and the electroweak BSM scale $\Lambda$ will nevertheless be treated here as independent.

The chiral effective Lagrangian HEFT~\cite{Feruglio:1992wf,Alonso:2012jc,Azatov:2012bz,Alonso:2012px,Alonso:2012pz,Buchalla:2013rka,Brivio:2013pma,Gavela:2014vra,Gavela:2014uta,Brivio:2014pfa,Alonso:2014wta,Hierro:2015nna,Brivio:2015kia, Brivio:2016fzo,Merlo:2016prs}, which
in the context of generic non-linear realizations of EWSB describes the interactions among SM gauge degrees of freedom, SM fermions and a light Higgs resonance,  
   consists of all operators invariant under Lorentz and SM gauge symmetries and written in terms 
of the SM spectrum with the only exception of the Higgs doublet, whose four degrees of freedom are distributed in two separate sets. On one side, a unitary matrix $\U(x)$ describes only the 
three SM would-be Nambu-Goldstone bosons~\cite{Appelquist:1980vg,Longhitano:1980iz,Longhitano:1980tm,Feruglio:1992wf} -- that become the longitudinal components of the gauge 
bosons after EWSB. On the other side, the physical Higgs particle $h$ is introduced as an independent field, a generic singlet of the SM 
 with arbitrary couplings~\cite{Feruglio:1992wf,Grinstein:2007iv,Azatov:2012bz, Alonso:2012px, Alonso:2012pz}. 
 For particular values of the latter parameters and correlations of the operator coefficients the usual SMEFT linear 
 formulation would be recovered~\cite{Alonso:2015fsp,Alonso:2016btr,Alonso:2016oah,Alonso:2012px,Brivio:2013pma,Brivio:2014pfa, Gavela:2014vra,Alonso:2014wta,Hierro:2015nna,Brivio:2015kia,Eboli:2016kko,Brivio:2016fzo,Merlo:2016prs}.

The HEFT building blocks can be chosen to be the gauge field strengths $G_{\mu\nu}$, $W_{\mu\nu}$ and $B_{\mu\nu}$  plus two $SU(2)_L$ covariant objects:
\beq
\begin{aligned}
\V_\mu(x)&\equiv \left(\DLR_\mu\U(x)\right)\U(x)^\dag\, , \qquad \qquad  &\,\T(x)&\equiv \U(x)\sigma_3\U(x)^\dag\,,\\
\end{aligned}
\eeq
with
\beq
\U(x)=e^{i\sigma_a \pi^a(x)/v}\, , 
\eeq
 where $\pi^a(x)$ denotes the longitudinal degrees of freedom of the gauge bosons and  $\sigma_a$  the Pauli matrices. In this notation, the   
 covariant derivative reads
\beq
\DLR_\mu \U(x) \equiv \derp_\mu \U(x) +igW_{\mu}(x)\U(x) - \dfrac{ig'}{2} B_\mu(x) \U(x)\sigma_3 \,.
\eeq
Under $SU(2)_{L,R}$ 
global transformations ($L$, $R$ respectively), the objects defined above transform as 
\beq
\begin{aligned}
\U(x) \rightarrow L\, \U(x) R^\dagger\, , \qquad  &\V_\mu(x) &\rightarrow L\, \V_\mu(x) L^\dagger\,,\qquad \T(x) &\rightarrow L\, \T(x) L^\dagger\,.\\
\end{aligned}
\eeq
The physical Higgs particle $h$ is then customarily introduced as a SM isosinglet  via generic polynomial  
functions $\cF_i(h)$~\cite{Grinstein:2007iv} expanded in powers of $h/v$,
\beq
\cF_i(h)=1+ a_i h/v + b_i (h/v)^2+\ldots\,,
\label{F_def}
\eeq
 where $a_i$, $b_i$ $\ldots$ are constant coefficients. Finally, the SM fermions are often grouped into doublets of $SU(2)_L$ 
and $SU(2)_R$, $Q_{L,R}\equiv(u_{L,R}\,, d_{L,R})$, $L_L\equiv(\nu_L\,, e_L)$ and $L_R\equiv(0\,, e_R)$. The 
notation chosen allows an easy identification of  terms breaking the custodial symmetry $SU(2)_C$ to which the global group $SU(2)_L\times SU(2)_R$ gets broken after  EWSB.  $SU(2)_C$ is explicitly broken by the gauging of the hypercharge $U(1)_Y$ and 
by the heterogeneity of the fermion masses; insertions of the scalar chiral field $\T(x)$, which is not invariant under transformations of the 
full $SU(2)_R$, account for breaking of the custodial symmetry in the effective operators.

The task now consists in the generalization of the HEFT Lagrangian  to include  insertions of derivatives of $a/f_a$. This could be approached via the  insertion in that  Lagrangian of  general polynomial functions  of the SM singlet  scalar $a$, $\cF_i(a/f_a)$, in analogy with the treatment given to the scalar $h$ in the HEFT Lagrangian. After all, the $\cF_i(h/v)$ polynomials  are reminiscent of the deformed exponential Nambu-Goldstone nature  of the Higgs particle in some non-linear EWSB realizations, such as ``composite Higgs'' models~\cite{Kaplan:1983fs,Kaplan:1983sm,Banks:1984gj}.  From this point of view, to restrict below to terms with a single $a(x)/f_a$ insertion is consistent with the assumption $f_a\ge \Lambda$. In summary, the effective Lagrangian can be written as
\beq
\LL_\text{eff}^\text{chiral}=\LL^{\text{LO}}+\, \delta\LL_a^\text{bosonic}\,,
\label{Lchiral}
\eeq
where now the LO Lagrangian includes the usual HEFT LO terms plus two ALP-dependent terms,
\beq
\LL^{\text{LO}}= \LL^{\text{LO}}_{\text{HEFT}} + \LL^{\text{LO}}_a
\label{Lchiral-LO}
\eeq
with
\beq
\begin{split}
\LL^{\text{LO}}_{\text{HEFT}}=& \frac{1}{2} (\derp_\mu h)(\derp^\mu h) -\dfrac{1}{4} G^a_{\mu\nu}G^{a\mu\nu}-\dfrac{1}{4}\WWd^a W^{a\mu\nu}-\dfrac{1}{4}\BBd\BBu- V (h)+\\
&-\dfrac{v^2}{4}\tr[\VL_\mu\VL^\mu]\,\F_C (h)+c_T\,v^2\tr[\T\V_\mu]\tr[\T\V^\mu]\cF_T(h)+ i\bar{Q}\slashed{D}Q+i\bar{L}\slashed{D}L+\\
&-\dfrac{v}{\sqrt2}\left(\bar{Q}_L\U \mathcal{Y}_Q(h) Q_R+\hc\right)-\dfrac{v}{\sqrt2}\left(\bar{L}_L\U \mathcal{Y}_L(h) L_R+\hc\right)+\\
&-\dfrac{g_s^2}{16\pi^2}\theta G^\alpha_{\mu\nu}\tilde{G}^{\alpha\mu\nu}\,,
\end{split}
\label{Eq:SMLO}
\eeq
where the dependence on $x$,   
as well as that on $v$ of $\cF(h/v)$, has been left implicit for  brevity.  The first line in Eq.~(\ref{Eq:SMLO}) accounts for the $h$ and gauge boson kinetic terms,  and a general scalar potential $V(h)$. The  first term in the second line describes the $W$ and $Z$ 
masses and their interactions with $h$, as well as the kinetic energy of their longitudinal components;  the second term in this line is a custodial-breaking term that we will disregard in what follows, being phenomenologically extremely suppressed (for this reason sometimes it is included instead among the NLO chiral terms even if it is a two-derivative coupling). The fermion kinetic energy and Yukawa-like terms written in the mass eigenstate basis come next, with 
\beq
\mathcal{Y}_{Q,L}(h)\equiv\cY_{Q,L}\cF_{Q,L}(h)\,,
\label{curly_Y}
\eeq
 where $\cY_{Q,L}$ are the $6\times6$ block-diagonal matrices containing the usual 
Yukawa couplings as defined in Eq.~(\ref{Yukawa_matrices}).
 This notation follows the assumption that the  Yukawa-type fermion-$h$ couplings are aligned with the fermion masses.
Finally, the last line contains the usual QCD $\theta$ term associated to the strong CP problem. 

\vspace{0.3cm} 
$\LL^{\text{LO}}_a$ contains two terms which are two-derivative couplings, 
\beq
\LL^{\text{LO}}_a=\frac{1}{2} (\derp_\mu a)(\derp^\mu a)+c_{2D}\A_{2D}(h)\,,
\label{La_LO}
\eeq
where $\A_{2D}(h)$ is a custodial breaking two-derivative operator with mass dimension three,
\beq
\A_{2D}(h)=iv^2\tr[\T\V_\mu]\de^\mu\frac{a}{f_a}\F_{2D}(h)\,.
\label{Eq:A2D}
\eeq
This operator appears then singled out at the LO in the chiral expansion, unlike the case of the linear expansion in which the only LO ALP 
term was the $a$ kinetic energy, see  Eq.~(\ref{Lbosonic-lin}) and Table~\ref{comparison-2point}.  In other words, 
if the EWSB is non-linearly realized $\A_{2D}(h)$ may well provide the dominant and distinctive signals.  
It induces a two-point function  of the form $Z^\mu \derp^\mu a$  which contributes to the longitudinal component of the $Z$ boson together with the usual 
would-be Nambu-Goldstone boson of the SM, and thus to the $Z$ mass.  
 Its  impact is in this respect analogous to that of the two-point function stemming  from the  $d=5$ NLO linear operator $\O_{a\Phi}$, see 
 Sect.~\ref{Sect:2pointF_linear} and Eq.~(\ref{OaPhi}).   
 Nevertheless, it will be shown  in Sects.~\ref{Sect:2pointF} and \ref{linvsnonlin} that  $\A_{2D}$ has additional physical consequences distinct from those induced by  
  $\O_{a\Phi}$, as illustrated  in Table~\ref{comparison-2point}.  

 \begin{table}[h!]
\begin{center}
\begin{tabular}{m{3.7cm}  C{2.7cm} C{2.7cm} C{2.7cm} C{2.7cm}}
& \input{Fdiagrams/apsipsi}& \input{Fdiagrams/apsipsih} & \input{Fdiagrams/Zha} & \input{Fdiagrams/Zhhanarrow} \\
&$\sim  {\bf Y}_\psi^\a \gamma_5 \sigma^3$&$\sim {\bf Y}_\psi^\a \gamma_5 \sigma^3$& $\sim  p_\m^a$ & $\sim p_\m^a$ \\ [.3cm]\hline 
&&&\\ [-.5cm]
Linear @ NLO ($d=5$)&   $\O_{a\Phi}$ &   $\O_{a\Phi}$ &  -- & -- \\
Chiral @ LO($2\de$) & $\A_{2D}$ &  $\A_{2D}$&  $\A_{2D}$  & $\A_{2D}$ 
\end{tabular}
\end{center}
\caption{ \it \small Couplings resulting from the bosonic axion NLO linear coupling $\O_{a\Phi}$ and from its LO chiral sibling $\A_{2D}$, as formulated 
in the Lagrangians Eqs.~\eqref{deltaLbosonic-lin} and \eqref{Complete}, respectively. Only fermionic vertices survive as physical impact from  
$\O_{a\Phi}$, as  in the linear expansion higher orders ($d\ge7$) are required for $aZh^n$ ($n\ne1$) couplings, while the latter are present in the chiral case at LO.   
For the complete Feynman rules see App.~\ref{App:feynman_rules}.
}
\label{comparison-2point}
\end{table}  
  
%\vspace{0.3cm}

\subsubsection*{A discussion of scales}
The normalization of the operators in Eqs.~(\ref{Eq:SMLO})-(\ref{Eq:A2D}) and in the NLO chiral corrections to be 
discussed below follows the Naive Dimensional Analysis (NDA) master formula for the HEFT 
Lagrangian as discussed in Refs.~\cite{Manohar:1983md,Luty:1997fk,Cohen:1997rt,Gavela:2016bzc}. 
With this convention the gauge boson kinetic terms appear canonically normalised. 
 In addition,  
  the strongly interacting regime would correspond to operator coefficients of $\sim\mathcal{O}(1)$.
  
Furthermore, the mass parameter in front of several operators in  Eqs.~(\ref{Eq:SMLO})  and in Eq.~(\ref{Eq:A2D}) 
should be a generic scale $f$, which in specific models  is that associated  to a Nambu-Goldstone ancestry for the Higgs resonance (alike to $f_\pi$ for QCD pions),  
such that $\Lambda\le4\pi f$~\cite{Manohar:1983md}. Instead,  $v$  --the electroweak scale-- is shown as explicit mass 
parameter for bosons and fermions in  Eqs.~(\ref{Eq:SMLO}) and (\ref{Eq:A2D}), with $v< f$: this inequality is the well-known fine-tuning 
of the chiral electroweak Lagrangian, necessary to recover the correct scale of the gauge boson masses.  It reflects as well the fine-tuning 
problems of specific ``composite Higgs'' scenarios. For consistency $v$ has been then chosen as weight in all mass-related terms in those equations; 
for instance  a factor of $f^2/v^2$ is thus implicitly embedded in the definition of  the coefficient $c_{2D}$ in Eq.~(\ref{La_LO}).

The same fine-tuning is at the origin of the $\cF_i(h)$ functions being customarily written as generic polynomials in $h/v$ instead of $h/f$, see Eq.~(\ref{F_def}).  
 It can be considered that in this parametrization factors of $v/f$ have been reabsorbed in the free parameters $a_i$, $b_i$, etc. in Eq.~(\ref{F_def}). Note as well that, in principle, a function $\cF_i(h)$ can be attached to any of the operators in 
Eqs.~(\ref{Eq:SMLO})  and  (\ref{La_LO}). However, those attachments  can be redefined away  
in both Higgs and fermionic kinetic terms  at the price of 
redefining $\cF_{Q,L}(h)$~\cite{Giudice:2007fh} and $\cF_{2D}(h)$. Moreover, $\F_i(h)$ insertions in the gauge bosons kinetic terms can be avoided assuming 
that the transverse components of the gauge fields do not couple at tree level to the Higgs sector, as it has been explicitly shown in 
Refs.~\cite{Alonso:2014wta,Hierro:2015nna} for composite Higgs models~\cite{Kaplan:1983fs,Kaplan:1983sm,Banks:1984gj}.  A similar 
assumption on the ALP sector prevents from writing terms of the type $a X_{\mu\nu}\tilde X^{\mu\nu}$ at LO. 

\vspace{0.3cm}

%%%%%%%%%%%%%%%% 2.1
\subsection{The NLO ALP Operators}
  The complete list of   HEFT CP-even bosonic operators at NLO is known~\cite{Alonso:2012px,Brivio:2013pma,Brivio:2014pfa} and will not be further discussed.  We address here  the NLO bosonic chiral interactions involving one insertion of  $a/f_a$, encoded in $\delta\LL_a^\text{bosonic}$ in Eq.~(\ref{Lchiral}). The additional inclusion of fermionic 
couplings and the construction of a complete and non-redundant CP-even basis, which will turn out to be composed of a total of 32 -- bosonic and fermionic -- operator structures (including the LO axionic operator $\A_{2D}$ and assuming one flavour), 
is deferred to App.~\ref{App:FermionicCouplings}.  The NLO Lagrangian $\delta\LL_a^\text{bosonic}$ consists instead of 20 independent bosonic operator structures (disregarding in the counting the different coefficients inside the $\cF_i(h)$ functions),
\beq
\delta\LL_a^{\text{bosonic}}=\sum_{X=\tilde{B},\tilde{W},\tilde{G}}c_{X}\A_{X}+\sum_{i=1}^{17}c_i\A_i(h)\,,
\label{La_NLO}
\eeq
where
\begin{equation}
\begin{aligned}
&\begin{rcases}
\A_{\tilde{B}} =-\BBd\tilde{B}^{\mu\nu}\dfrac{a}{f_a}\\
\A_{\tilde{W}} =-\WWd^a\tilde{W}^{a\mu\nu}\dfrac{a}{f_a}\\
\A_{\tilde{G}} =-G^a_{\mu\nu}\tilde{G}^{a\mu\nu}\dfrac{a}{f_a}\\
\\
\A_1(h) = \dfrac{i}{4\pi} \tilde{B}_{\mu\nu} \tr[\T\V^\mu] \de^\nu \dfrac{a}{f_a} \,\F_1(h)\\
\A_2 (h)= \dfrac{i}{4\pi} \tr[\tilde{W}_{\mu\nu}\V^\mu] \de^\nu \dfrac{a}{f_a} \,\F_2(h)\\
\A_3(h)= \dfrac{1}{4\pi} B_{\mu\nu}\de^\mu \dfrac{a}{f_a}\de^\nu \,\F_3(h) \\
\end{rcases} \textrm{Custodial symmetry preserving} \\
\\
&\,\, \A_4 (h) = \frac{i}{(4\pi)^2} \tr[\V_\mu\V_\nu]\tr[\T\V^\mu]\de^\nu \frac{a}{f_a }\,\F_4(h)\\
&\,\, \A_5 (h) = \frac{i}{(4\pi)^2} \tr[\V_\mu\V^\mu]\tr[\T\V^\nu] \de_\nu \frac{a}{f_a}\, \F_5(h)\\
&\,\, \A_6(h)  = \frac{1}{4\pi} \tr[\T[\WWd,\V^\mu]]\de^\nu \frac{a}{f_a} \,\F_6(h)\\
&\,\, \A_7(h)  = \frac{i}{4\pi}  \tr[\T\tilde{W}_{\mu\nu}]\tr[\T\V^\mu]\de^\nu \frac{a}{f_a} \,\F_7(h)\\
&\,\, \A_8(h)  = \frac{i}{(4\pi)^2} \tr[[\V_\nu,\T]\D_\mu\V^\mu] \de^\nu \frac{a}{f_a} \,\F_8(h)\\
&\,\, \A_9(h)  = \frac{i}{(4\pi)^2} \tr[\T\V_\mu]\tr[\T\V^\mu]\tr[\T\V_\nu]\de^\nu \frac{a}{f_a} \,\F_9(h)\\
& \A_{10}(h) =\frac{1}{4\pi} \tr[\T W_{\mu\nu}]\de^\mu \frac{a}{f_a} \de^\nu \,\F_{10}(h)\\
& \A_{11}(h) = \frac{i}{(4\pi)^2} \,\tr[\T\V_\mu]\square \frac{a}{f_a}\de^\mu\,\F_{11}(h)\\
& \A_{12}(h) = \frac{i}{(4\pi)^2} \,\tr[\T\V_\mu]\de^\mu\de^\nu \frac{a}{f_a}\de_\nu \,\F_{12}(h)\\
& \A_{13}(h) = \frac{i}{(4\pi)^2} \,\tr[\T\V_\mu]\de^\mu \frac{a}{f_a} \square\,\F_{13}(h)\\
& \A_{14}(h) = \frac{i}{(4\pi)^2} \,\tr[\T\V_\mu]\de_\nu \frac{a}{f_a}\de^\mu \de^\nu \F_{14}(h)\\
& \A_{15}(h) = \frac{i}{(4\pi)^2} \,\tr[\T\V_\mu]\de^\mu \frac{a}{f_a} \de_\nu \,\F_{15}(h)\de^\nu\, \F_{15}^\prime(h)\\
& \A_{16}(h) = \frac{i}{(4\pi)^2} \,\tr[\T\V_\mu]\de_\nu \frac{a}{f_a} \de^\mu \,\F_{16}(h)\de^\nu \,\F_{16}^\prime (h) \\
& \A_{17} (h) = \frac{i}{(4\pi)^2} \,\tr[\T\V_\mu]\de^\mu\frac{\square a}{f_a}\,\F_{17}(h)\,.
\end{aligned}
\label{bosonic_basis}
\end{equation}
The requirement that all ALP couplings respect a (continuous or discrete) shift symmetry prevents the insertion of $\cF_i(h)$ functions in the  three first couplings in this list.   The first block of six operators are those invariant under custodial symmetry, 
assuming as customary  no sources 
of custodial symmetry breaking other than those  present in the SM.  

The ``penalization'' of the operator coefficients by inverse powers of $4\pi$ is a most conservative choice of their possible value, which
reflects the NDA normalization of the chiral sector~\cite{Manohar:1983md,Luty:1997fk,Cohen:1997rt,Gavela:2016bzc} in which $~\mathcal{O}(1)$ operator coefficients 
indicate the strong regime.   
A particular case is that of vertices 
involving one Higgs leg, for which  the overall amplitude will be proportional in practice to the product 
\beq
\tilde{a}_i\equiv c_i\,a_i\,,
\label{atilde}
\eeq
see Eq.~(\ref{F_def}). Given the  $f/v$ factor absorbed in the definition of $a_i$, in the strong coupling limit $\tilde{a}_i$  is expected to be somewhat smaller 
than $1$ for all $i\ne2D$. Conversely,  $\tilde{a}_{2D}$ as defined here  is expected to be larger than $1$ by a factor $~\mathcal{O}(f/v)$ in that limit, 
see the discussion at the end of Sect.~\ref{Sect:basis}. 
 Analogous reasoning applies to vertices with more than one Higgs leg.

%%%%%%%%%%%%%%%%%%%% 2.2
\subsection{Two-point functions}
\label{Sect:2pointF}
The last NLO operator in Eq.~(\ref{bosonic_basis}), $\A_{17}(h)$, introduces a $Z$-$a$ two-point function alike to that from the LO coupling $\A_{2D}(h)$, albeit with a higher momentum dependence. That is, both operators feed derivatives of the ALP field into the longitudinal components of the $Z$ boson, in addition to the usual derivative of the SM would-be Nambu-Goldstone neutral field:   
\begin{equation}
\begin{aligned}
 c_{2D}&\A_{2D}(h)+c_{17}\A_{17}(h) \supset\\ 
 \supset& -\frac{i}{f_a}\,
 \tr\big(\T\,\left(\de_\mu\de^\mu\U\right)\U^\dag\big)\,
 \left(c_{2 D}\,v^2\,a + \frac{c_{17}}{16\pi^2}\, \Box a\right)+ \label{two-point} \\
& +
  \frac{i}{2}\,g'\,B^\mu\,
   \left\{v^2\tr\bigg(\left(\de_\mu\U\right)\,\tau_3\,\U^\dag-\U\,\tau_3 \left(\de_\mu
    \U^\dag\right)\bigg) 
    - \dfrac{2i}{f_a}\left[c_{2D}\,v^2\,\de_\mu a + \frac{c_{17}}{16\pi^2}\,\de_\mu\left(\Box a\right)\right]\right\} \\
&+ \frac{i}{2}\,g\,W^i_\mu
    \left\{v^2\tr\bigg(\left(\de^\mu\U^\dag\right)\tau^i\U-\U^\dag\tau^i \left(\de^\mu\U\right)\bigg) 
    + \frac{i}{f_a}\left[c_{2D}\,v^2\,\de_\mu a
    + \frac{c_{17}}{16\pi^2}\,\de_\mu\left(\Box a\right)\right]\tr\left(\T\,\tau^i\right)\right\}\,.
\end{aligned}
\end{equation}
The physical impact can be illustrated best  via a field redefinition which trades completely 
this combination of two-point functions by interaction vertices, alike to the procedure applied to the linear operator $\O_{a\Phi}$  in Sect.~\ref{Sect:2pointF_linear},
\beq
\U(x) \to \U(x) \,\exp\left\{\frac{2i}{ f_a} \left(c_{2D}\, a(x)+c_{17}\frac{1}{16\pi^2 v^2}\,\square a(x)\right) \s^3\right\}\,,
\label{Eq:Uredef}
\eeq
 which translates also in contributions to the definition of the gauge fixing terms, the mass term for the gauge bosons and  the 
Yukawa couplings (see also Ref.~\cite{Gavela:2014vra} for a similar 
discussion in the context of CP-odd effective operators within non-linearly realised EWSB). The net physical impact is: 
\begin{itemize}
\item  The introduction of new fermionic couplings, alike to those  fully equivalent  in the linear case to the  bosonic  operator $\O_{a\Phi}$, see  Eqs.~(\ref{Oaphiequiv})-(\ref{deltaLbosonic-lin}).
\item The presence in addition of 
$aZh$ and other vertices of the form $(Z_\mu \de^\mu a) h^n,\, n\geq 1$, which are {\it not} redefined away in the non-linear case. The reason is that  the    functional dependence on $h$ of  $\cF_i(h)$ differs generically from that characteristic of the linear regime (in powers of $(v+h)^2$). 
\end{itemize}
The purely bosonic couplings cannot be thus completely traded by fermionic ones in the generic case of non-linear EWSB.  
This is remarkable, as it implies that $aZh$ couplings could be then expected among the dominant signals of ALPs, at variance with 
linear realizations in which they are only expected at NNLO (as argued in Sect.~\ref{linvsnonlin} below). This comparison is illustrated in Table~\ref{comparison-2point}.

The fermionic couplings, 
stemming from  $\A_{2D}(h)$ and $\A_{17}(h)$ after the field redefinition discussed,    
 will be denoted by $\A^\psi_{2D}$ and $\A^\psi_{17}$ and defined as: 
\beq
\begin{aligned}
\A^\psi_{2D}&=-i\sqrt{2}v \, \frac{a}{f_a}\,\sum_{\psi=Q,L}\left(\bar\psi_{L}\mathcal{Y}_\psi(h) \U \sigma^3 \psi_{R}\right) +\hc\,,\\
\A^\psi_{17}&=-\dfrac{i\sqrt{2}v}{16\pi^2}\, \frac{\Box a}{v^2 f_a}\,\sum_{\psi=Q,L}\left(\bar\psi_{L}\mathcal{Y}_\psi(h) \U \sigma^3 \psi_{R}\right)+\hc\,,
\end{aligned}
\label{2-point-psi}
\eeq
see Eq.~(\ref{curly_Y}) and App.~\ref{App:field_redef} for details. These expressions   
are the non-linear equivalent of the linear interaction $\O_{a \Phi}^\psi$ in Eq.~(\ref{Axion-Yukawa-compact}).  
 Alternatively, the part of $\A_{2D}$ and $\A_{17}$ that can be traded by fermionic couplings could be written as chirality-conserving transitions, e.g.
\begin{equation}
\begin{aligned}
 \A^\psi_{2D} &\rightarrow  \frac{\de_\mu a}{f_a}\sum_{ \psi=Q,\,L}\left(\bar\psi \g^\mu\g_5\sigma^3 \psi\right)\F_\psi(h)\,,\\
\A^\psi_{17} &\rightarrow \frac{1}{16\pi^2 v^2}\left(\frac{\de_\mu \Box a}{ f_a}\right)\sum_{ \psi=Q,\,L} \left(\bar\psi \g^\mu\g_5\sigma^3 \psi\right)\F_\psi(h)\,, 
\end{aligned}
\label{Apsi_2D_da}
\end{equation} 
 which are the chiral equivalent of Eq.~(\ref{Axion-fermion-from-bosonic}).  In this work, when analyzing the non-linear EWSB scenario we will use the 
 formulation of chirality-flipping fermionic couplings in Eq.~(\ref{2-point-psi}).\footnote{The Feynman rules for all bosonic and fermionic vertices 
 stemming from $\A_{2D}(h)$ and $\A_{17}(h)$, up to four-field couplings, can be found in~\ref{FR.Zah} - \ref{FR.aeeh} and~\ref{FR.add} - \ref{FR.aee} 
 of App.~\ref{App:feynman_rules}, respectively.}

\subsection{The  bosonic  chiral ALP basis}

In summary, the resulting bosonic ALP Lagrangian up to NLO couplings can  be written, after the redefinition in Eq.~(\ref{Eq:Uredef}), 
as the sum of 23 terms, besides the kinetic term:
\beq
\LL_a^\text{chiral}=\frac{1}{2}(\partial_\mu a)(\partial^\mu a)+c_{2D}\A'_{2D}(h)+\sum_{X=\tilde{B},\tilde{W},\tilde{G}}c_X \A_X + 
\sum_{i=1}^{16}c_i\A_i+c_{17}\A'_{17}(h)+\sum_{i=2D,17}c_{i}\A_i^{\psi}\,,
\label{Complete}
\eeq
where $\A'_{2D}(h)$ and $\A'_{17}(h)$ are defined as the  operators $\A_{2D}(h)$ and $\A_{17}(h)$ without their $h$-independent 
terms, which have been traded instead by
the fermionic $\A_i^{\psi}$ couplings as defined in Eq.~(\ref{2-point-psi}). 
The rest of the operators have been defined in Eq.~(\ref{bosonic_basis}). All Feynman rules  
stemming from $\LL_a^\text{chiral}$ can be found in App.~\ref{App:feynman_rules}, up to four-leg interactions. 
 
$\A_{\tilde{B}}$,  $\A_{\tilde{W}}$, $\A_{\tilde{G}}$ and $\A_{2D}^\psi$  
  are identical to the operators found in the framework of the  linear EWSB Lagrangian. In consequence, the bounds on ALP-photon and  ALP-gluon vertices in Eqs.~\eqref{agg_constraint} and  \eqref{agg_constraint_kev} apply. This would also hold, restricted to the indicated mass ranges,  for the $aW^+W^-$ coupling in Eq.~(\ref{rarebound}), if $\A_{\tilde W}$ was considered just by itself. Nevertheless, the caveats to that approach discussed in the linear case are even stronger here in the sense that the $aW^+W^-$ couplings may receive contributions in the non-linear case from the set (see \ref{FR.WWa})
 \begin{equation}\label{ja}
 \{c_{\tilde W}, c_2, c_6, c_8\}\,.
 \end{equation}
 Analogously, the ALP-fermion vertices in Eqs.~(\ref{a-f-2})-(\ref{boundaphifermions}) would constrain the magnitude of $\A_{2D}$, if the latter is taken just by itself, to 
 \beq
 \label{boundc2Dfermions}
|c_{2D}|/f_a\,<\,\unit[\left(1.7\cdot 10^{-8} - 1.4\cdot10^{-6}\right)]{GeV^{-1}} \quad(90\cl)   \quad \text{for}\quad m_a \lesssim \unit[3]{GeV}\,.
\eeq 
Again, in the non-linear EWSB setup many other couplings may contribute in addition %--at one loop level-- 
to rare meson decay processes than in the linear case, see \ref{FR.auu}-\ref{FR.aee}, to wit
 \begin{equation}\label{ayayaya}
 \{c_{2D}, c_{\tilde W}, c_2, c_6, c_8, c_{17}, \{c_{\B^q_i}\}\}\,,
 \end{equation}
 In this ensemble, the subset $\{c_{\B^q_i}\}$ of operator coefficients refers to the flavour-changing operators of the general ALP-fermion 
 couplings $\B^{q}_i$ in the complete Lagrangian, see Eq.~(\ref{fermionicoperators}), which can contribute either at tree-level or at one loop via $W$, $Z$ or $h$ exchange, and thus on the same footing than for instance $c_{\tilde W}$ or $c_2$, $c_6$, $c_8$ and $c_{17}$.
 Even if the data analysis was restricted for simplicity to bosonic couplings (the focus of this work), a six-dimensional parameter space would still remain, which means that a large freedom remains for the possible value of one given coupling. In consequence,  
 consistently with the complementarity perspective, in the second -phenomenological- part of this work we will explore  the independent impact that the bosonic non-linear operator coefficients in Eq.~(\ref{ayayaya}) may have on LHC signals, which they impact via a different combination than in rare decays. Those couplings will be thus considered   there first one at a time and occasionally in some combinations.

\section{Linear vs non-linear expansions}
\label{linvsnonlin}
The results in the previous sections on bosonic ALP-SM interactions uncovered a plethora of effective couplings in the bosonic sector of the 
chiral expansion, in contrast with the mere four  operator structures  of the linear one shown in Eq.~(\ref{general-NLOLag-lin}), when both Lagrangians are considered up to NLO. 
All ALP couplings are NLO ones in the linear case, while one of the chiral set ($\A_{2D}$) stands out at LO.

Three operators are exactly the same in both expansions. They are those with an ``anomalous-type'' structure of the form  $a X_{\mu\nu} \tilde X^{\mu\nu}$,  where $X_{\mu\nu}$ stands for a SM field strength: $\A_{\tilde{B}}, \A_{\tilde{W}}$ and $\A_{\tilde{G}}$.  The total number of independent interactions has to be equal in both expansions when all orders are considered, though. It is thus pertinent to  identify which are the effective operators of the linear expansion that lead to the same  interaction vertices than the  
 chiral (up to) NLO couplings. This is accomplished in App.~\ref{App:linear_siblings}, which identifies the linear siblings with mass dimension: 
\begin{itemize}
\item  $d=5$, corresponding to $\A_{\tilde{B}}, \A_{\tilde{W}}$ and $\A_{\tilde{G}}$ and to the fermionic couplings induced by 
$\A_{2D}$ with no attached Higgs leg (these are identical in both expansions),  as well as other fermionic vertices.
\item $d=7$, corresponding to $\A_1$-$\A_6$, $\A_8$,  $\A_{10}$-$\A_{12}$ and  $\A_{15}$-$\A_{17}$.
\item $d=9$, corresponding to $\A_7$,   $\A_{13}$ and $\A_{14}$. \item $d=11$, corresponding to $\A_{9}$.
\end{itemize}
Furthermore, the  siblings of the vertices induced by $\A_{2D}$ with one or more Higgs legs are  linear effective operators with dimension $d=7$~\cite{Bauer:2016zfj} or 
 higher, depending on their Lorentz structure.
 
\subsubsection*{Common/distinctive phenomenological signals}
 
 Interaction vertices predicted by both expansions include the well-known ALP-photon and ALP-gluon couplings, and in addition the yet mainly unexplored $a\gamma Z$, $aZZ$,  $aW^+W^-$, $a\gamma W^+W^-$ and $aZW^+W^-$ signals. 
 
 Distinctive signals are those only present in the chiral EWSB Lagrangian at the order considered, which are:  i)  an extra ALP-gauge boson vertex, $aZZZ$, and new Lorentz structures in others such as $aZZ$, $aW^+W^-$, $a\gamma W^+W^-$  and $aZW^+W^-$  ; ii)  ALP-Higgs interactions stemming from $\A_{2D}$, which include $a\gamma h$, $aZh$, $a\gamma Zh$, $aZZh$, $aW^+W^-h$, $a\gamma h h$ and  $aZh h$ interactions,  among others.  All these signals are thus putatively important pointers of non-linear realizations of EWSB.

A natural question about the bosonic ALP-Higgs interactions is how come those $(Z_\mu \de^\mu a) h^n$ couplings with $n\geq 1$  appear at LO in the  non-linear expansion while they are instead very suppressed in the linear one, as after all the latter is a limit of the former. The gist lies in the generality of the $\F_i(h)$ functions, and more specifically in the difference between $\cF_C(h)$ and $\cF_{2D}(h)$, see Eqs.~(\ref{F_def}),    
 (\ref{Eq:SMLO}) and (\ref{Eq:A2D}). Would those two functions be equal, as it happens in the linear expansion, all bosonic ALP vertices involving the Higgs would also be redefined away completely  in the chiral expansion at LO and NLO. Furthermore, even if  the difference between the $a_i$, $b_i$ etc.  coefficients for those two $\cF_i(h)$ functions was considered to be qualitatively  a NLO effect, all $(Z_\mu \de^\mu a) h^n,\, n\geq 1$ couplings would still be phenomenologically considered NLO effects, which means in any case higher strength expected than in linear realizations of EWSB (where they start to appear only at NNLO). 

The phenomenology of the ALP couplings to heavy SM bosons will be explored in  Sects.~\ref{Sect:PhenoAnalysisPresent} and \ref{Sect:PhenoAnalysisFuture}  below.

%%%%%%%%%%%%%%%%%%%%%%%%%%%%%%%%%%%%%%%%%%%%%%%%%%
%%%%%%%%%%%%%%%%%%%%%%%%%%%%%%%%%%%%%%%%%%%%%%%%%% HERE GO THE 2 SECTIONS OF PHENO
%%%%%%%%%%%%%%%%%%%%%%%%%%%%%%%%%%%%%%%%%%%%%%%%%%

\section{Assumptions and Validity of the EFT}
\label{Sect:Validity}
The theoretical results in the previous sections focused on a generic Nambu-Goldstone boson,  singlet under the SM,  identifying all bosonic derivative couplings at 
LO  in the linear and chiral expansions (a complete set including fermionic ones was also derived and for the chiral case they can be found 
in App.~\ref{App:FermionicCouplings}). They hold independently of whether the --unknown-- underlying global symmetry is exact or slightly 
and explicitly broken, that is of whether the ALP is indeed exactly massless or not, as far as its mass is negligible compared to the typical momenta considered.
A few considerations are nevertheless in order before moving to the phenomenological analysis of ALPs signatures at colliders.

\subsubsection*{Validity of the EFT} 

For the effective Lagrangian description to be valid, the relevant suppression scale, in this case $f_a$, must be significantly larger than the typical 
energy scale of the process under study. 
In order to strictly ensure the validity of the EFT, one should require 
$\sqrt{\hat{s}} < f_a$ for each event ($\sqrt{\hat{s}}$ corresponding to the invariant mass of the event).
However, $\sqrt{\hat{s}}$ is not experimentally observable in processes with invisible particles in the final state.
In this case, the comparison to $f_a$ may be naively performed using either the missing transverse energy of a given event 
$\ETmiss$ or the transverse mass $m_T$, defined as (in events characterized by the presence of a lepton and significant $\ETmiss$)
\begin{equation}
m_T^2  = 2 p_T^{\ell} \slashed{E}_T(1-\cos\phi) ,
\end{equation}
where $p_T^{\ell}$ is the transverse momentum of the lepton and $\phi$ is the azimuthal angle between the lepton  
 and the 
missing transverse momentum vector $\vec{\ETmiss}$ (note that $m_T$ encompasses contributions from both the visible and invisible 
parts of the final state). We use these two variables in the analysis below, depending on the process, and require that 
the maximum values allowed for those variables obey

\begin{itemize}

\item $m_T^\text{max}<f_a$ for mono-$W$ analyses (see Sect.~\ref{Sect:monoW_monoZ}), as the ATLAS search we reinterpret uses $m_T$ as discriminating variable.  
$m_T^\text{max}$ corresponds to the highest $m_T$ data bin 
in a given analysis, for each value of $f_a$ considered.

\item $2\ETmiss^\text{max}<f_a$ for all other processes analyzed.  
$\ETmiss^\text{max}$ is the highest $\ETmiss$ data bin in a given analysis for each value of $f_a$ considered.
\end{itemize}

The effect of imposing the strict validity criterium $\sqrt{\hat{s}} < f_a$ can be assessed through the correlation between $\ETmiss,\,m_T$ and $\sqrt{\hat{s}}$ 
for each analyzed signal, obtained from Monte Carlo. For binned analyses, the signal event fraction for 
which $\sqrt{\hat{s}}> m_T^\text{max},\,\ETmiss^\text{max}$ in different bins may then be discarded. We will explicitly use this procedure 
for the mono-$W$ and mono-$Z$ analyses in Sects.~\ref{Sect:monoW_monoZ} and~\ref{Sect:monoWZ_Future}, 
and discuss the impact of the strict validity criterium on the bounds/sensitivities on $f_a/c_i$ obtained from the rest of analyses. \footnote{
See also Ref.~\cite{Pobbe:2017wrj}, where a similar method has been applied to DM searches with the added feature of marginalizing over the unknown contribution of new physics beyond the cutoff.}

On a different note, we stress that as the chiral expansion has an implicit BSM electroweak scale $\Lambda\le 4\pi f$, there is an underlying assumption 
that $f_a\ge \Lambda$. This $\Lambda/f_a$ hierarchy sustains the choice of restraining the analysis to vertices involving only one ALP.

\subsubsection*{ALP stability at the LHC and its mass} 
In the LHC phenomenological exploration to follow, it will be assumed that the ALP is stable on collider scales, thus escaping  
the detector as missing transverse energy $\ETmiss$. This further restricts the range of values of  $m_a$, $f_a$, 
appropiate for the concrete numerical analysis below, given the various interactions of $a$ that could allow its decay -- see Eqs.~(\ref{FR.AAa}) - (\ref{FR.Zah})
and~(\ref{FR.auu}) - (\ref{FR.aee}) in App.~\ref{App:feynman_rules}.  The valid $m_a$ range  should be specified for a correct interpretation of the collider results: 
because of the assumed stability, all phenomenological results to be obtained below hold for ALP masses $m_a\le1$ MeV, without any additional assumption about which channels may be open.  
The ratio between the ALP mass $m_a$ and $f_a$  is then safely small, $m_a/f_a\le\,$MeV/TeV, for characteristic $f_a$ scales  of at least 
a few TeV. 

For ALP masses above the MeV, the signals to be studied below may also be present even if the pattern is altered, accompanied by new ones which can be used to 
precisely test the couplings through which the ALP may decay within the detector (e.g. leptonic couplings).~\footnote{As an example, the decay 
channel $a\to e^+e^-$ can produce collimated signals of $e^+ e^-$ signals; we thank Jos Vermaseren for this comment.}  
  This would require an extended dedicated study.

 In this work,  an ALP mass $m_a\sim \unit[1]{MeV}$ is used in the numerical simulations,  light enough to avoid altogether 
$a \to \ell^{+}\ell^{-}$ and $a \to \nu\bar\nu \ell^{+}\ell^{-}$ decays. The decay channels which then remain {\it a priori} available are: 

\vspace{2mm}
  \noindent $- \quad {\uline{a\to \nu\bar{\nu}\nu\bar{\nu}}}\quad$ 
  As neutrinos are undetectable at the LHC, this decay doesn't have any impact on our phenomenological analysis. It would simply become part of the $\ETmiss$ contributions.
  
  \vspace{2mm}
 \noindent $- \quad {\uline{a\to\gamma\gamma}}\quad$   This decay is constrained by astrophysical observations, as detailed at the 
 end of Sect.~\ref{Sect:2pointF_linear}. The distance $d$ covered in the laboratory 
 frame by an ALP before decaying can be estimated as 
 \begin{equation}
  d = \tau \beta c = \frac{\hbar}{\Gamma(a)}\frac{|\vec{p}_a|}{m_a}\, c\,,
 \end{equation} 
 where $\tau$, $\Gamma(a)$ and $\vec{p}_a$ are, respectively, the proper lifetime, width and three-momentum of the $a$ particle, and $c$ denotes the speed of light. 
 Restricting the width to $\Gamma(a\to\gamma\gamma)$ and using the coupling strength $g_{a\gamma\gamma}$ as defined in Eq.~\eqref{gagammagamma-def},  it follows that
 \begin{equation}
  d = \frac{16\pi \hbar c}{m_a^4}\frac{1}{g_{a\gamma\gamma}^2}|\vec{p}_a|\,,
 \end{equation} 
which can be rewritten as
 \beq
 d\simeq10^8 \left( \frac{\unit[]{MeV}}{m_a}\right)^4 \left(\frac{\unit[10^{-5}]{GeV^{-1}}}{g_{a\gamma\gamma}}\right)^2 \left(\frac{|p_a|}{\unit{GeV}}\right)\unit[]{m}. 
\eeq
   For $m_a=\unit[1]{MeV}$, given the experimental constraint (see Eqs.~\eqref{def_agg}-\eqref{agg_constraint}), it results
 \begin{equation}
 d> 4 \cdot \unit[10^8]{m} \times \left(\frac{|\vec{p}_a|}{\unit{GeV}}\right) \,.  
 \end{equation} 
  The ALP momentum $|\vec{p}_a|$ is typically of the order of the missing energy of the candidate signals, selected imposing a minimum $\ETmiss$ cut, which for instance 
  using ATLAS and CMS data is  $\gtrsim {\cal O}( 100)$ GeV.  Thus, within the allowed range for $g_{a\gamma\gamma}$ and $\ETmiss$, the ALP always covers an enormous 
  distance -- many orders of magnitude larger than the LHC detectors size ($\sim \unit[10]{m}$) -- before decaying into two photons. For lighter ALPs, the situation 
  is even  safer given the inverse quartic dependence of $d$ with $m_a$.  ALP masses above the MeV range and up to hundreds of MeV could be considered without risking 
  two-photon ALP decay in the data analyzed~\footnote{Near the GeV range and further up there are barely any constraints~\cite{Mimasu:2014nea}  on 
  the value of $g_{a\gamma\gamma}$. A more elaborate study of the ALP signals involving SM gauge bosons could be pertinent for that scenario, allowing for the 
  corresponding decay channels to be taken into account. This is beyond the scope of this work.} by raising the minimum $\ETmiss$ cut imposed on data, but this 
  would open the $e^+\,e^-$ leptonic decay channels. 
 
 \vspace{2mm}
  \noindent $- \quad {\uline{a\to\gamma\nu\bar{\nu}}}\quad$ Analogously, this process does not affect the stability of the ALP particle at the LHC.  It could be mediated by the ALP-$Z$-$\gamma$ interaction parametrized by $g_{aZ\gamma}$,
 \beq
 \delta\LL_a \supset  - \frac{1}{4}g_{aZ\gamma} \,a\,F_{\m\n}\tilde{Z}^{\m\n}\,,
 \label{gaZgamma-def}
 \eeq
 where $Z^{\m\n}$ denotes the $Z$-boson field strength.  
  The decay width shows a very strong dependence on the mass of the ALP, due to a 
 peculiar cancellation occurring in the phase space integration. In the limit $m_a\ll m_Z$ (and neglecting  
  the $Z$ boson width for simplicity) 
 we find
\begin{equation}
\Gamma(a\to\gamma\nu\bar{\nu}) = \frac{g^2\,g_{aZ\gamma}^2\,m_Z^3}{1024\, (2\pi)^3\,c_{\theta}^2} \times \left( \frac{13}{20}\frac{m_a^7}{m_Z^7} + \mathcal{O}(m_a^{9}/m_Z^{9})\right)\,.
\end{equation} 
For $m_a=\unit[1]{MeV}$,  this corresponds to a distance covered by the ALP before decaying
\begin{equation}
d\simeq \unit[10^{22}]{m} \times \left(\frac{|\vec{p}_a|/g_{aZ\gamma}^2}{\unit{GeV^3}}\right)>  3.3 \cdot \unit[10^{27}]{m} \times \left(\frac{|\vec{p}_a|}{\unit{GeV}}\right)\,,
\end{equation} 
 where  on the last inequality the constraint on $g_{aZ\gamma}$ derived   further below has been used (see Eq.~\eqref{aZg_constraint}).

 \begin{table}[t!]\centering
\hspace*{-11mm}
\begin{tabular}{|c|l >{$}l<{$}|*2{>{$}p{2mm}<{$}}>{$}l<{$}| *{2}{>{$}p{2mm}<{$}}>{$}p{3.5mm}<{$}*3{>{$}p{1.5mm}<{$}}*3{>{$}p{1.5mm}<{$}}>{$}p{3mm}<{$}>{$}p{8mm}<{$}>{$}p{3.5mm}<{$} |}
\cline{2-18}
\multicolumn{1}{c|}{}&\multicolumn{2}{c|}{\multirow{2}{*}{\bf Observables/Processes}}& \multicolumn{15}{c|}{\bf Parameters contributing}\\
\cline{4-18}
\multicolumn{1}{c|}{}& &  &\multicolumn{3}{c|}{\bf Linear} & \multicolumn{12}{c|}{\bf Non-Linear}\\
\hline
&Astrophysical obs.& g_{a\gamma\gamma} & 
\bf{c_{\tilde{W}}}& \bf{c_{\tilde{B}}}& &
\bf{c_{\tilde{W}}}& \bf{c_{\tilde{B}}}& & & & & & & & &  & \\

&Rare meson decays&  & 
\bf{c_{\tilde{W}}}& &\bf{c_{a\Phi}} &
\bf{c_{\tilde{W}}}& &\bf{c_{2D}} & &c_2 & &c_6 & &c_8 & &  &c_{17} \\

\hline
\parbox[t]{4mm}{\multirow{5}{4mm}{\rotatebox[origin=r]{90}{\bf New constraints}}}& 
\multicolumn{2}{l|}{\bf LEP data} & \multicolumn{3}{l|}{}& \multicolumn{12}{l|}{}\\
&BSM $Z$ width&  \Gamma(Z \to a\gamma)  &   
\bf{c_{\tilde{W}}}& \bf{c_{\tilde{B}}}& &
\bf{c_{\tilde{W}}}& \bf{c_{\tilde{B}}}& &  c_1& c_2& & & c_7 & &  & & \\
& \multicolumn{2}{l|}{\bf LHC processes} & \multicolumn{3}{l|}{}& \multicolumn{12}{l|}{}\\
& Non-standard $h$ decays & \Gamma(h \to a Z)&
& & &
& & \bf{\at_{2D}} & & & \at_{3} & & & & \at_{10} & \at_{11-14}& \at_{17}\\
& Mono-$Z$ prod.& pp\to a\,Z&
\bf{c_{\tilde{W}}}& \bf{c_{\tilde{B}}}& c_{a\Phi}&
\bf{c_{\tilde{W}}}& \bf{c_{\tilde{B}}}& c_{2D}& c_1& c_2& c_3& &c_7& &c_{10}& c_{11-14}& c_{17}\\	
& Mono-$W$ prod.& pp\to a\,W^{\pm}&
\bf{c_{\tilde{W}}}& c_{\tilde{B}}& c_{a\Phi}&
\bf{c_{\tilde{W}}}& c_{\tilde{B}}& c_{2D}& &c_2& &\bf{c_6}& &c_8& c_{10}&  & \\
\hline
\parbox[t]{3mm}{\multirow{3}{3mm}{\rotatebox[origin=c]{90}{\bf Prospects}}}& 
Associated prod.& pp\to a W^\pm \gamma &
\bf{c_{\tilde{W}}}& c_{\tilde{B}} &c_{a\Phi}&
\bf{c_{\tilde{W}}}& c_{\tilde{B}}& c_{2D}& c_1 & c_2 & &\bf{c_6}& c_7& c_8&  & & \\
&VBF prod.&  pp\to a j j (\gamma) &
c_{\tilde{W}}& c_{\tilde{B}}& c_{a\Phi}&
c_{\tilde{W}}& c_{\tilde{B}}& c_{2D}& c_1& c_2&  &c_6& c_7& c_8&  & & \\
&Mono-$h$ prod.& pp\to h\,a&
& & & 
& & \bf{\at_{2D}}& & &\bf{\at_{3}}& & & &\bf{\at_{10}}& \at_{11-14}& \at_{17} \\
&$at\bar{t}$ prod.&  pp\to a t\bar{t} &
& & \mathbf{c_{a\Phi}}&
& & \mathbf{c_{2D}}& & &  &&&&  & & \\
\hline
\end{tabular}
 \caption{ \it \small Couplings contributing to the observables considered here, stemming from the purely bosonic operators  in the linear and non-linear scenarios. 
 The block of new constraints 
 explores the sensitivity of LEP and present LHC data to different operators, see Sect.~\ref{Sect:PhenoAnalysisPresent}. The last block 
 corresponds instead to the sensitivity analysis 
 from Sect.~\ref{Sect:PhenoAnalysisFuture}, which assumes both LHC prospects with $300$ fb$^{-1}$ of data 
 and projections to the HL-LHC phase with $3000$ fb$^{-1}$ of data. 
  The operator coefficients to which present or expected measurements are found to be sensitive appear in bold. 
 }\label{tab:contributions}
\end{table} 
%%%
%%%%%%%%%%%%%%%%%%%%%%%%%% 3. 
%%%
\section{Phenomenological Analysis I: New Bounds}
\label{Sect:PhenoAnalysisPresent}

In this section we derive new constraints on the operator coefficients using LEP and  LHC Run I and II  data. Table~\ref{tab:contributions} summarizes the 
observables/processes which are sensitive to the various effective operator coefficients, to be considered in this and the next section.

Unless otherwise specified,  we will consider the effect of one operator  at a time. Note that the   dependence of the signal cross section $\sigma$ or partial 
width $\Gamma$ on an  operator coefficient $c_i$ is $(c_i/f_a)^2$, hence the ratio $c_i/f_a$ is the relevant combination of parameters throughout the analysis.

For the operator coefficients we will use the notation of the chiral expansion,  as its couplings outnumber and include those of the 
linear expansion -- see Sect.~\ref{Sect:basis}.  Whenever pertinent, the applicability of a given bound or a sensitivity prospect to both expansions will be specified.
   Special attention will be paid overall to the comparison between the expectations based on the linear and non-linear effective Lagrangians.

\subsection{ALP coupling to \tpdf{$Z$}{Z}-photon}
\label{Sect:aAZ}
\begin{figure}[t]\centering
\includegraphics[width=\textwidth]{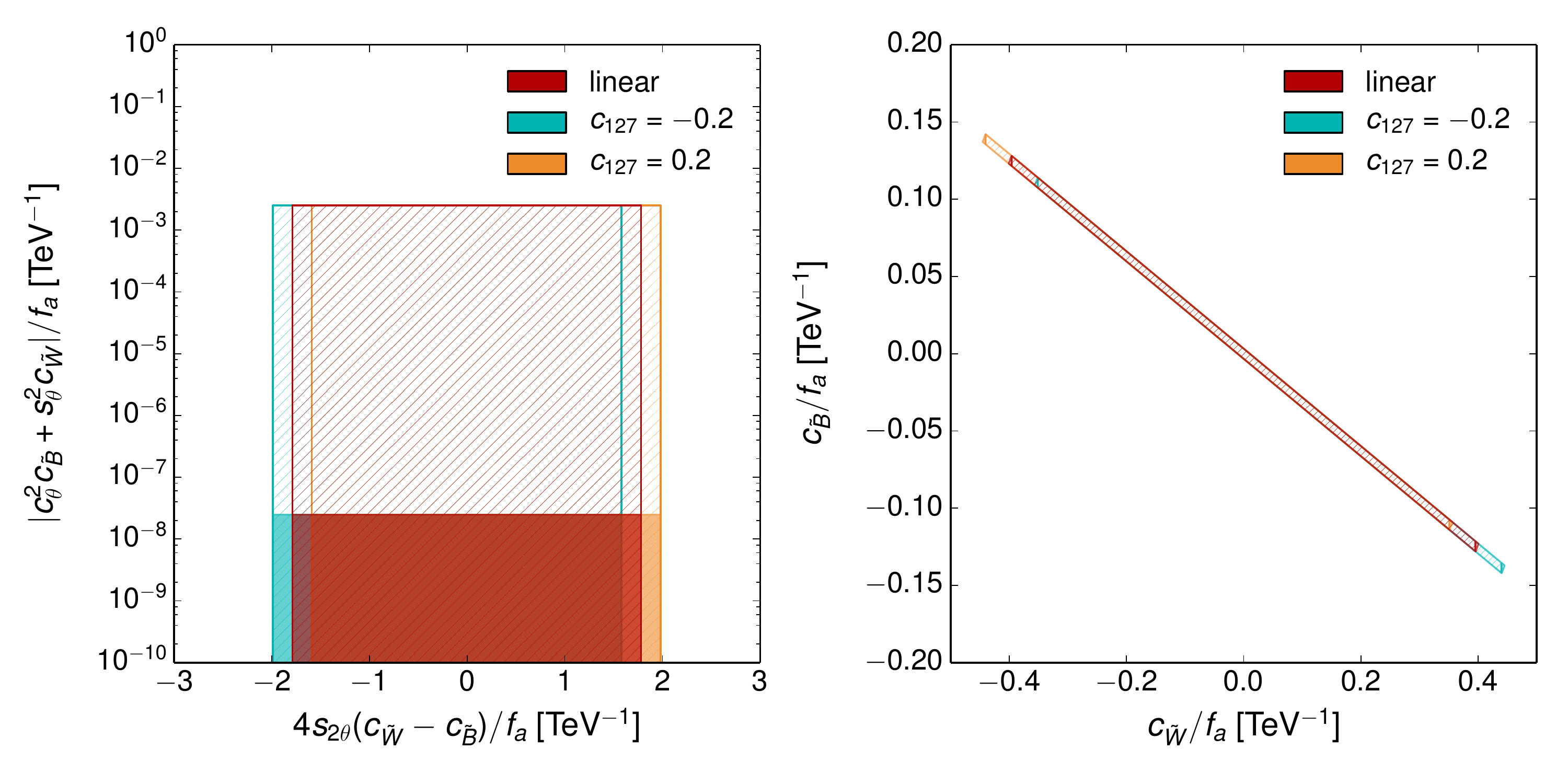}

\vspace{-4mm}

\caption{\it \small Left: Constraints on the parameters $c_{\tilde{B}}/f_a$ and $c_{\tilde{W}}/f_a$ derived from  the tree-level bounds on the combinations 
$g_{a\g\g}$ ($y$-axis) and $g_{aZ\g}$ ($x$-axis) defined in Eqs.~\eqref{def_agg} and~\eqref{def_azg}. The hatched (solid) region is obtained with 
the benchmark mass $m_a\simeq\unit[1]{MeV\, (keV)}$. The different colors show how the allowed region is shifted in the non-linear setup, 
depending on the parameter $c_{127} = \frac{g}{4\pi}(2c_1+t_\theta(c_2+2c_7))$. The value $c_{127}=0.2$ is about maximal, as it is obtained 
fixing $c_1=c_2=c_7=1$ typical of the strongly interacting regime. The linear case corresponds to $c_{127}=0$. Right: the 
rotated figure shows only the region allowed for $m_a=\unit[1]{MeV}$.
}
\label{Fig:cWcB_allowed} 
\end{figure}
In the non-linear expansion, the effective  $aZ\gamma$ coupling
\beq
 \delta\LL_a \supset  - \frac{1}{4}g_{aZ\gamma} \,a\,Z_{\m\n}\tilde{F}^{\m\n}\,
 \label{gaZgamma-def}
 \eeq
 takes the form:
\begin{equation}\label{def_azg}
g_{aZ\g} =f_a^{-1}\left[4\sdt(c_{\tilde{W}}-c_{\tilde{B}})+\frac{g}{4\pi}\left(2c_1+t_\theta(c_2+2c_7)\right) \right]\,,
\end{equation} 
with  the custodial-preserving limit recovered for $c_7=0$ and the linear limit at NLO recovered for $c_1=c_2=c_7=0$.~This 
interaction can be constrained from various sets of experimental data:
\begin{itemize}
  \item The uncertainty on the $Z$ boson width~\cite{Agashe:2014kda}, $\Gamma(Z\to\text{BSM}) \lesssim \unit[2]{MeV}$ at $95\cl$, 
 allows to set a conservative bound on the process $Z\to a\g$. The latter would contribute to the $Z$ width as
 \begin{equation}
  \Gamma(Z\to a \g)=\frac{M_Z^3}{384\pi}\, g_{aZ\g}^2\left(1-\frac{m_a^2}{M_Z^2}\right)^3\,.
 \end{equation} 
In consequence, we use for the first time the $Z$ boson width to obtain a bound on this coupling,
constraining the combination of coefficients in Eq.~(\ref{def_azg}) within the limit (with basically no dependence on $m_a$ for $m_a\lesssim\unit[1]{GeV}$)
\begin{equation}\label{aZg_constraint}
 |g_{aZ\g}|<\unit[1.8]{TeV^{-1}}\quad(95\cl)\,.
\end{equation}

\item LEP limits on $Z \to 3 \g$~\cite{Jaeckel:2015jla} constrain  the product $g_{a\gamma\gamma} \, g_{aZ\gamma}$.
However, given the bounds on $g_{a\gamma\gamma}$ reported at the end of Sect.~\ref{Sect:2pointF_linear}, the inferred bound on $g_{aZ\gamma}$ is weaker than 
that in Eq.~(\ref{aZg_constraint}). 

\end{itemize}
As illustrated in Fig.~\ref{Fig:cWcB_allowed}, the $Z$ boson width is able to probe regions in the parameter space orthogonal to those tested by $g_{a\gamma\gamma}$. 
In the linear EWSB setup, those two bounds constrain $c_{\tilde{B}}/f_a$ and $c_{\tilde{W}}/f_a$ to take values within a limited area:  imposing Eq.~\eqref{cWB} 
leads to $|c_\Wt /f_a|< \unit[0.42]{TeV^{-1}}$. 
In the non-linear EWSB case, that region can be shifted depending on the value taken by the 
combination $c_{127}\equiv\frac{g}{4\pi}(2c_1+t_\theta(c_2+2c_7))$, as shown in Fig.~\ref{Fig:cWcB_allowed}. 
Overall, the constraints on the quantities $c_i/f_a$ are of order $\unit{TeV^{-1}}$ and thus correspond to a loose 
$\mathcal{O}(1)$ bound on the coefficients $c_i$ for $f_a=\unit[1]{TeV}$.

\subsection{ALP coupling to \tpdf{$Z$}{Z}-Higgs: Non-standard Higgs decays }
\label{Sect:hZax_Br}
As shown in Sect.~\ref{Sect:2pointF} and App.~\ref{App:field_redef}, the presence of the coupling $aZh$ is a characteristic feature of 
the non-linear effective Lagrangian, as in the linear expansion it would only be expected at NNLO. 
We propose here for the first time to use non-standard Higgs channels to bind couplings of the ALP to the Higgs particle.
Consider a range of ALP masses such that it allows
the Higgs particle to decay into $Za$.
The presence of non-standard decay modes of the Higgs is constrained by ATLAS and CMS global 
fits to Higgs signal strengths. Current constraints on the Higgs non-standard branching fraction  $\Br (h\to {\rm BSM})$  from 
LHC $7$ and $8$ TeV data yield~\cite{Khachatryan:2016vau}
\begin{equation}
\Br (h\to {\rm BSM}) = 
\frac{\Gamma_\text{BSM}}{\Gamma_\text{BSM}+\Gamma_\text{SM}} \leq 0.34 \quad\,\, 
(95\cl)\,,
\label{H_NS_Width_ATLAS}
\end{equation}
where the SM Higgs width is $\Gamma_\text{SM} = \unit[(4.07\pm 0.16)]{MeV}$~\cite{Denner:2011mq} and 
$\Gamma_\text{BSM}$ denotes 
the non-standard Higgs partial width stemming in this case from the presence of the ALP, 
\begin{eqnarray} \label{H_NS}
 \Gamma_\text{BSM} & = & \Gamma_{h \to aZ} + \Gamma_{h \to aZ\gamma} + \Gamma_{h \to a\,\text{f}\bar{\text{f}}}\,. 
 \end{eqnarray}
The interaction vertices contributing to $\Gamma_{h \to aZ}$, $\Gamma_{h \to a\,\text{f}\bar{\text{f}}}$  and $\Gamma_{h \to aZ\gamma}$ are shown in \ref{FR.Zah}, 
 \ref{FR.auuh} - \ref{FR.aeeh} and ~\ref{FR.ahAZ} in App.~\ref{App:feynman_rules}, respectively.  The last two terms are 
three-body phase-space suppressed and yield negligible contributions to the Higgs total width; \footnote{ $\Gamma_{h \to a\,f\bar{f}}$ is 
further suppressed by factors of 
$(m_f/v)^2 \ll 1$, while the interaction $ahZ\gamma$ is linked to the  $aZ\gamma$ vertex (see~\ref{FR.ahAZ} and~\ref{FR.ZAa} in App.~\ref{App:feynman_rules}), 
whose strength is bounded from the $Z$ width (see Sect.~\ref{Sect:aAZ}).}
they will be then discarded in what follows. Using then $\Gamma_\text{BSM} \simeq \Gamma_{h \to aZ}$ in Eq.~\eqref{H_NS_Width_ATLAS}
yields the present bound
\begin{equation}\label{HtoZaWD_bound}
 \Gamma_{h \to aZ}  < \unit[2.1]{MeV} \quad\,\, 
(95\cl)\,.
\end{equation}

$\Gamma_{h \to aZ}$ receives contributions from the chiral LO operator $\A_{2D}$, Eq.~(\ref{Eq:A2D}), and from several NLO  ones in Eq.~(\ref{bosonic_basis}),  
\beq
\begin{aligned}
\Gamma_{h \to aZ} &= \frac{m_h^7}{1024\pi^5 v^4 f_a^2} \left(\left(1-\frac{m_a^2}{m_h^2}-\frac{m_Z^2}{m_h^2}\right)^2-\frac{4m_a^2m_Z^2}{m_h^4}\right)^{\nicefrac{3}{2}} 
\left(\kappa_h+\kappa_Z \frac{m_Z^2}{m_h^2}+\kappa_a \frac{m_a^2}{m_h^2} \right)^2\\
&\simeq \frac{m_h^7}{1024\pi^5 v^4 f_a^2}\left(1-\frac{m_Z^2}{m_h^2}\right)^3\left( \kappa_h+\kappa_Z \frac{m_Z^2}{m_h^2}\right)^2+\mathcal{O}(m_a/m_h)\,,
\end{aligned}
\eeq
with
\beq
\begin{aligned}
\kappa_h&=\at_{13}+\frac{1}{2}(\at_{12}-\at_{14})-\frac{2\pi\sdt}{e}(\at_3\st-\at_{10}\ct)-16\pi^2\tilde{a}_{2D}\frac{v^2}{m_h^2} \label{Hinv_bound_kh}\,,\\
\kappa_Z &= -\frac{1}{2}(\at_{12}-\at_{14})\,,\\
\kappa_a&=\at_{17}-\at_{11}+\frac{1}{2}(\at_{12}-\at_{14})+\frac{2\pi\sdt}{e}(\at_3\st-\at_{10}\ct)\,,
\end{aligned}
\eeq
and where the coefficients $\tilde{a}_i$ for the couplings involving one Higgs leg  have been defined in Eq.~(\ref{atilde}). The bound in Eq.~(\ref{HtoZaWD_bound}) 
translates into the constraint 
\begin{equation}
\label{Limit_aZh_HiggsDecay}
\dfrac{1}{f_a} \left|\kappa_h+\dfrac{m_Z^2}{m_h^2} \kappa_Z +\dfrac{m_a^2}{m_h^2} \kappa_a\right| \lesssim \unit[0.22]{GeV^{-1}}  \, 
  \longrightarrow   \frac{f_a}{\tilde{a}_{2D}} \gtrsim \unit[2.78]{TeV}\,  \quad \text{for}\quad  m_a \lesssim \unit[34]{GeV}\,,
\end{equation} 
where we use the fact that  the inequality on the left is generically dominated by the $\tilde{a}_{2D}$ contribution, as it enters weighted by a large factor. If 
the constraint in Eq.~(\ref{boundc2Dfermions}) is considered, the impact of $\A_{2D}$ on $h \to aZ$ decay is negligible for ALP masses below $3$ GeV, and in consequence 
the bound in Eq.~(\ref{HtoZaWD_bound}) would apply to the combination of  $\at_3$ and $\at_{10}$. However, present LHC sensitivity does not allow to constrain these operators.

The above limits are expected to improve significantly at the high luminosity phase of LHC (HL-LHC). For example, Ref.~\cite{ATL-PHYS-PUB-2014-016} estimates that a bound
\begin{equation}
\Br (h\to {\rm BSM}) \leq 0.1\quad\,\, (95\cl)\,,
\end{equation} 
will be reached for $\unit[3000]{fb^{-1}}$ of data at $\sqrt s=\unit[14]{TeV}$ (neglecting here theoretical uncertainties). 
This would roughly translate into a sensitivity $\Gamma_{h \to aZ}^{\unit[3]{ab^{-1}}} \lesssim \unit[0.45]{MeV}$ ($f_a/\at_{2D} \gtrsim \unit[6]{TeV}$ for the case in 
Eq.~(\ref{Limit_aZh_HiggsDecay})).

\vspace{2mm}

An alternative approach to tackle $\Gamma_{h \to aZ}$ is to use the constraints from direct searches for invisible Higgs decays, since $h \to a Z$ yields an invisible 
Higgs decay for $Z\to\nu\bar{\nu}$.
  Current experimental searches by ATLAS \cite{Aad:2015txa,Aad:2015pla} and CMS \cite{Chatrchyan:2014tja} 
constrain the branching ratio for Higgs decay into invisible states $\Br (h\to \text{inv})$ to~\cite{Aad:2015pla}
\begin{equation}
\Br (h\to \text{inv}) < 0.23 \quad\,\, 
(95\cl)\,.
\label{H_Invisible_Width_ATLAS}
\end{equation}
Nevertheless, no constraint on $\Gamma_{h \to aZ}$ follows from this present bound, since 
$\Br(Z \to \nu \bar\nu) = 0.2 \pm 0.006$~\cite{Olive:2016xmw}. In the future, given the improvement on the sensitivity to $\Br (h\to \text{inv})$ foreseen 
at HL-LHC with $\unit[3000]{fb^{-1}}$ of data at $\sqrt s=\unit[14]{TeV}$~\cite{ATL-PHYS-PUB-2013-014},
\begin{equation}
\label{H_Invisible_Width_Future}
\Br (h\to \text{inv}) < 0.08\quad\,\, (95\cl)\,,
\end{equation} 
 direct searches of the invisible decays of the Higgs resonance may be sensitive to $\Gamma_{h \to aZ}$. Indeed,    
in the ALP scenarios under discussion 
\begin{equation}
\Br (h\to \text{inv})\simeq\frac{\Gamma_{h \to a Z} \times \Br(Z \to \nu \bar\nu)}{\Gamma_{h \to a Z}+\Gamma_\text{SM}}\,,
\end{equation}
and in consequence, barring a positive signal in future data, Eq.~\eqref{H_Invisible_Width_Future} may translate into 
$\Gamma_{h \to aZ}^{\unit[3]{ab^{-1}}} \lesssim \unit[2.71]{MeV}$, setting new limits on the operator coefficients 
participating in this decay.
This expected sensitivity is however weaker than the present bound obtained from global fits to Higgs signal strengths in Eq.~(\ref{Hinv_bound_kh}), and  
the latter will be used in 
the remainder of the paper.

\subsection{Mono-W and mono-Z searches at $\sqrt{s}=\unit[13]{TeV}$}
\label{Sect:monoW_monoZ}

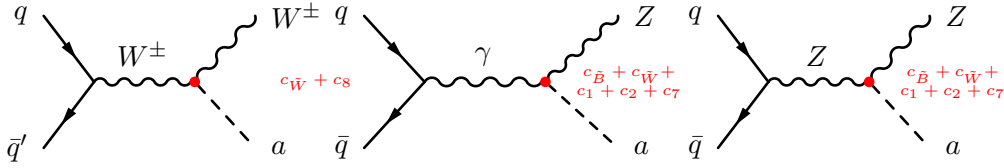
\begin{figure}[t]\centering
\input{Fdiagrams/monoZW_nonlinear}
\caption{  \it \small Feynman diagrams contributing to mono-$W$ and mono-$Z$ production.}\label{diagram_monoWZ}
\vspace{2mm}
\end{figure}

We now study the production of $a$ in association with a $W$ and a $Z$ boson, as illustrated in Fig.~\ref{diagram_monoWZ}. Since the ALP escapes the LHC 
detectors as missing transverse energy $\ETmiss$, this yields respectively 
 ``mono-$W$''~\cite{Bai:2012xg} and  ``mono-$Z$''~\cite{Bell:2012rg,Carpenter:2012rg,Alves:2015dya,No:2015xqa,Neubert:2015fka} signatures. 
 Both channels are being currently searched for by the ATLAS and CMS experimental collaborations. In this 
 section we use their studies from public Run II data to set limits on the presence of different ALP effective operators that contribute to 
 these signals. 

 \vspace{-0.2cm} 

\subsubsection*{Analysis tools}

 \vspace{-0.2cm}
 
All signals and backgrounds to be discussed below in this and the next section will be generated using  {\tt MadGraph5$\_$aMC@NLO}~\cite{Alwall:2014hca}. 
For this section it is enough to consider a parton-level analysis as the final states considered involve only leptons in addition to the ALP.
 
 \vspace{-0.2cm} 
 
\subsubsection*{Statistical tools}

 \vspace{-0.2cm} 
 
In order to set limits on $c_i/f_a$ for each effective operator,  a binned likelihood analysis will be performed. The likelihood function for 
a given lepton flavour in the final state $\ell = e, \,\mu$, 
is built as a product of bin Poisson probabilities
\begin{eqnarray}
L^{\ell}(\mu_i) = \prod_k \,e^{-(\mu_i\,s^i_k +\, b_k)}\, \frac{(\mu_i\,s^i_k + b_k)^{n_k}}{n_k !} \,,   
\label{likelihood_NS}
\end{eqnarray}
where 
\beq
\mu_i \equiv (c_i/f_a)^2
\label{mui}
\eeq
 and $b_k$ and $s^i_k$ are respectively the background prediction and the signal prediction for $c_i = 1$ and $f_a = 1$ TeV in a given  
bin $k$. The significance is estimated via the test statistic $Q^{\ell}_{\mu_i}$, 
\begin{eqnarray}
Q^{\ell}_{\mu_i} \equiv -2\, \mathrm{Log} \left[\frac{L^{\ell}(\mu_i)}{L^{\ell}(\hat{\mu_i})} \right]\,,
\label{likelihood_1}
\end{eqnarray}
with $\hat{\mu_i}$ being the value of $\mu_i$ which maximizes $L^{\ell}(\mu_i)$. Alternatively, we may include the effect of systematic uncertainties on 
the background prediction (which for the mono-$W$ searches can be obtained from Refs.~\cite{ATLAS-CONF-2015-063,Aaboud:2016zkn}  and for the mono-$Z$ 
searches from Ref.~\cite{CMS:2016yfc}) by convoluting each bin Poisson probability 
with a Gaussian prior,\footnote{The Gaussian normalization in Eq.~\eqref{likelihood_S} is consistent as long as $\sigma_i \ll 1$, which is the case in our present analysis.} 
such that the likelihood function is given by 
\begin{eqnarray}
L^{\ell}_{S}(\mu_i) = \prod_k \int_{0}^{\infty} dr\, \frac{e^{\frac{-(r-1)^2}{2\sigma_k^2}}}{\sqrt{2\pi}\sigma_k}  
\,e^{-(\mu_i\,s^i_k +\, r\, b_k)}\, \frac{(\mu_i\,s^i_k + r\,b_k)^{n_k}}{n_k !}  \,,
\label{likelihood_S}
\end{eqnarray}
with $\sigma_k$ being the background systematic uncertainty in each 
bin $k$. Our test statistic accounting for background systematic uncertainties $Q^{\ell}_{S\,\mu_i}$ is then defined as
\begin{eqnarray}
Q^{\ell}_{S\,\mu_i} = -2\, \mathrm{Log} \left[\frac{L^{\ell}_S(\mu_i)}{L^{\ell}_S(\hat{\mu_i})} \right]\,.
\label{likelihood_2}
\end{eqnarray}
The value of $\mu_i$ that can be excluded at $95\cl$  
corresponds to 
$Q^{\ell}_{\,\mu_i} = 3.84$ ($Q^{\ell}_{S\,\mu_i} = 3.84$) if background systematic uncertainties are not (are) included.

\subsubsection{Mono-\tpdf{$W$}{W} signatures: \tpdf{$pp \to a \, W^{\pm}$}{pp -> a W}}
\label{Sect:monoW-ATLAS}
 
We are targeting in this paper bosonic couplings of the ALP particle,  and here in particular ALP couplings to electroweak gauge bosons, as illustrated 
in Fig.~~\ref{diagram_monoWZ}. 
 Let us first concentrate on the ALP production in association with a $W$ boson, as illustrated in Fig.~\ref{diagram_monoWZ} (left). 
It is possible to derive limits on the coefficient of each effective
operator contributing to this process  from LHC Run II data at $\sqrt s = \unit[13]{TeV}$, 
by reinterpreting the ATLAS search for $W'$ decaying to $\ell + \ETmiss$ final states with $3.3\,\mathrm{fb}^{-1}$ of integrated 
luminosity~\cite{ATLAS-CONF-2015-063} (with $\ell = e,\,\mu$).  The backgrounds will be taken from Ref.~\cite{ATLAS-CONF-2015-063}, considering independently the 
electron and muon samples and selecting events with transverse momentum 
$p_T > 65$ GeV ($55$ GeV) as well as $\ETmiss > 65$ GeV ($55$ GeV) and transverse mass 
$m_T > 130$ GeV ($110$ GeV) in events with electrons (muons).

\begin{figure}[h!]
\centering

\vspace{2mm}

\includegraphics[width=\textwidth]{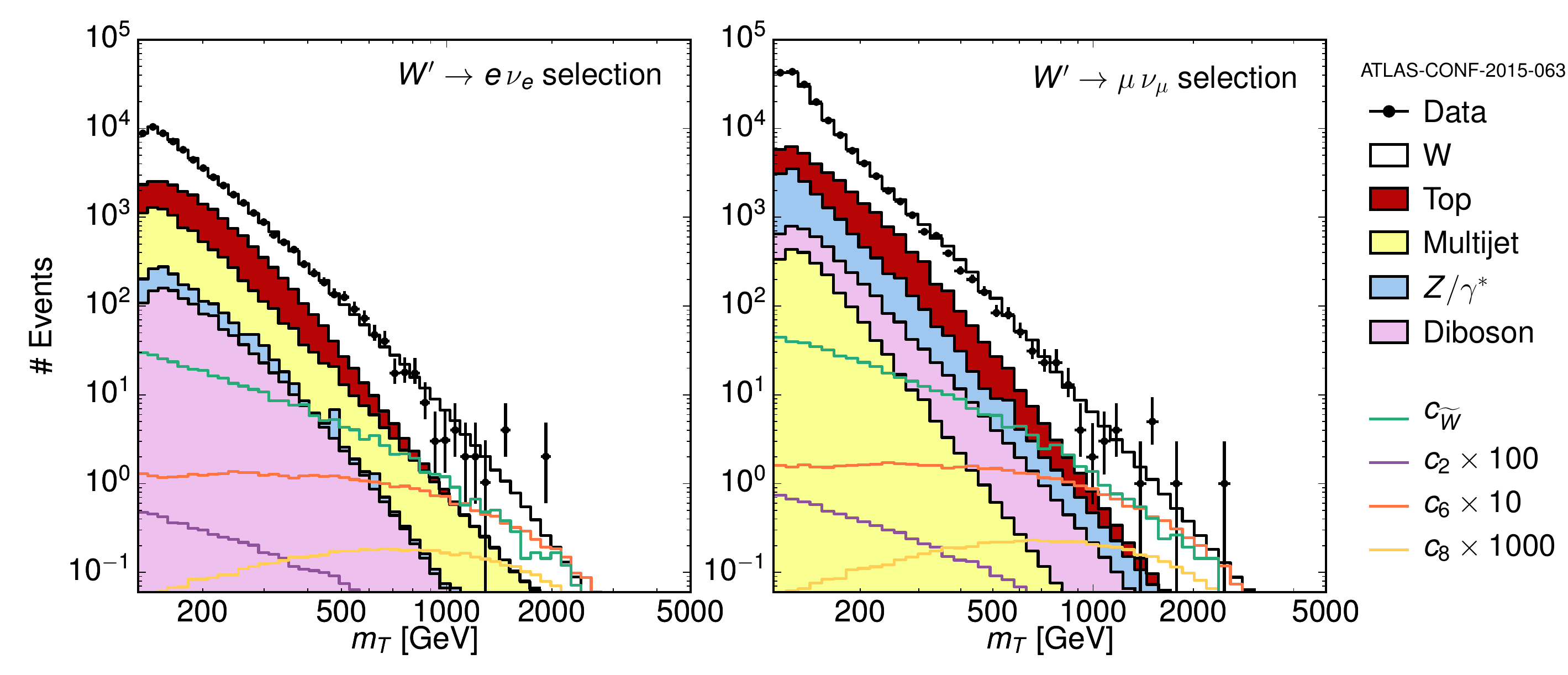}
\caption{ \it \small Transverse mass $m_T$ distribution for $a\,W^{\pm}$ ($W^{\pm} \to \ell^{\pm} \nu_{\ell}$) production 
in the $e + \ETmiss$ final state (Left) and $\mu + \ETmiss$ final state (Right), generated from $\mathcal{A}_{\tilde W}$ (green), $\mathcal{A}_2$ (purple), 
$\mathcal{A}_6$ (orange) and $\mathcal{A}_8$ (yellow). Also shown are the binned experimental data and dominant backgrounds 
from the 13 TeV  $(3.3\,\mathrm{fb}^{-1})$  ATLAS analysis~\cite{ATLAS-CONF-2015-063}.} 
\label{Fig_aWW_Events}
\end{figure}

\begin{figure}[h!]
\centering

\vspace{-2mm}
\includegraphics[width=0.98\textwidth]{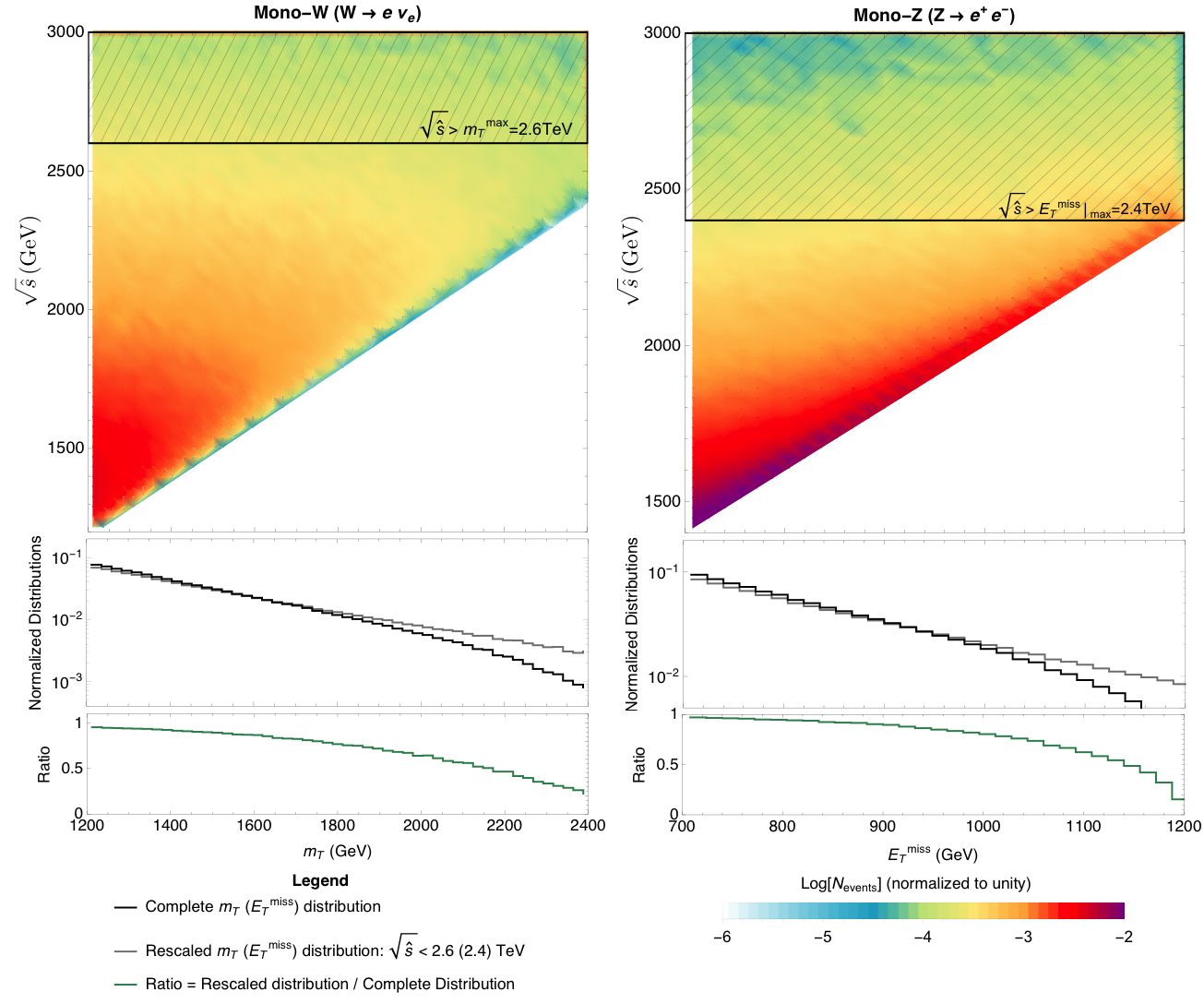}
\caption{\it\small The left (right) top panel shows the correlation between $m_T$ ($\ETmiss$) and 
$\sqrt{\hat{s}}$ for mono-$W$ (mono-$Z$) through a double-differential Monte Carlo distribution for the $\A_{\tilde{W}}$ signal. 
The centre plots show the normalized $m_T$ ($\ETmiss$) distribution before/after discarding the events 
for which $\sqrt{\hat{s}}> m_T^\text{max}$ (\,$2\,\ETmiss^\text{max}$) (grey/black). The bottom panel displays the ratio between the two distributions above.}
\label{Validy_Double_Dist}
\end{figure}

The couplings that may contribute to this process are the custodial-invariant $\A_{\tilde{W}}$ and $\A_2$ operators, and the custodial-breaking ones $\A_6$ and $\A_8$ 
in Eq.~(\ref{bosonic_basis}), as  illustrated in Fig.~\ref{diagram_monoWZ} and shown in 
the Feynman rules \ref{FR.GGa}. Fig.~\ref{Fig_aWW_Events} depicts the $m_T$ spectrum of the SM background contributions, as well as 
the various signals corresponding to the $c_2$, $c_6$, $c_8$, $c_{\tilde{W}}$ Wilson coefficients, 
for $f_a = 1$ TeV and $c_i = 1$ (with $c_{\tilde{B}}$ obeying Eq.~(\ref{cWB})). The bins used in this figure are those for which there is experimental information 
on the background~\cite{ATLAS-CONF-2015-063}, corresponding to $m_T < m_T^\text{max} = 2.6$ TeV for electrons and $m_T < m_T^\text{max} =3$ TeV for muons.  
As discussed in Sect.~\ref{Sect:Validity}, the strict EFT validity condition 
$\sqrt{\hat{s}} < f_a$ can be imposed by computing the fraction of events in each bin for which 
$\sqrt{\hat{s}}> m_T^\text{max}$, and discarding it. In Fig.~\ref{Validy_Double_Dist} (left) we show the correlation between $m_T$ and 
$\sqrt{\hat{s}}$ for mono-$W$ through a double-differential Monte Carlo distribution for the $\A_{\tilde{W}}$ signal. 
We also show the normalized $m_T$ distribution (again, for the $\A_{\tilde{W}}$ signal) before/after discarding the events 
for which $\sqrt{\hat{s}}> m_T^\text{max}$.

We note that although  the $c_6$ and $c_8$ signatures in Fig.~\ref{Fig_aWW_Events} exhibit a kinematical shape {\it a priori} much more favorable to 
be distinguished from background than those proportional to $c_{\tilde{W}}$ and $c_2$, at the end the most prominent impact on this purely LHC analysis is that 
of $c_{\tilde{W}}$ (followed by that of $c_6$) due to 
suppression factors in the cross sections.\footnote{The impact of $\mathcal{A}_8$ is  suppressed with respect to that 
from $\mathcal{A}_6$ well beyond what suggests the $\sim(g/4\pi)^{2}$  factor in 
the Feynman rule \ref{FR.GGa}, as the squared matrix element of its contribution $q \bar{q}' \to W^{\pm} a$ 
 vanishes with the quark mass  as $\sim m_{q}^2/m_{W}^2$ ($\sim 2 \times 10^{-4}$ for the charm quark).} Mono-$W$ signatures from 
 the operators $\A_6$, $\A_8$ and $\A_2$ are  buried in the backgrounds of present LHC data, and they will remain out of reach with future HL-LHC data, 
 except for $\A_6$, see Sect.~\ref{Sect:PhenoAnalysisFuture} below. 
 
The loop-level bound obtained  in Eq.~(\ref{rarebound}) would imply  (if taken at face value) that $\A_{\tilde{W}}$ is out of 
reach of foreseen LHC prospects, for light enough ALPs; however, as previously discussed, because more than one operator contributes to those rare 
process --see Eq.~(\ref{ayayaya})-- the data only constrain a combination of operator coefficients which differs from that in LHC signals, see Eq.~(\ref{ja}); 
it is thus pertinent to analyze the impact of $\A_\Wt$ on LHC independently. 

The results obtained, for which the LHC sensitivity in $f_a/c_{\tilde{W}}$ extends up to significant values, 
are listed in Table~\ref{Likelyhood_Table}. They show an important impact of 
the systematic uncertainties on the background and also indicate that present 
LHC Run II limits on $f_a/c_{\tilde{W}}$ from mono-$W$ signals would {\it a priori} be sensitive to  $c_{\tilde{W}}$ only in the region of 
strong coupling $c_{\tilde{W}} \gtrsim 1$ (possible in non-linear EWSB constructions), for values of $f_a$ compatible with the validity of the EFT. These bounds have been computed in compliance with the strict validity criterium ($\sqrt{\hat{s}}<f_a$) by discarding the fraction of events in each bin for which $\sqrt{\hat{s}}>m_T^\text{max}$ (recall the discussion 
in Sect.~\ref{Sect:Validity}). We note that here the effect of considering a strict validity criterium instead of the milder $f_a > m_T^\text{max}$ one  is of the order of the few percent on the numbers in Table~\ref{Likelyhood_Table}. 
The bound which suffers the most from applying the strict validity criterium is the present constraint from the $W\to e\n$ final state, 
where applying the naive validity criterium would imply overestimating the bound in $\sim 20\%$. However, 
this is not a problem since it is the muon channel with yields a more constraining result.

%This is in contrast to the results with the stronger reach of mono-$Z$ signals to be discussed next.

%
\begin{table}
 \begin{center}
 \begin{tabular}{ | l l|*4{c |}}
 \hhline{~~----}
 \multicolumn{2}{c|}{}& \multicolumn{2}{ c | }{$c_{\tilde{W}}$ (mono-$W$)} &  \multicolumn{2}{ c | }{$c_{\tilde{W}}$ (mono-$Z$)}  \\
 \hline
\multicolumn{2}{|c|}{$\ell$} & e  & $\mu$ & e &  $\mu$ \\
[0.5ex] 
\hline\hline

\hline
$\left(f_a/c_{\tilde{W}}\right)_{\min}$ $[\unit{TeV}]$ & & 0.94 & 1.63  & 3.77 & 2.54   \\
\hline\hline

$\left(f_a/c_{\tilde{W}}\right)_{\min}$ $[\unit{TeV}]$ &[No Syst.]  & 1.62 & {2.44} &  3.79 & 2.54 \\
\hline
\end{tabular}
\caption{\it \small Present $95\cl$ $f_a/c_{\tilde{W}}$ exclusion limits for the effective operator $\mathcal{A}_{\tilde W}$ from mono-$W$ (left),
inferred from the search presented in Ref,~\cite{ATLAS-CONF-2015-063} as detailed in Sect.~\ref{Sect:monoW-ATLAS} and mono-$Z$ (right)
inferred from the search presented in Ref.~\cite{CMS:2016yfc} as detailed in Sect.~\ref{Sect:monoZW_correl}. Values obtained 
without including background systematics are labeled [No Syst.].} \label{Likelyhood_Table}
\end{center}
\end{table}

\vspace{-2mm}

\subsubsection{Mono-\tpdf{$Z$}{Z} signatures: \tpdf{$pp \to a \, Z$}{pp -> a Z}}
\label{Sect:monoZ}
Consider now  ALP production in association with a $Z$ boson, in hadronic collisions, as illustrated in Fig.~\ref{diagram_monoWZ} (center and right). The recent CMS 
$Z + \ETmiss$ search~\cite{CMS:2016yfc} with $\sqrt s = \unit[13]{TeV}$
and  integrated luminosity $2.3\,\mathrm{fb}^{-1}$ will be used to estimate present sensitivities to various Wilson coefficients. Table~\ref{tab:contributions} summarizes 
the couplings which may {\it a priori} contribute to a mono-$Z$ signal among those in the chiral basis,  Eqs.~(\ref{Eq:A2D}) and (\ref{bosonic_basis}). It will be 
argued next that only $c_{\tilde{W}}$ may be expected to be seriously tested by this signal.

The $\ETmiss$ distribution for signal and background will be used as kinematic discriminator, applying the same tools and procedure described at the beginning of 
Sect.~\ref{Sect:monoW_monoZ}. In order to optimize the search, the following preselection and selection cuts are 
applied: $p^{\ell}_T > 20$ GeV, $\left|\eta_\ell\right| < 2.5$, $p^{\ell\ell}_{T} > 50$ GeV, 
$m_{\ell\ell} \in [80,\,100]$ GeV, $\ETmiss > 80$ GeV, $\left|\ETmiss - p_T^{\ell\ell}\right|/p_T^{\ell\ell} < 0.2$, $\Delta \phi_{\ell\ell,\vec{\ETmiss}} > 2.7$ (rad), 
an furthermore $3^{\mathrm{rd}}$-lepton and extra high-$p_T$ jets vetoes are implemented. 
The cut $\ETmiss > 80$ GeV ensures that a contamination from the gluon-fusion initiated 
signal leading to  $s$-channel Higgs mediation can be safely neglected: for an on-shell Higgs the maximum $\ETmiss$ is $\sim 30$ GeV. Furthermore,  for 
a higher $\ETmiss$ cut  the fraction of the cross section contributed by this channel may be estimated as the integral of the 
Breit-Wigner distribution of the Higgs resonance for $\hat{s} > m_Z^2+ 2\ETmiss^2 [1 + (1 + m_Z^2/\ETmiss^2)^{1/2}]$, which for $\ETmiss > 80$ GeV gives a suppression 
factor of $5\times10^{-6}$. Given that the on-shell Higgs production via gluon-fusion is $\sigma(gg \to h) = \unit[48.6]{pb}$~\cite{deFlorian:2016spz},   
the Higgs-mediated contribution  is completely negligible. Similarly, 
contributions involving a quark in the $t$-channel are not relevant in the kinematic region considered.
 In summary, with present data the signal cross-sections for $p p \to Z a$ have a negligible dependence on 
the Wilson coefficients parameterizing the $qqa$ and $hZa$ vertices, {\it i.e.} $c_{a\Phi}$ in the linear case and 
$c_{2D},\,c_3,\,c_{10-14},c_{17}$ in the non-linear one (see App.~\ref{App:feynman_rules}).

The remaining ALP-gauge boson interactions which may induce a mono-$Z$ signal are the custodial invariant operators  $\A_{\tilde{W}}$, $\A_{\tilde{B}}$, 
$\A_1$ and $\A_2$, and the custodial-breaking coupling $\A_7$, see Fig.~\ref{diagram_monoWZ} (center and right). $\A_{\tilde{B}}$ will not be considered 
independently all through the rest of this work, given the constraint in Eq.~(\ref{cWB}). The contribution from $\A_7$  does not need to be considered 
separately either, as $c_7$ enters exclusively through the combination $c_2+ 2c_7$, see the Feynman rules \ref{FR.ZAa} and \ref{FR.ZZa}. The analysis 
focuses thus on $c_{\Wt}$, $c_1$ and $c_2$.

\begin{figure}[h!]
\centering

\vspace{-2mm}

\includegraphics[width=\textwidth]{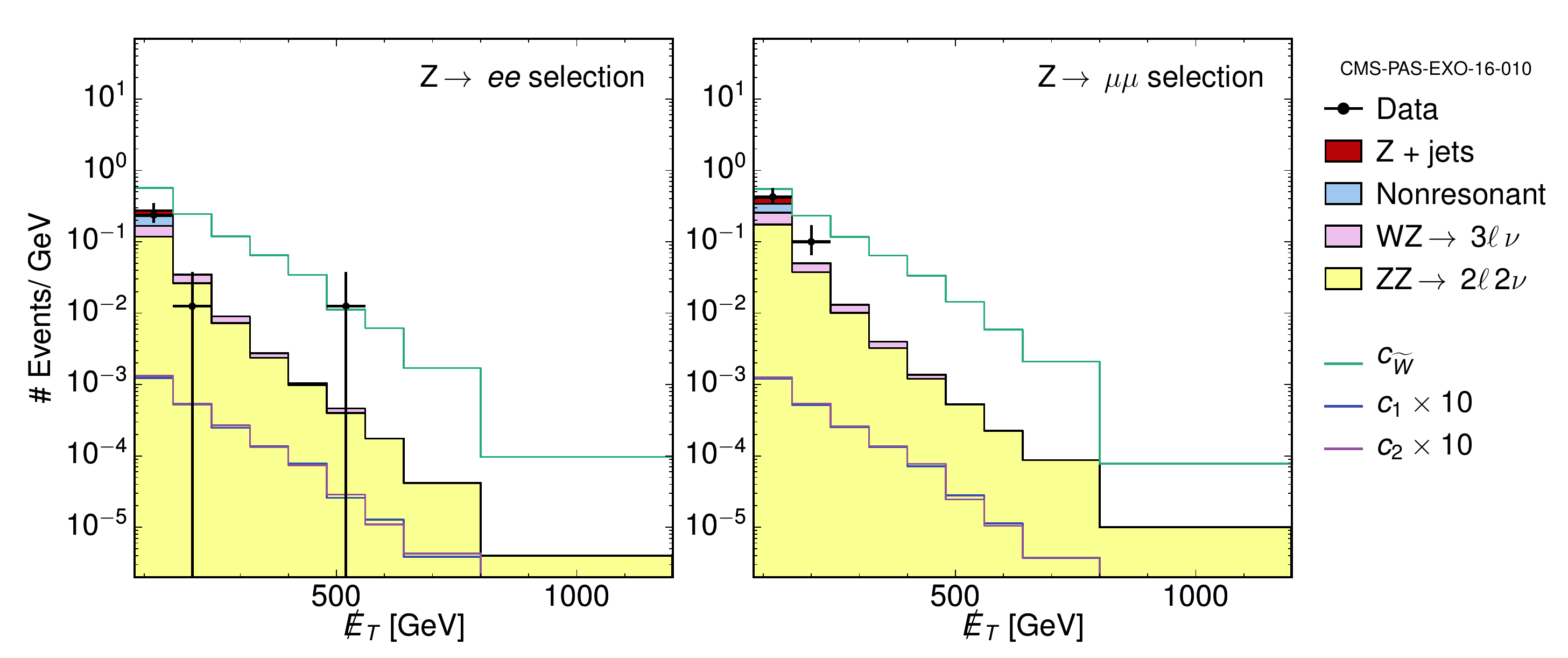}
\caption{\it\small $\ETmiss$ distribution for $a\,Z$ ($Z \to \ell^{+} \ell^{-}$) production 
in the $ee + \ETmiss$ final state (Left) and $\mu\mu + \ETmiss$ final state (Right), generated from $\mathcal{A}_{\tilde W}$ (green), $\mathcal{A}_1$ (blue), 
and $\mathcal{A}_2$ (purple). Also shown are the binned experimental data and dominant backgrounds 
from the 13 TeV ($2.3\,\mathrm{fb}^{-1}$) CMS analysis~\cite{CMS:2016yfc}.} 
\label{Fig_monoZ_Events}
\end{figure}

The comparison of signals and background $\ETmiss$ distributions for $\ell = e, \mu$ is shown in Fig.~\ref{Fig_monoZ_Events}. 
The highest-energy bin considered is $\ETmiss^\text{max} = 1.2$ TeV. 
In analogy with the previous mono-$W$ analysis, the EFT validity condition $\sqrt{\hat{s}} < f_a$ 
is implemented by discarding the fraction of events in each bin for which $\sqrt{\hat{s}}> 2 \ETmiss^\text{max}$ (see Sect.~\ref{Sect:Validity}).
The correlation between $\sqrt{\hat{s}}$ and $\ETmiss$ is shown in Fig.~\ref{Validy_Double_Dist} (right), as well as the 
normalized $\ETmiss$ distributions before/after discarding the invalid event fraction in each bin.

The results obtained for %the impact of 
$\mathcal{A}_{\Wt}$ are listed in Table~\ref{Likelyhood_Table}: the present mono-$Z$ search turns out to 
be significantly more powerful in constraining $c_{\Wt}/f_a$ than the ATLAS mono-$W$ search previously analyzed. Furthermore, the impact of systematic errors is 
negligible in this case.  An interesting fact is the different discriminating power of electrons and muons in mono-$Z$ signals induced by ALP emission with 
respect to the coupling strength: while  present muon data data could {\it a priori} be sensitive to  $c_{\tilde{W}}$ only in the region of strong 
coupling $c_{\tilde{W}}\ge 1$, a signal in electron data would be compatible as well with $c_{\tilde{W}}$  values in the perturbative regime, $c_{\tilde{W}}\le 1$. 
It is relevant to point out that these results, obtained imposing $\sqrt{\hat{s}}< m_T^{\text{max}} < f_a$, are equal up to the permille level to the ones which are obtained if the naive validity criterium (only $m_T^{\text{max}} < f_a$) is used instead.

The contributions to mono-$Z$ signals from $\A_{1,2}$ are shown in Fig.~\ref{Fig_monoZ_Events} for illustration only, as the 
corresponding values for $c_{1,2}$ would lie outside the region of validity of the EFT in present data and also if assuming the $3000$ fb$^{-1}$ 
integrated luminosity foreseeable at HL-LHC, see next section. The mono-$Z$ analysis with present and projected data is thus only sensitive to 
the $\mathcal{A}_{\Wt}$ operator, which is common to the NLO of the linear and of the non-linear expansion. It follows that mono-$Z$ searches alone are not 
sensitive to a possible non-linear component in the nature of EWSB, unlike mono-$W$ future searches at HL-LHC.

\begin{figure}[h!]
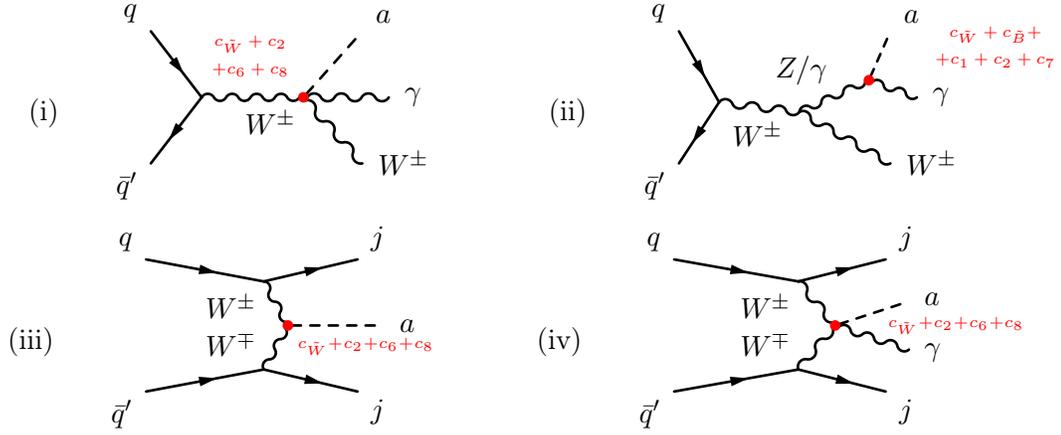

\include{Fdiagrams/AP_VBF_diagrams}
 \caption{\it \small  
Main diagrams contributing to the processes analysed in Sect.~\ref{Sect:AP_gamma}. Upper line: $a\gamma W$  associated production. 
Lower line: VBF-type interaction producing $ajj$ (iii) and $ajj\gamma$ (iv). The proportionality of each diagram to the non-linear parameters is indicated in the 
figure (overall factors and relative coefficients are not displayed).} \label{diagram_AP_VBF}
\end{figure}

%%%%%%%%%%%%%%%%%%%%%%%%%%%%%%%%%%%%%%%%%%%%%%%%%%
%%%%%%%%%%%%%%%%%%%%%%%%%%%%%%%%%%%%%%%%%%%%%%%%%% PHENO SECTION II
%%%%%%%%%%%%%%%%%%%%%%%%%%%%%%%%%%%%%%%%%%%%%%%%%%

\section{Phenomenological Analysis II: $\sqrt{s}=\unit[13]{TeV}$ LHC Prospects}
\label{Sect:PhenoAnalysisFuture}
This section explores the sensitivity prospects for constraining the effective ALP couplings to SM bosons at the HL-LHC, as well as the analysis strategy sensitive to the 
linear/non-linear character of the underlying EWSB mechanism. Assuming thus proton-proton collisions at c.o.m. energy $\sqrt{s}=\unit[13]{TeV}$ and successive 
integrated luminosities of  $300$ fb$^{-1}$ and $3000$ fb$^{-1}$, the following channels will be analyzed:
\begin{itemize}
\item Mono-$W$ and mono-$Z$ signatures, see Fig.~\ref{diagram_monoWZ}, projecting the analysis in Sect.~\ref{Sect:PhenoAnalysisPresent} onto  future data. 
A qualitative discussion of  their ratio
 as a probe of non-linear character will be added. 
\item $Wa\gamma$ associated production, see Fig.~\ref{diagram_AP_VBF}.
\item Mono-Higgs signatures, see Fig.~\ref{Plots_monoh_dih}.
\end{itemize}
Table~\ref{tab:contributions} summarizes the set of  operator coefficients  that could contribute to these signals  and  be tested in LHC prospects, among those 
defined in the Lagrangians Eqs.~(\ref{ABtilde})-(\ref{OaPhi}) and (\ref{Eq:A2D})-(\ref{La_NLO}).  The corresponding Feynman rules are shown in  
App.~\ref{App:feynman_rules}.

Mono-$W$ and $Wa\g$ associated production, together 
with $ajj(\gamma)$ production through vector boson fusion (VBF) -- also shown in Fig.~\ref{diagram_AP_VBF} -- are intimately related processes, 
as they probe the same limited set of effective operator coefficients~\footnote{\label{footnote_c2D}$c_{2D}$ --or its linear sibling {$c_{a\Phi}$}-- is 
also mentioned in Table~\ref{tab:contributions} in connection to these channels, contributing through the fermionic vertices that it induces. Nevertheless, 
this contribution is in any case much suppressed by ratios of quark mass over momentum.
} $c_{\tilde{W}}$, $c_{\tilde{B}}$, $c_1$, $c_2$, $c_6$, $c_7$ and $c_8$. 

A special role 
is played by the Higgs-ALP couplings associated to $c_{2D}$ which, barring extreme fine-tunings, may be only expected among the leading signals 
if the underlying EWSB enjoys a non-linear character,  and within a certain ALP mass range, as previously discussed. 
This coupling will be shown to be {\it a priori}  testable through mono-Higgs searches at HL-LHC, that exhibit a sensitivity reach well 
beyond the bounds obtained  in Sect.~\ref{Sect:hZax_Br} from the limits on the non-standard Higgs decay width.

 Relevant information about the structure of the ALP couplings can be inferred both by analyzing the different signatures independently and by 
studying their interplay. As some effective operators contribute to several processes, a combined analysis may be necessary in order 
to access the individual Wilson coefficients. Furthermore, the study of (de)correlations between the various putative signals serves as a 
good probe of the degree of EWSB non-linearity.

\subsection[Mono-$W$ and mono-$Z$ signatures]{Mono-$W$ and mono-$Z$ signatures} 
\label{Sect:monoWZ_Future}

\begin{table}[t!]
 \begin{center}
\begin{tabular}{ | l l | c | c | c | c |}
\hhline{~~----}
\multicolumn{2}{c|}{}& \multicolumn{4}{ c | }{$c_{\tilde{W}}$ (mono-$Z$)} \\
\hhline{------}
\multicolumn{2}{|c|}{$\ell$} & \multicolumn{2}{ c | }{e} &  \multicolumn{2}{ c | }{$\mu$} \\
\hhline{------}
\multicolumn{2}{|c|}{Luminosity [fb$^{-1}$]} & 300  & 3000 & 300 & 3000 \\
[0.5ex] 
\hline\hline
\hhline{------}
$f_a/c_{\tilde{W}}$ $[\unit{TeV}]$ &  & 10.47 & 15.81 & 9.79  & 14.33 \\
\hline\hline
$f_a/c_{\tilde{W}}$ $[\unit{TeV}]$ &[Syst.$\times1/2$]  &11.10 &18.40  & 10.39 & 16.67 \\
\hline\hline
$f_a/c_{\tilde{W}}$ $[\unit{TeV}]$ &[No Syst.]  & 11.64 & 21.47  & 10.91 & 19.64 \\
\hhline{------}
\end{tabular}
\caption{ \it\small Projected $95\cl$ $f_a /c_i$ reach at LHC, with $\mathcal{L} = 300\,\,\mathrm{fb}^{-1}$ and $\mathcal{L} = 3000\,\,\mathrm{fb}^{-1}$ 
for $\mu_{\tilde{W}}=(c_{\tilde{W}}/f_a)^2$ %for the effective operators relevant to
 from mono-$Z$ production, as detailed in Sect.~\ref{Sect:monoZ}. Top row: Assuming future systematic uncertainties on the background scale as present ones.  
 Middle row: Assuming systematic uncertainties are reduced by a factor 2 w.r.t. present ones. Bottom row: Assuming no background 
 systematic uncertainties.} \label{Likelyhood_Table3}
\end{center}
\end{table}

The result of extending the analysis  in Sect.~\ref{Sect:monoW-ATLAS} to the projected  sensitivity in $(c_i/f_a)^2$ for LHC 
13 TeV with 300 $\mathrm{fb}^{-1}$ and 3000 $\mathrm{fb}^{-1}$ is summarized  in 
Table~\ref{Likelyhood_Table3} (for mono-$Z$) and Table~\ref{Likelyhood_Table4} (for mono-$W$), considering electrons and/or muons in the final state.

 They show that mono-$Z$ searches will be stronger than mono-$W$ ones in probing at LHC the effective operator $\mathcal{A}_{\tilde{W}}$. Both electron and 
 muon channels will access the perturbative regime $c_{\tilde{W}}<1$. Mono-$Z$ searches would reach ALP scales 
 up to $f_a \sim 20$ TeV (for $c_{\tilde{W}} = 1$) 
with 3000 $\mathrm{fb}^{-1}$ disregarding background systematics -- see Table~\ref{Likelyhood_Table3}. Assuming instead future background systematics as 
($1/2$ of) the present ones, the mono-$Z$ reach is somewhat milder,  
up to $f_a \sim 15$ TeV ($\sim 18$ TeV).   Table~\ref{Likelyhood_Table4} shows that instead the limits on $c_{\tilde W}/f_a$ from LHC mono-$W$ searches 
are systematics dominated.\footnote{The mono-$W$ results are shown for electrons in the final state. Muon final states display similar sensitivities.}

Future mono-$W$ searches appear instead of special interest in order to uncover the $\mathcal{A}_{6}$ coupling, which is a signal of non-linearity  up to NLO. 
Table~\ref{Likelyhood_Table4} shows that with 300 $\mathrm{fb}^{-1}$ and 3000 $\mathrm{fb}^{-1}$ it is possible to either discover it or derive a consistent 
projected limit. The sensitivity to $c_6$ turns out to be mainly limited by statistical uncertainties, being 
less dependent than $\mathcal{A}_{\tilde{W}}$ on SM background systematics. Nevertheless a significant reduction of the latter is shown to have a 
significant impact also on tackling $\mathcal{A}_6$, particularly with 3000 $\mathrm{fb}^{-1}$:  scales up to $f_a/c_6 \le \unit[3.44]{TeV}$ ($\unit[4.68]{TeV}$) 
would be then attainable if systematic errors were reduced by $1/2$ (completely) with respect to their present value (see Table~\ref{Likelyhood_Table4}), 
leading to $c_6$ being testable within the perturbative region. 

Finally, mono-$W$ and mono-$Z$ signals may turn out to be especially prominent as phenomenological signals of the complete NLO ALP basis, 
in particular of ALP-fermion couplings
in the chiral EWSB case.  For instance, the $aZ\bar{\psi}\psi$ couplings   $\B^q_3$, $\B^q_4$, $\B^q_6$, $\B^q_7$, $\B^q_8$ and $\B^q_{10}$ 
in Eq.~(\ref{fermionicoperators}) may have a large impact on the very sensitive mono-$Z$ channel, while  the $aW\bar{\psi}\psi$ 
vertices in $\B^q_3$, $\B^q_5$, $\B^q_6$, $\B^q_7$, $\B^q_9$ and $\B^q_{10}$ may induce mono-$W$ signals; these couplings are not Yukawa 
suppressed and will be explored in a future study.

\begin{table}[t!]
 \begin{center}
\begin{tabular}{ | l l | c | c | c | c |}
\hhline{~~----}
\multicolumn{2}{c|}{}& \multicolumn{2}{ c | }{$c_6$ (mono-$W$)} &  \multicolumn{2}{ c | }{$c_{\tilde{W}}$ (mono-$W$)} \\
\hhline{------}
\multicolumn{2}{|c|}{Luminosity [fb$^{-1}$]} & 300  & 3000 & 300 & 3000 \\
[0.5ex] 
\hline\hline
\hhline{------}
$f_a/c_i$ $[\unit{TeV}]$ & & 2.00  & 2.53 & 1.83 & 2.20       \\
\hline\hline
$f_a/c_i$ $[\unit{TeV}]$ &[Syst.$\times 1/2 $]  & 2.24  & 3.25  & 2.23 & 2.90 \\
\hline\hline
$f_a/c_i$ $[\unit{TeV}]$ &[No Syst.]  & 2.51  & 4.51  & 3.40 & 6.05  \\
\hhline{------}
\end{tabular}
\caption{\it\small Projected  $95\cl$   $f_a/c_i$  LHC  reach  for $\ell = e$ final states, 
with $\mathcal{L} = 300\,\,\mathrm{fb}^{-1}$ and $\mathcal{L} = 3000\,\,\mathrm{fb}^{-1}$  for the effective operators relevant to 
 mono-$W$ production, as detailed in Sect.~\ref{Sect:monoW-ATLAS}. Top row: Assuming future systematic uncertainties on the background scale as present ones. 
 Middle row: Assuming systematic uncertainties are reduced by a factor 2 w.r.t. present ones. Bottom row: Assuming no background systematic uncertainties.} 
 \label{Likelyhood_Table4}
\end{center}
\end{table}

\subsubsection{Strategy for a combined analysis}
\label{Sect:monoZW_correl}

As is apparent from the discussion above, the interplay between mono-$Z$ and mono-$W$ signatures may be relevant as a way of disentangling 
the presence of non linearity in the Higgs sector.  Up to NLO in both expansions and barring extreme fine-tunings of operator coefficients, the cross sections 
for those two processes are:
\begin{itemize}
\item Strongly correlated in the linear case, being both controlled 
by the coefficient $c_{\tilde{W}}$ ($c_{\tilde{B}}$ is not independent, see Eq.~(\ref{cWB})). 
\item Less correlated in the non-linear case, as   operators other than $\A_{\tilde{W}}$ and $\A_{\tilde{B}}$ are expected to contribute to those mono-signals. For instance the purely chiral $\A_6$ operator may contribute visibly to mono-$W$ production within the projected HL-LHC prospects, as shown above.
\end{itemize}
A combined analysis of mono-$Z$ and mono-$W$ appears thus to be a valid method 
to shed light on the nature of the EWSB dynamics, once  a positive detection occurs. 
 Here we illustrate the (de)correlations of those signals in a purely qualitative way. The cross-sections for $pp\to Za$ and $pp\to W^\pm a$ at a c.o.m.~energy $\sqrt{s}=\unit[13]{TeV}$ are computed using {\tt MadGraph5$\_$aMC@NLO}, and subject to no other constraint than Eqs.~\eqref{agg_constraint} and \eqref{aZg_constraint} and a kinematical cut $\ETmiss > 200$ GeV.  A random scan of  Wilson coefficients $c_i \in [-1,1]$ has been performed, along three scenarios: i) the {\it linear} setup, which in practice reduces to the custodial-preserving  $\mathcal{A}_{\Wt}$ and $\mathcal{A}_{\Bt}$ operators, see Eq.\eqref{ABtilde} and Eq. \eqref{AWtilde}; ii)  
the {\it non-linear custodial} case, involving operators $\mathcal{A}_{\Wt}$, $\mathcal{A}_{\Bt}$, $\mathcal{A}_{1}$ and $\mathcal{A}_{2}$; iii) the non-linear case including both custodial preserving and non-custodially invariant couplings,  here denominated {\it non-linear} for short, which adds  to the previous set $\mathcal{A}_{6}$ and $\mathcal{A}_{7}$, see Eq.~\eqref{La_NLO}. 

\begin{figure}[t!]\centering
 \includegraphics[height=7.5cm]{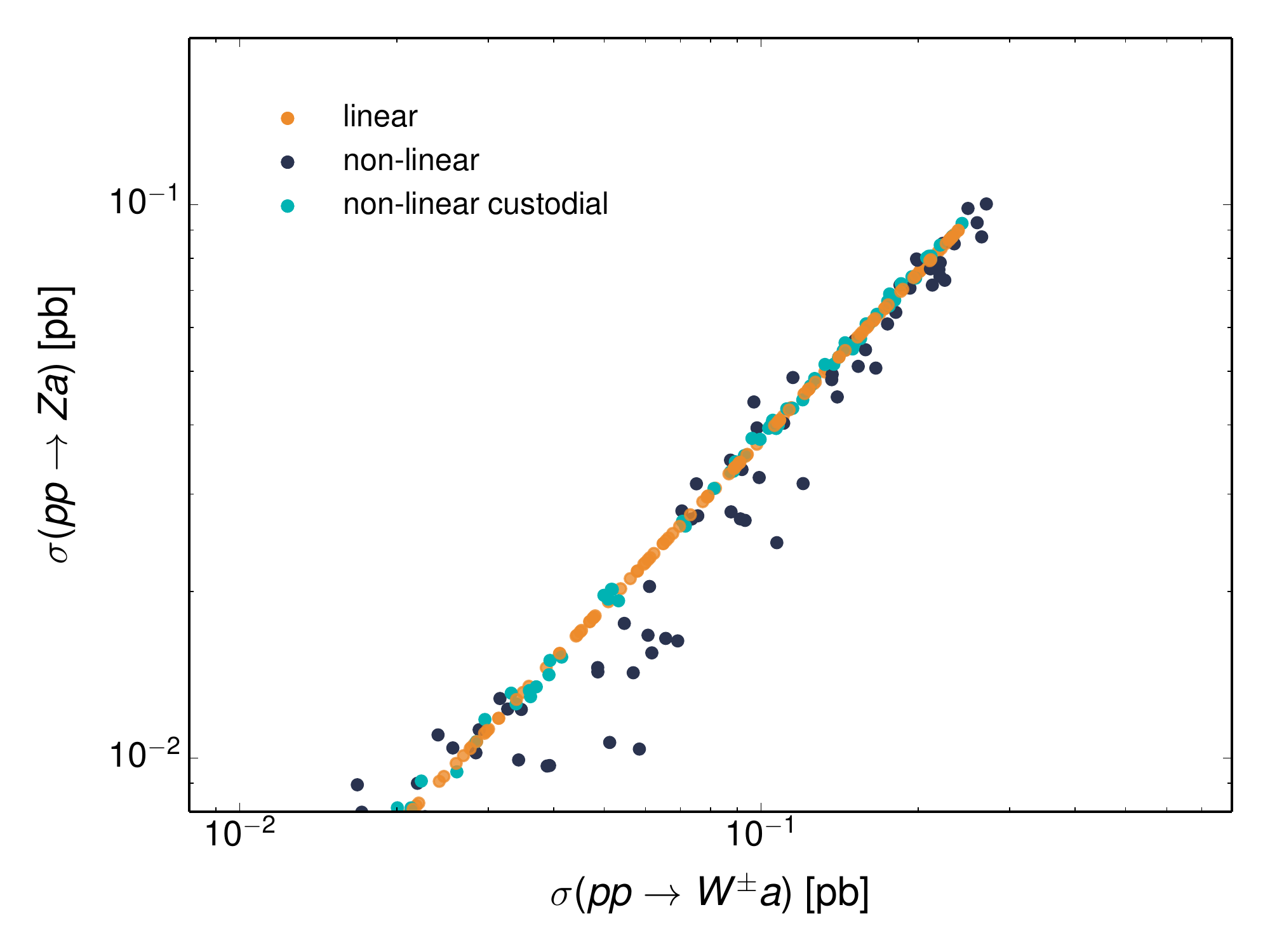}
 \includegraphics[width=\textwidth]{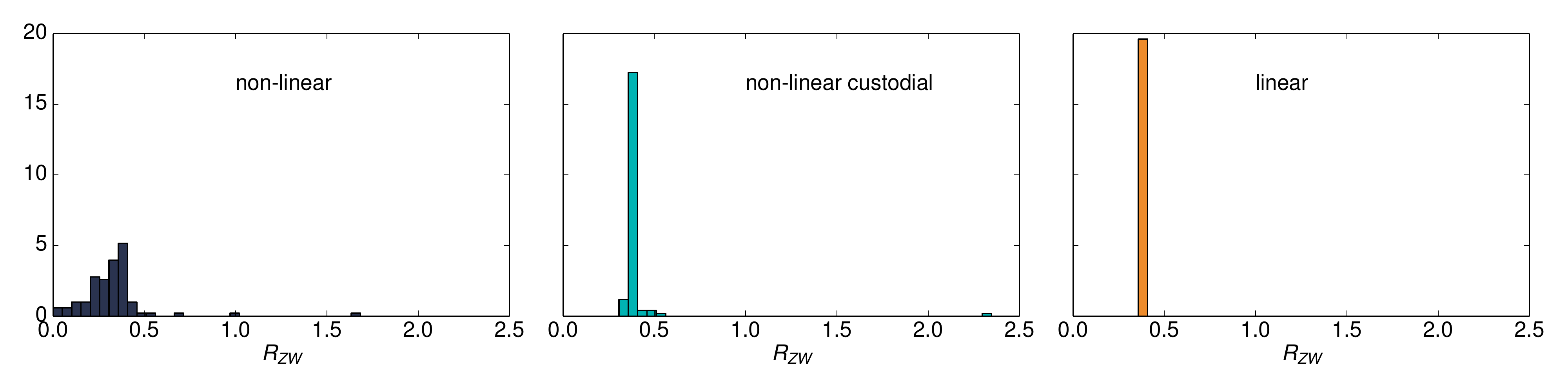}
 \caption{\it\small Top: Cross sections for $pp\to Z a$ and $pp\to W^\pm a$ at $\sqrt{s}=\unit[13]{TeV}$ with ${\ETmiss\geq\unit[200]{GeV}}$, 
 computed with {\tt Madgraph5$\_$aMC@NLO} for a random scan of Wilson coefficients $c_i \in [-1,1]$ (see text for details) within the region 
 allowed by Eq.~\eqref{agg_constraint} and~\eqref{aZg_constraint}, and using $m_a=\unit[1]{MeV}$ and $f_a=\unit[1]{TeV}$ for illustration. 
 Bottom: Distribution of the ratio $R_{ZW}$ defined in Eq.~\eqref{RZW_def}, with the area of the histograms normalized to 1.
 In both cases, orange, cyan and dark blue  correspond respectively to linear, non-linear custodial and non-linear non-custodial setups.}\label{plot_monoZW}
\end{figure}

The results  for the cross-sections are summarized in Fig.~\ref{plot_monoZW} (top) in the $\sigma(p p \to W^{\pm}a)$, $\sigma(p p \to Z a$) plane,  for linear (orange), cyan (non-linear custodial) and dark blue (non-linear),  
for $f_a = 1$ TeV.  
The strong correlation characteristic of the EWSB linear scenario is clearly seen.  In contrast, in the non-linear setup  deviations from the sharp linear pattern emerge as expected as they stem  from the non-linear operators $\A_{1,2,6,7}$. Those  deviations are necessarily small, though,  as the contribution from any of the coefficients  $c_{1,2,6,7}$  is suppressed by a factor $g/(16\pi)$  -- see Feynman rules in App.~\ref{App:feynman_rules} -- compared to that of $c_{\tilde{W}}$, $c_{\tilde{B}}$ 
(this conclusion may however be somewhat modified  if a harder $\ETmiss$ cut is  
imposed on the signal). 
In any case, in the event of a mono-$W$ and/or mono-$Z$ excess in future data, the ratio
\begin{equation}\label{RZW_def}
 R_{ZW} = \frac{\s(pp\to Za)}{\s(pp\to W^\pm a)}
\end{equation} 
may be used to discern among possible physical explanations.  This observable has the advantage of being in principle independent of the scale $f_a$ as well as of the ALP 
mass $m_a$ (provided that $m_a\ll m_Z$, as is the case assumed in this analysis for ALP stability reasons). Fig.~\ref{plot_monoZW} (bottom) shows, for  the three sets of operators considered, the $R_{ZW}$ distributions
 obtained letting the coefficients of each operator considered assume random values in the interval [-1,1].  Note that neither the slope in the upper plot in Fig.~\ref{plot_monoZW}, nor the numerical values in the other plots in this figure, are  meaningful {\it per se},  but rather  strongly dependent 
on the specifics of the analysis (e.g. on the kinematical cuts applied, here $\ETmiss > 200$ GeV). Therefore, 
the strategy to follow in a realistic experimental analysis would be to look for a coincidence/tension between the expected value of $R_{ZW}$ in the linear 
scenario and the measured one which, if detected, could indicate the presence of non-linearity in the Higgs sector.

We stress that the considerations in this subsection are aimed at discussing the expected relative strength of the mono-$W$ and mono-$Z$ observables in a purely 
qualitative way, having in mind future hadronic machines in general. Indeed, besides the strong dependence of the results on the kinematical cuts chosen, no 
consideration of backgrounds has been taken into account here. This is in contrast to the detailed phenomenological analysis at the 
beginning of the subsection, where it was shown that  only the deviations stemming from   $\A_{6}$ have a chance of being visible within the foreseen HL-LHC prospects. 

\subsection[Associated production: $pp\to a W^\pm \gamma$]{Associated production: $pp\to a W^\pm \gamma$} 
\label{Sect:AP_gamma}
Consider next ALP production in association with both a $W$ boson and a photon, as illustrated in Fig.~\ref{diagram_AP_VBF} i) and ii). Examining  the 
interactions in the chiral effective Lagrangian Eq.~\eqref{bosonic_basis}, it is easy to see that those  
couplings exhibit  a particularly interesting combined potential for disentangling the presence of different effective operators: 
\begin{eqnarray}
\label{aWW_aWWgamma}
a\, W^{+}\, W^{-}  &\rightarrow & \frac{g}{4\pi f_a}\left[c_6\,g^{\mu\nu}(p_+^2-p_-^2)+
\left(\frac{g}{4\pi}c_8-c_6\right)\left(p_+^\mu p_+^\nu -p_-^\nu p_-^\nu\right)\right]+
\nonumber \\ 
& & -\frac{4i}{f_a}\left(c_{\tilde{W}}+\frac{g}{16\pi}c_2\right)p_{+\a}p_{-\b}\e^{\mu\nu\a\b}\,,
\label{eq_APVBFcoupl} \\
a\, W^{+}\, W^{-}\, \gamma & \rightarrow& \frac{ge}{4\pi f_a}\left[\left(\frac{g}{4\pi}c_8-c_6\right)(g^{\m\r}p_a^\n+g^{\n\r}p_a^\m)
+2c_6g^{\mu\nu}p_a^\rho\right]-\frac{4ig}{f_a}\left(c_\Wt +\frac{g}{16\pi}c_2\right)\e^{\m\n\r\a}p_{a\a}\,, \nonumber
\end{eqnarray}
as illustrated respectively in  \ref{FR.WWa} and \ref{FR.WWAa} of App.~\ref{App:feynman_rules} and summarized in  Table~\ref{tab:contributions}. 
Both processes are thus {\it a priori} sensitive~\footnote{The sensitivity to $c_{2D}$ ($c_{a\Phi}$ in the linear expansion) remains in 
practice negligible even for $\mathcal{L}=\unit[3000]{fb^{-1}}$ for the same reasons explained in Footnote~\ref{footnote_c2D}. }
 to  $\A_6$, $\A_8$, and to a fixed combination of $\A_{\tilde{W}}$ and $\A_2$ which therefore singles out  a flat direction.  In contrast, in the linear scenario only the Wilson coefficient $c_\Wt$ contributes significantly to both interaction vertices. 

The first process in Eq.~(\ref{aWW_aWWgamma}) leads to the striking mono-$W$ signal being already  searched by LHC collaborations and whose physics impact 
has been explored in Sects.~\ref{Sect:monoW-ATLAS} and~\ref{Sect:monoWZ_Future}.  The second process leads to $a W^\pm \gamma$  associated production, a search not being yet performed by the ATLAS and CMS collaborations. We will explore its prospects next,  focusing
on final states characterized by leptonic $W$ decays.  
 It is necessary to take into account, though, that  the $p p \to a W^\pm \gamma$ channel may be induced also by $a\,Z\,\gamma$-mediated contributions, to which 
 the set $\{\mathcal{A}_{\tilde{W}}, \mathcal{A}_{\tilde{B}},\mathcal{A}_1,\, \mathcal{A}_2,\,\A_7\}$ may contribute as illustrated 
 in Fig.~\ref{diagram_AP_VBF} (ii),\footnote{ The   $a\,\gamma \,\gamma$ contribution to $pp\to a W^\pm \gamma$ is  proportional to  $\left(\ct^2c_{\tilde{B}} + \st^2c_{\tilde{W}}\right)$ and thus irrelevant, see   Eq.~(\ref{cWB}).}
\begin{eqnarray}
\label{AP_pho_verts}
a\,Z\,\gamma&\rightarrow & \frac{i}{f_a}p_{Z\a}p_{A\b}\e^{\mu\nu\a\b}\left(-2t_\theta c_{\tilde{W}}-\frac{g}{8\pi}(2c_1+t_\theta(c_2+2c_7))\right)\,,
\end{eqnarray}
where the constraint  in Eq.~(\ref{cWB}) has been applied. 
The contribution of $\A_7$  is equivalent to that of $\mathcal{A}_1$ and it is not necessary to consider it independently.  In summary, the analysis is done on five distinct operators: $\{ \A_1,\,\A_2,\, \A_6, \A_8\}$ and the combination of $\{\A_\Wt,\,\A_\Bt\}$ orthogonal to the $a\g\g$ coupling. They are studied next, one at a time and keeping our analysis at parton 
level.\footnote{An analysis of the associated $aW\gamma$ channel including parton shower and a detector simulation is beyond the scope of 
this work and it is left for the future once the viability of the searches proposed is established.}

\begin{figure}[t!]\centering
\includegraphics[width=.75\textwidth]{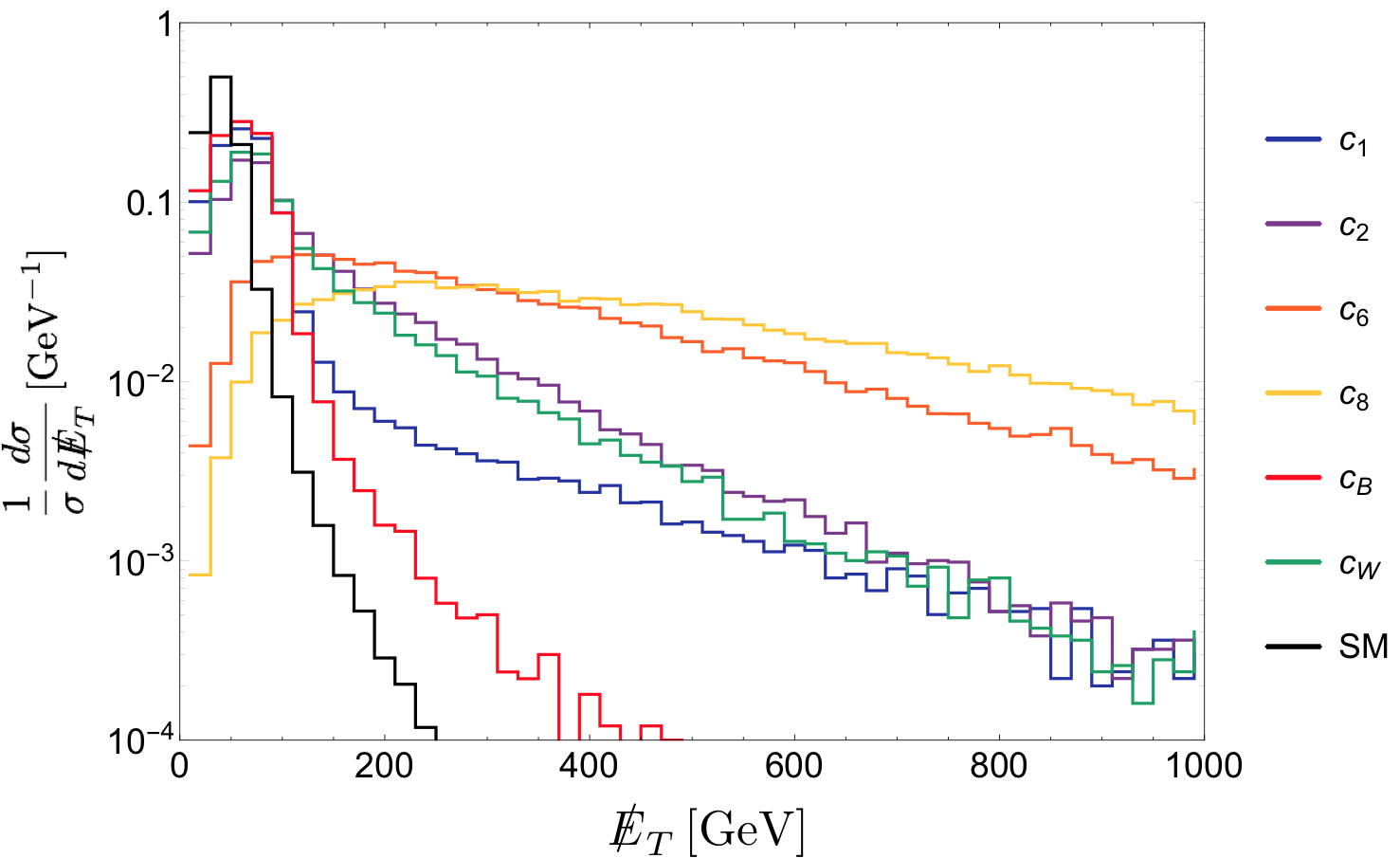}
\caption{\it\small  Missing transverse energy distributions 
for $pp\to aW^{\pm}\gamma$ ($W^{\pm}\to \ell^{\pm}\nu$) at $\sqrt{s}=\unit[13]{TeV}$ LHC 
normalised to unity, for signals generated one by one for operators  $\A_1$ (blue), $\A_2$ (violet), $\A_6$ (orange), $\A_8$ (yellow), $\A_{\tilde{W}}$ (green) and $\A_{\tilde{B}}$ (red). } 
\label{AP_pho_metdists}
\end{figure}

The main irreducible SM background is $pp\to W^{\pm}\gamma \,( \text{with}\,\, W^{\pm}\to \ell^{\pm}\n)$, a process which has been measured by 
ATLAS~\cite{Aad:2013izg} and CMS~\cite{Chatrchyan:2013fya} during the LHC $\sqrt{s} = 7$ TeV Run. 
The reducible backgrounds are subdominant with respect to the direct $W^\pm\g$ production and consist of $(i)$ $W^{\pm}$+jets (with a jet misidentified as a photon), $(ii)$ $Z\,\ell^+ \ell^-$ (with one of the leptons 
misidentified as a photon or unidentified and the $Z$ decaying into neutrinos), $(iii)$ $\gamma$+jets (with a lepton originating from a heavy quark decay) and $(iv)$
$t\bar{t}$ (with a semileptonic decay of the top pair and a misidentification of a field as a photon). Their combined effect is to approximately increase the size of the $W^{\pm}\gamma$ background by 15-25\% depending on 
kinematics and the flavour of the lepton~\cite{Aad:2013izg,Chatrchyan:2013fya}. For the present analysis, we simply account for this by 
scaling up our dominant SM $W^{\pm}\gamma$ background by 20\%.

The event selection requirements for photons and leptons for both signal and background are 
$p_T^\gamma > \unit[20]{GeV}$, $p_T^{\ell} > \unit[20]{GeV}$, $|\eta^\gamma| < 2.5$ and $|\eta^{\ell}| < 2.5$. 
$\ETmiss$ will be employed as kinematic variable for distinguishing signal from background, as we find that this variable has significantly more signal discrimination 
power than the $p_T$ of the lepton, because it receives contributions directly from the ALP in the signal set. The $\ETmiss$ distributions (normalized to unity) 
for the various effective operators and the SM background are shown in 
Fig.~\ref{AP_pho_metdists}. The harder momentum dependence of the effective couplings explored compared to the SM contribution are illustrated. 
In practice, we simulate events only up to $\ETmiss = 1$ TeV, as we find that signal cross sections for  $\ETmiss > 1$ TeV are negligible.

The significance $\Bsigma_i$ of a signal associated to one given operator $\A_i$ is defined here as~\cite{Cowan:2010js} 
\[ \Bsigma_i=\sqrt{2\left[(\mu_i s_i+b)
\ln\left(1+\frac{\mu_i s_i}{b}\right)
-\mu_i s_i\right]}\,,\label{Asimov_sensitivity}\]
where $\mu_i$ was defined in Eq.~(\ref{mui}), and $\mu_i s_i$ and  $b$ denote respectively the number of events in the signal and the background, 
alike to the definitions used  in Eq.~(\ref{likelihood_NS}) with, in this case, 
\beq
s_i= \L\times
 \int_{\ETmiss^{\text{min}}}^{\ETmiss^\text{max}}\frac{d\s_i}{d\ETmiss}d\ETmiss \qquad \text{and} \qquad
 b=\L\times\int_{\ETmiss^{\text{min}}}^{\ETmiss^\text{max}}\frac{d\s_{\text{SM}}}{d\ETmiss}d\ETmiss\,,
 \label{sib}
\eeq
where $\L$ is the integrated luminosity, $\sigma_i$ stands for the cross-section induced by $\A_i$ and $\s_{\text{SM}}$ for the SM one.   
The kinematical cuts are taken as follows:
\begin{itemize}
\item $\ETmiss^\mathrm{min}=\unit[200]{GeV}$, as it optimizes the sensitivity by removing most of the background, see Fig.~\ref{AP_pho_metdists}.
Higher $\ETmiss^\mathrm{min}$ values do not improve the signal-to-background ratio.
\item $\ETmiss^\mathrm{max}={f_a}/{2}$ for a given $f_a$ value, as required by the EFT validity considerations, see Sect.~\ref{Sect:Validity}. 
\end{itemize} 
A very slight improvement in sensitivity to $c_6$ is 
found for $\ETmiss^\mathrm{min} = \unit[300]{GeV}$, as illustrated in Table~\ref{AP_pho_cuts} where the optimal cuts in $\ETmiss^\mathrm{min}$ and  the 
corresponding sensitivity reach are shown. Nevertheless, for comparison purposes it is more appropiate to use one single cut for all operators, 
and the value $\ETmiss^\mathrm{min}=\unit[200]{GeV}$ indicated above will be used in the $aW\gamma$ analysis for all operators. 
 
\begin{figure}[t!]\centering 
\includegraphics[height=0.35 \textwidth]{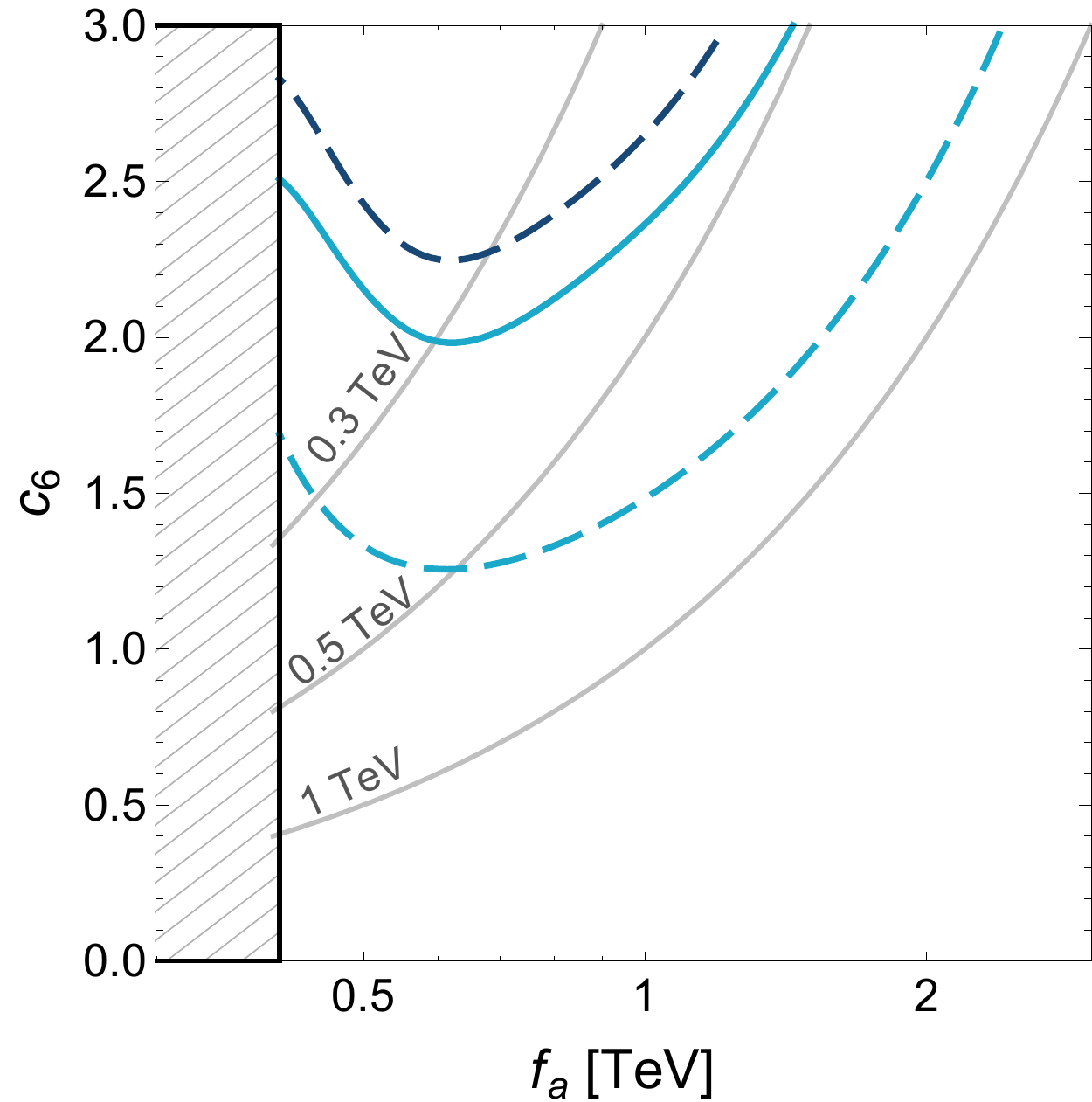} \hspace{0.4cm}
\includegraphics[height=0.35 \textwidth]{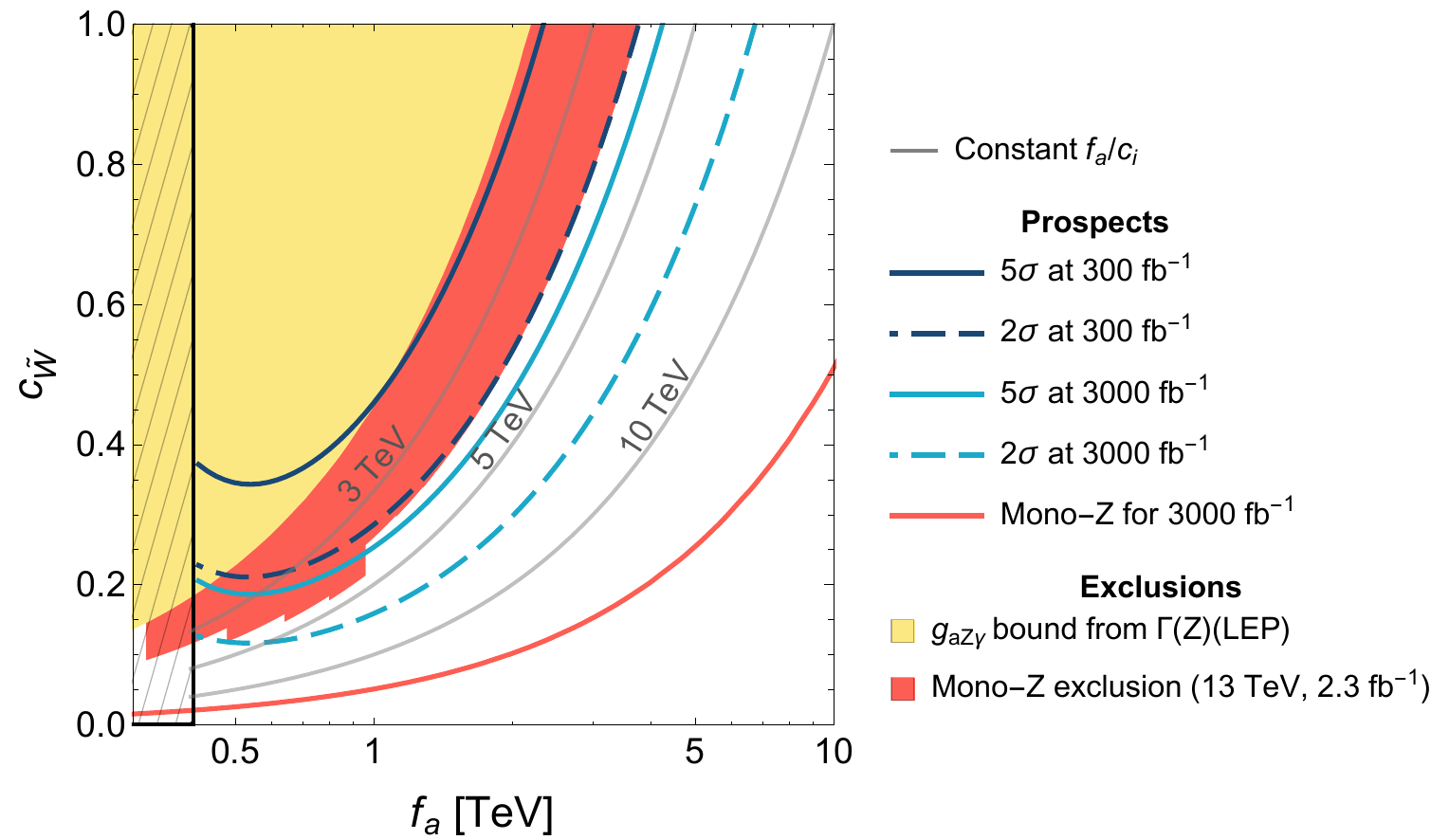}
\caption{\it\small   
Contours for $\Bsigma=2$ (dashed) and $\Bsigma=5$ (solid) sensitivity to  $p p \to aW^{\pm}\gamma$ ($W^{\pm}\to \ell^{\pm}\nu$) signal 
at the  LHC with $\sqrt{s}=\unit[13]{TeV}$ and for an integrated luminosity of $\unit[300]{fb^{-1}}$ (dark blue) and $\unit[3000]{fb^{-1}}$ (light blue), as 
a function of $\{f_a,c_i\}$. The left (right) panel shows the results obtained assuming that only the operator $\A_{6}$ (the combination of operators $\left( \mathcal{A}_\Wt-t_\theta^2 \mathcal{A}_\Bt \right)$) is contributing.  The hatched region corresponds to $f_a < 2\ETmiss^\mathrm{min}$, and is excluded by the EFT validity. The yellow region is excluded by the 
 bound on  $g_{aZ\gamma}$ reported in Eq.~\eqref{aZg_constraint}. The mono-Z exclusion region from $\unit[\sqrt{s}=13]{TeV}$ LHC 
 with $\unit[2.3]{fb^{-1}}$ of data is depicted by the red region. The gray reference lines correspond to constant values of $f_a/c_i$. 
 The region explored for $c_{\tilde W}$ would be superseeded by the bound from rare decays in Eq.~(\ref{rarebound}), within their 
 range of applicability, if the correlation between operators contributing simultaneously was disregarded.}
\label{AP_pho_figs}
\end{figure}
\begin{table}[t!]
 \begin{center}
 \begin{tabular}{| l  | c  c | c  c | }
 \hhline{~----}
 \multicolumn{1}{c|}{}&\multicolumn{2}{ c | }{$c_6$} &\multicolumn{2}{ c | }{$c_{\tilde{W}}$} \\
 \hline
Luminosity [fb$^{-1}$] & $300 $ & $3000$  & $300$ &$3000$ \\
[0.5ex] 
 \hline\hline
Optimal $\ETmiss^{\text{min}}$ [$\unit[]{GeV}$] & 300 & 330 & 220 & 220 \\ 
\hline
$(f_a/c_i)_{\max}$ [GeV]  & 470 & 950 & 3800 & 6800 \\
  \hline
\end{tabular}
\caption{\it\small  Optimal missing transverse energy cut $\ETmiss^{\text{min}}$, 
and $(f_a/c_i)_{\max}$ $2\Bsigma$ projected sensitivity reach   for 
$aW\gamma$ production,  for $\sqrt{s}=\unit[13]{\,TeV}$ and integrated luminosities 
$\unit[300]{\,fb^{-1}}$ and $\unit[3000]{\,fb^{-1}}$.}
\label{AP_pho_cuts}
\end{center}
\end{table}

Fig.~\ref{AP_pho_figs} shows the $2\Bsigma$ and $5\Bsigma$ sensitivity to the Lagrangian terms $c_\Wt\left( \mathcal{A}_\Wt-t_\theta^2 \mathcal{A}_\Bt \right)$ (right) 
and $c_6 \mathcal{A}_6$ (left), depicted in the $\{f_a,\,c_i\}$ plane and for $\unit[300]{\,fb^{-1}}$ and $\unit[3000]{\,fb^{-1}}$. 
The hatched area is excluded as it would correspond to $f_a \le 2\ETmiss^\mathrm{min}$ (corresponding to all signal events being outside the range 
of validity of the EFT).  
The $2\Bsigma$ exclusion sensitivity reaches $f_a / c_\Wt \lesssim \unit[3.8]{\,TeV}$ ($ \unit[6.8]{\,TeV}$) 
and $f_a / c_6 \lesssim \unit[0.4]{\,TeV}$ $(\unit[0.8]{\,TeV})$  for an integrated luminosity of $\unit[300]{\,fb^{-1}}$ ($\unit[3000]{\,fb^{-1}}$) of data, 
assuming the naive EFT validity criterium $f_a > 2 \ETmiss^\mathrm{max}$. \footnote{We warn the reader that, 
while the sensitivity to $c_\Wt$ is expected not to appreciably change 
if the strict EFT validity criterium $\sqrt{\hat{s}} < f_a$ were required, the sensitivity to $c_6$ 
could be significantly modified. We leave a more precise assessment of this effect for future work.}
We also note that when $f_a$ drops below $\unit[2]{TeV}$ (twice the energy of the highest bin in the $\ETmiss$ distribution in Fig.~\ref{AP_pho_metdists}) the 
reach in $f_a/c_i$ is diminished: this can be seen from Figure~\ref{AP_pho_figs} comparing the sensitivity curves with the gray reference 
lines which correspond to constant $f_a/c_i$. The rightmost parts of the sensitivity curves is ``parallel'' to the latter lines, signaling that 
the reach in $f_a/c_i$ is constant in this region. For $f_a\leq\unit[2]{TeV}$  instead, the sensitivity lines drift upwards compared to the reference lines, 
meaning that in that region the analysis is sensitive only to smaller values of $f_a/c_i$ than for the regions to the right. 
This effect is due to the fact that, as $f_a$ is diminished, the EFT validity gradually excludes the high-energy bins from the analysis, thus losing discrimination power
(note also that considering the strict EFT validity criterium $\sqrt{\hat{s}} < f_a$ would amplify this effect).
 
Alike to the conclusions in Sect.~\ref{Sect:monoWZ_Future} based on mono-$W$ and mono-$Z$ searches, associated $aW\gamma$ production at the LHC exhibits 
thus some (weaker but complementary) sensitivity to $c_{\tilde{W}}$ and to $c_6$ (for large values of the latter), and may potentially reach stronger 
constraints on $c_\Wt$ than those obtained from LEP data, see Sect.~\ref{Sect:aAZ}, but weaker than the limits from rare-decays-- see discussion 
around Eqs.~(\ref{rarebound}).
It should be possible to further increase the reach of the analysis
by using a sophisticated version of the transverse mass instead of $\ETmiss$, the so-called  $m_{T2}$ variable~\cite{Lester:1999tx,Barr:2003rg}.

\vspace{-0.3cm}

\subsubsection*{Decorrelating power}

\vspace{-0.2cm}

The vertices in Figs.~\ref{diagram_monoWZ} (left) and \ref{diagram_AP_VBF} contribute simultaneously to
the production of $a$ and $a\gamma$ in association with $W^{\pm}$, as well as to  
 VBF processes.  Eqs.~(\ref{AP_pho_verts}) and (\ref{eq_APVBFcoupl}) show a dependence of these processes on certain combinations of coefficients 
 and thus correlation effects are {\it a priori} expected. A combined analysis of $a/a\gamma$ production in association with $W^{\pm}$ and through VBF would enlarge the amount of kinematical information available, helping to disentangle their respective contributions.\footnote {The study of $a/a\gamma$ production in VBF is  significantly more involved, 
due to the difficulty in accurately modelling the important 
multijet background, and is left for the future.}

We consider for illustration the simultaneous action of  $c_2$ and $c_{\tilde{W}}$ on mono-$W$ signals and on $aW\gamma$ production. Note that it is precisely the contribution to the latter process of the $aZ\gamma$ vertex in Eq.~(\ref{AP_pho_verts})  which would allow to separate the contributions of those two operator coefficients if data were sensitive to both,  
 while from Eqs.~(\ref{eq_APVBFcoupl}) alone the two coefficients would have been tied in a  blind direction. Similar considerations  apply to other combinations of coefficients. 

 Nevertheless, our results indicate that, within the foreseen experimental prospects, no sensitivity is 
 expected via mono-$W$ and $aW\gamma$ production to couplings other than $c_{\tilde{W}}$ and $c_6$,  that is to $\{ \A_1,\,\A_2,\, \A_7, \A_8\}$ as these yield very suppressed contributions. The combination of mono-$W$ data and $aW\gamma$ production data will therefore allow to disentangle the measurement/constraint of $c_{\tilde{W}}$   (a custodial-invariant signal common to linear and non-linear EWSB)  from that of $c_6$ (only expected if the EWSB mechanism enjoys non-linear aspects and violates custodial symmetry).

\vspace{2mm}

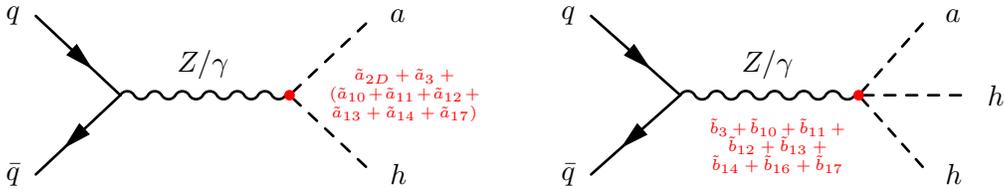
\begin{figure}[t]\centering
\input{Fdiagrams/mono_di_Higgs}
\caption{\it\small Main diagrams contributing to mono-Higgs (left) and di-Higgs (right) production in association with an ALP. The non-linear 
parameters entering each vertex are reported in the figure. Note that for the case of mono-Higgs, the contributions of $\at_{2D}$ and $\at_3$ are 
phenomenologically dominant, as the other coefficients enter the coupling with an extra suppression. The compact notation 
$\at_i\equiv c_i a_i$, $\bt_i\equiv c_i b_i$ has been adopted.}\label{Plots_monoh_dih}
\end{figure}

\vspace{-2mm}

\subsection{Higgs signatures}
\label{Sect:Higgs_signatures}
 Bosonic ALP-Higgs couplings are an interesting class of new signals which may be observable only within non-linear realizations of EWSB. Indeed, in the 
 latter case $aZh^{n}$ vertices with $n \geq 1$ are expected at LO, while they do not appear in the linear expansion below NNLO, as discussed 
 in Sects.~\ref{Sect:2pointF} and~\ref{linvsnonlin}.  They could induce especially interesting ALP signals:  non-standard Higgs decay ($h\rightarrow a Z$) including invisible Higgs decay  ($h\rightarrow a \nu\bar{\nu}$), associated ALP-Higgs production 
yielding an $h + \ETmiss$ ``mono-Higgs'' signature at the LHC, or even a $ hh+\ETmiss$ di-Higgs signature, see Fig.~\ref{Plots_monoh_dih}. 
The possibility that $aZh$ couplings of heavy pseudoscalars with masses in the $\unit[0.5-2]{TeV}$ range may yield observable signals 
in $pp \to a \to Zh  \,(h\to b\bar{b})$ was recently considered in the context of the linear expansion~\cite{Bauer:2016zfj}
(while the ALP signatures in Higgs and $Z$ decays are presented in this paper for the first time), 
stemming from one loop corrections to the NLO linear Lagrangian and from $d=7$ operators.

The set of operators in the Lagrangian  Eq.~\eqref{La_NLO} contributing {\it a priori} to those signals are 
$\{\A_{2D},\,\A_{3},\,\A_{10},\,\A_{11},\,\A_{12},\,\A_{13},\,\A_{14},\,\A_{17}\}$, see  Table~\ref{tab:contributions}. Nevertheless, only 
the first three will be phenomenologically relevant within the LHC prospects, as the contributions from the rest are comparatively much suppressed by extra 
powers of $1/(4\pi)$ and/or $m_a^2/v^2$ in the case of $\A_{17}$, see Feynman Rules~\ref{FR.Aah} and~\ref{FR.Zah}. This section focuses thus on the prospects for detecting $\A_{2D}$, $\A_3$ and $\A_{10}$, both taken one by one and in a combined analysis. 
The vertices relevant to the mono-$h$ signal and the non-standard Higgs decays are
\begin{equation}\label{anonima}
\begin{aligned}
a Z h \quad\to\quad&-\frac{4e v}{ \sdt  f_a}
\at_{2D}\, p_a^\mu+
\frac{1}{2\pi v f_a}(\at_3\st-\at_{10}\ct)(p_Z^\mu p_h\cdot p_Z-p_h^\mu p_Z^2)\,,\\
a \g h \quad\to\quad&\frac{1}{2\pi v f_a}(\at_3\ct+\at_{10}\st)\left(p_A^\mu p_a\cdot p_A-p_A^2 p_a^\mu\right)\,,
\end{aligned}
\end{equation}
showing that $a_3$ and $a_{10}$ enter  in two different combinations into the processes considered:
\begin{itemize}
\item   the mono-$h$ (and di-Higgs) signatures depend on the combination $\at_3\st-\at_{10}\ct$ via $Z$ exchange, and also on the orthogonal 
one $\at_3\ct+\at_{10}\st$ via $\gamma$ exchange -- see Fig.~\ref{Plots_monoh_dih};
\item  in contrast, the non-standard Higgs decays depends only on $\at_3\st-\at_{10}\ct$.
\end{itemize}

We are thus contemplating three coefficients and two distinct processes.  For the range of ALP masses used in the present numerical simulations ($m_a\le 1$ MeV), the fermionic-induced bound  on $c_{2D}$ in Eq.~(\ref{boundc2Dfermions}) would lead to disregard the impact of $\A_{2D}$ on   LHC data if that coupling were considered by itself. Nevertheless, given that a different combination of couplings is at work in rare decays and in LHC signals, for consistency with the perspective of exploring complementary approaches, and given that for larger ALP masses  the LHC signals would still be present in a refined analysis, the contributions of $\A_{2D}$ must be retained in the analysis to follow.
 With this strategy, the impact of $\A_3$ and $\A_{10}$  on the non-standard Higgs decay width is subdominant 
with respect to that of $\A_{2D}$, given the different $v$ dependence, see Eq.~(\ref{anonima}).
On the contrary, LHC data are instead quite sensitive to $c_3$ and $c_{10}$, 
in addition to $c_{2D}$, given the stronger momentum dependence of $\A_3$ and $\A_{10}$. 
 This suggests that, in order to disentangle the  contributions  from $\A_3$ and $\A_{10}$,  a detailed study of the kinematic 
 distributions of the mono-Higgs channel would be necessary, together with the combination of these results with those stemming from bounds on $h\to {\rm BSM}$  from Higgs signal strength measurements. On the other side, $a_3$ and $a_{10}$ have a similar overall impact on the total mono-$h$ cross section. For the sake of simplicity, we will then 
 consider here only the impact of $a_{2D}$ and $a_3$, separately and combined, deferring the detailed study of $\A_{10}$  to a future work.

A remark on the range of values of the operator coefficients is pertinent. Generally speaking, large values  
correspond to strongly interacting regimes, and NDA suggests  $c_{i}\leq 1$, with the bound saturated in the strong regime. Nevertheless, as discussed in 
Sect.~\ref{Sect:basis}, a factor $(f/v)$ has been implicitly absorbed  in the definition of the parameter $\at_{2D} = c_{2D}a_{2D}$, where $a_{2D}$ is 
the coefficient of the one-Higgs contribution in the polynomial $\F_{2D}(h)$.  The ratio $\xi\equiv v^2/f^2$ is not a parameter from the effective theory point of 
view, but it is currently bounded to be $\lesssim0.2$ in concrete models~\cite{deFlorian:2016spz} such as composite Higgs scenarios. Numerically, 
this would translate into an enhancement of a factor $f/v\gtrsim 2.3$, which implies that the absolute value of the parameter $\at_{2D}$ can 
naturally exceed by at least 2-3 units the bare NDA constraint $\at_{2D}\leq 1$. In this section we will assume a maximum absolute value $\tilde{a}_\max = 3$ 
for both $\at_{2D}$ and $\at_3$, along the same lines as the analysis presented in Sect.~\ref{Sect:AP_gamma}.

\begin{figure}[t!]
\centering
\includegraphics[width=\textwidth]{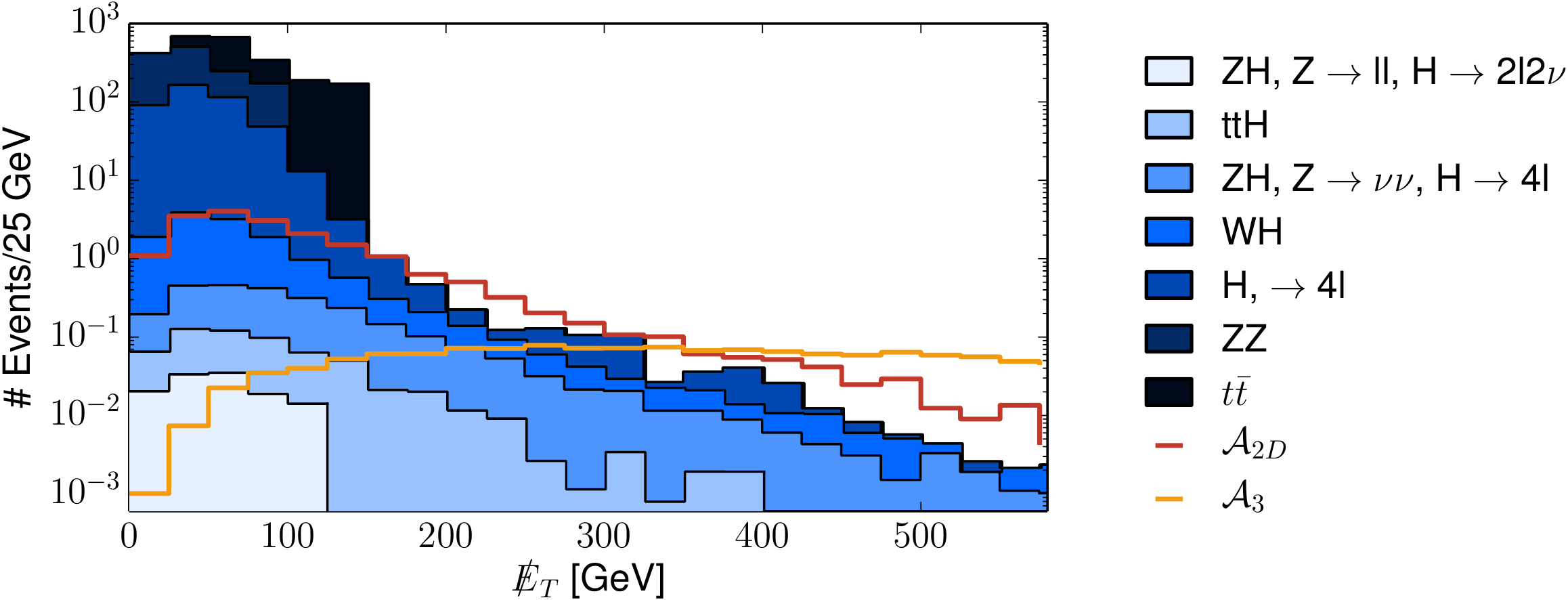}
 \caption{\it\small $\ETmiss$ distributions for $4\ell + \ETmiss$ signal and background for $\sqrt{s}=\unit[13]{TeV}$ and $\unit[3000]{fb^{-1}}$ of integrated 
 luminosity, after applying the selection cuts from~\cite{Brooijmans:2016vro}. SM $\ETmiss$ background distributions are obtained 
 directly from~\cite{Brooijmans:2016vro}, and the signal $pp \to a h$ ($h \to 4 \ell$) $\ETmiss$ distribution is shown for 
 $\A_{2D}$ (red) and $\A_{3}$ (orange).}\label{plot.monoH_MET}
\end{figure}

\subsubsection{Mono-Higgs: $p p \to a\, h$}
\label{Sect:monoH}

\noindent The process $ p p \to a h \,(h \to 4\ell)$ is considered next at 13 TeV LHC, and it follows  the mono-Higgs analysis from the ``Les Houches 2015'' 
report~\cite{Brooijmans:2016vro}, considering both $300$ fb$^{-1}$ and  $3000$ fb$^{-1}$ integrated luminosity. Our signal sample is produced with 
\texttt{MadGraph5\_aMC@NLO}~\cite{Alwall:2014hca}, passed on to \texttt{Pythia 8}~\cite{Sjostrand:2007gs} for showering and hadronization 
and then to \texttt{FastJet}~\cite{Cacciari:2011ma} for jet reconstruction. 
The reconstructed events are finally filtered imposing the selection cuts from Ref.~\cite{Brooijmans:2016vro}, for a consistent comparison 
with SM backgrounds which are taken precisely from that reference.

The $\ETmiss$ spectrum can be used to disentangle 
the new interactions from the SM background. This applies in particular to $\A_3$,  which induces a strong momentum dependence  
through both the $aZh$ and the $a\gamma h$ contributions to the mono-$h$ signal.  This is  illustrated in Fig.~\ref{plot.monoH_MET}  for 
an integrated luminosity of $\unit[3000]{fb^{-1}}$. As expected, the $\ETmiss$ spectrum 
produced by $\A_3$ (orange line) is harder compared to that produced by $\A_{2D}$ (red line) while, at the same time, the total (no cuts) 
integrated cross section for the signal 
generated with $\A_3$ is manifestly lower than the one induced by $\A_{2D}$.

In order to quantify the potential for observing in future LHC data a mono-Higgs signal generated by either of the two operators 
$\A_{2D}$ and $\A_3$,  the analysis is done in two different stages.\footnote{Only tree level insertions of the operators $\A_{2D}$ and $\A_3$  will be considered below.}  

\vspace{-3mm}

\subsubsection*{One operator at a time}

\vspace{-2mm}

In a first stage, each of the two relevant operators, $\A_{2D}$ and $\A_3$, is considered individually, {\it i.e.} assuming that only one of the 
coefficients $c_{2D}$ and $c_3$ has a non-zero value. With this choice, 
the procedure already described in Sect.~\ref{Sect:AP_gamma}, Eqs.~(\ref{Asimov_sensitivity}) and (\ref{sib}), is applied. 
The significance  is computed as a function of $f_a/c_i$, integrating the distributions in Fig.~\ref{plot.monoH_MET}  from a 
chosen $\ETmiss^\mathrm{min}=\unit[150]{GeV}$ (which removes most of the background contribution) up to $\ETmiss^\mathrm{max}=f_a /2$, according to the naive validity criterium (recall the discussion in Sect.~\ref{Sect:Validity}).

\begin{figure}
\includegraphics[height=0.37\textwidth]{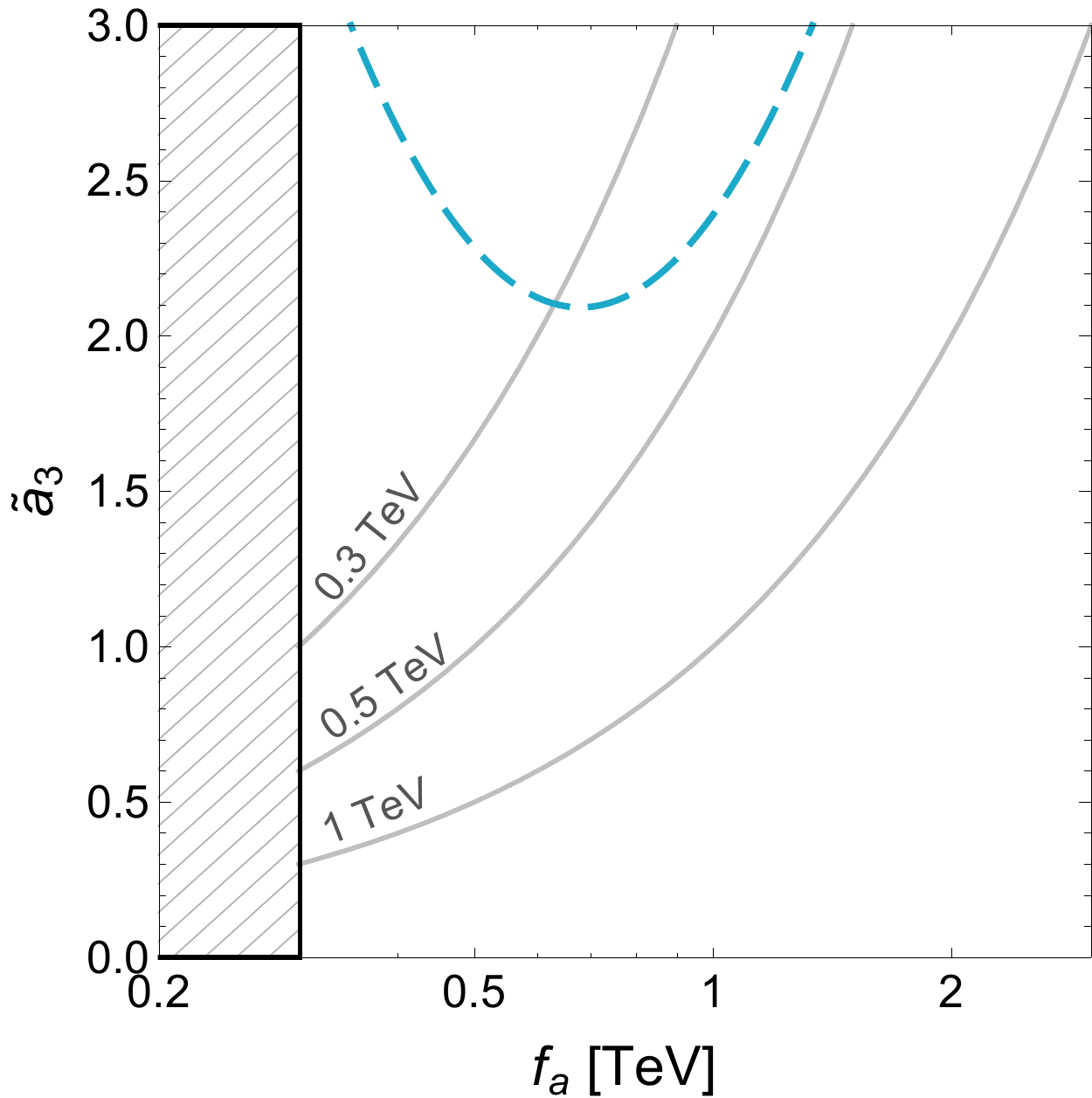}\hspace{0.3cm}
\includegraphics[height=0.37\textwidth]{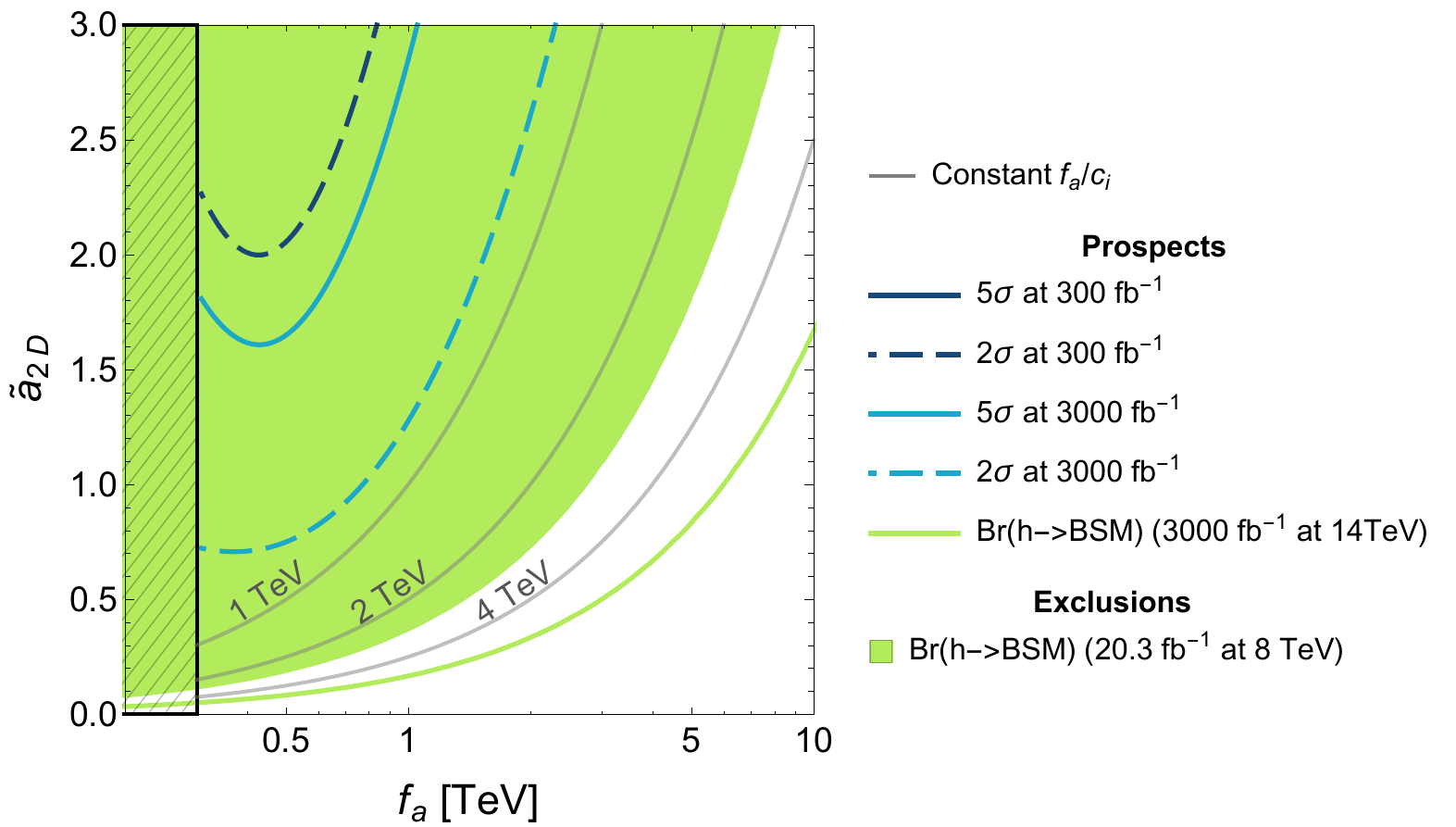}\\
 \caption{\it \small Contours for $\Bsigma=2$ (dashed) and $\Bsigma=5$ (solid) sensitivity to mono-H signal at the  LHC with $\sqrt{s}=\unit[13]{TeV}$ and for 
 an integrated luminosity of $\unit[300]{fb^{-1}}$ (dark blue) and $\unit[3000]{fb^{-1}}$ (light blue), as a function of $\{f_a,c_i\}$ . The left (right) panel shows the 
 results obtained assuming that only the operator $\A_3$ ($\A_{2D}$) is contributing.  The hatched region corresponds to $f_a < 2\ETmiss^\mathrm{min}$, and is 
 excluded by the EFT validity, while the green region is excluded by the 
 bound on $\Br(h\to {\rm BSM})$ reported in Eq.~\eqref{Limit_aZh_HiggsDecay} 
 (for the left-panel, this bound is not visible). The gray reference lines correspond to constant values of $f_a/c_i$. 
 }\label{plot.sensitivity_c2D_c3}
\end{figure}

Fig.~\ref{plot.sensitivity_c2D_c3} shows the $\Bsigma=2$ and $\Bsigma=5$ sensitivity regions obtained for the two coefficients $\at_{2D}$ and $\at_3$ 
individually (see Eq.~\eqref{Asimov_sensitivity}), and integrated luminosities of $\unit[300]{fb^{-1}}$ and $\unit[3000]{fb^{-1}}$. 
As shown in Fig.~\ref{plot.sensitivity_c2D_c3}, only a very restricted region of the parameter space for $\A_3$ is accessible within 
$\unit[3000]{fb^{-1}}$ at the LHC, due to its very small cross-section: it results in a $2\Bsigma$ sensitivity to $f_a / \at_3 \lesssim \unit[470]{\,GeV}$, which is 
expected to further degrade if the strict EFT validity criterium $\sqrt{\hat{s}} < f_a$ would be considered.

In contrast, Fig.~\ref{plot.sensitivity_c2D_c3} illustrates 
that mono-Higgs signatures in the $h \to 4 \ell$ final state at HL-LHC have the potential to explore some region of parameter space 
for $\A_{2D}$ within the range of EFT validity. The $2\Bsigma$ exclusion sensitivity 
reaches $f_a / \at_{2D} \lesssim \unit[340]{\,GeV}$ ($ \unit[780]{\,GeV}$) for an integrated luminosity of $\unit[300]{\,fb^{-1}}$ ($\unit[3000]{\,fb^{-1}}$) of data.
While considering the strict EFT validity criterium would somewhat degrade these limits, we also stress that considering other final states,
e.g. $h \to \gamma\gamma$, $h \to b \bar{b}$, would significantly increase the sensitivity of this search, and we leave such a study for the future.

These results can be contrasted with the bounds on $f_a/\at_i,\,(i=2D,\,3)$ inferred from the current upper 
limit on $\Br(h\to {\rm BSM})$ in Sect.~\ref{Sect:hZax_Br}, which is depicted as a green region in Fig.~\ref{plot.sensitivity_c2D_c3} (right). 
If only $\A_{2D}$ is considered, the area of parameter space which is to be probed by LHC with $\unit[3000]{fb^{-1}}$ is already ruled out by that limit. 
This is not the case when only $\A_3$ is considered, since its contribution to $h\to {\rm BSM}$ is very suppressed. Nevertheless, cancellations 
might exist amongst the contributions of those two operators to  non-standard Higgs decays, in regions of the parameter space where a mono-Higgs signal could be expected at 
a testable level. This is the motivation for the second stage in the analysis: a combined study where both operators are considered simultaneously. 

\vspace{-3mm}

\subsubsection*{Combination of the two operators $\A_{2D}$ and $\A_3$}

\vspace{-2mm}

In this case of simultaneous consideration, the shape of the $\ETmiss$ 
distribution after applying the analysis cuts can be estimated, for an arbitrary choice of $f_a$, $\at_{2D}$ and $\at_3$, 
as
\begin{equation}
(f_a/\unit{TeV})^{-1}\left[\at_{2D}^2 \,x_k +  \at_3^2 \,y_k + \at_{2D} \at_3 \,(z_k - x_k - y_k)\right]\,,
\end{equation} 
where the index $k$ runs over the distribution bins, and $x_k,\, y_k$, and $z_k$ represent the $\ETmiss$ prediction in the $k$-th bin 
obtained with $f_a=\unit[1]{TeV}$ and for the configurations $(\at_{2D} = 1, \at_3=0)$, $(\at_{2D}=0, \at_3=1)$ and $(\at_{2D}=1, \at_3=1)$, respectively.  
With this estimate of the $\ETmiss$ distribution one can easily compute the maximal projected sensitivity to mono-Higgs signals, varying the lower cut in 
missing transverse energy, $\ETmiss^\mathrm{min}$, in order to maximise the sensitivity $\Bsigma$ at each   $\{ f_a / \at_{2D},\,f_a/\at_3 \}$ point.

\begin{figure}
\includegraphics[width=.9\textwidth]{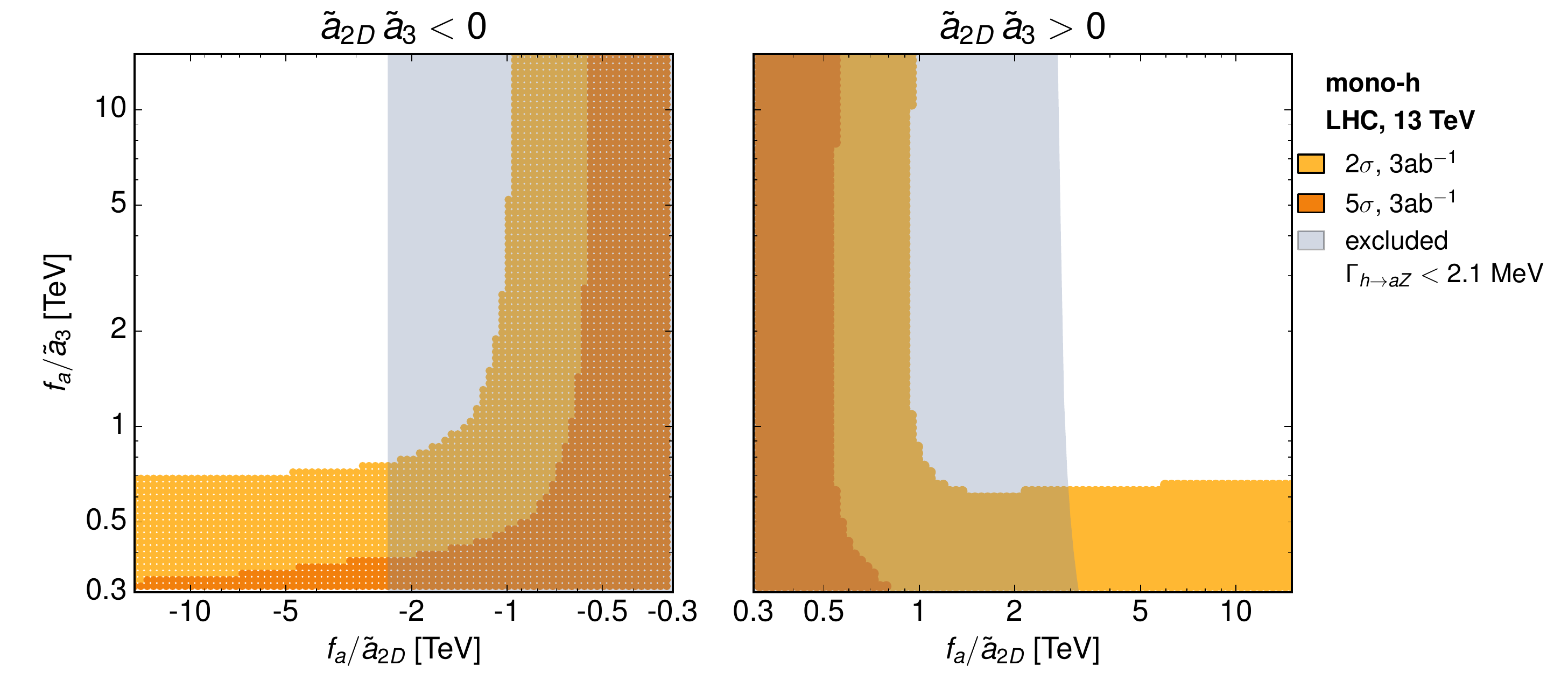}
\caption{\it \small Contours for $2\s$ and $5\s$ sensitivity to the mono-H signal at the LHC with $\sqrt{s}=\unit[13]{TeV}$ and for an integrated luminosity 
of $\unit[3000]{fb^{-1}}$, for different values of the parameters $(f_a/\at_{2D})$ and $(f_a/\at_3)$. The left (right) panel shows the result obtained 
for opposite-sign (same-sign) scaling factors.
The gray shaded region is excluded by the bound on $\Br(h\to {\rm BSM})$ reported in Eq.~\eqref{Limit_aZh_HiggsDecay}. }\label{plot.monoH_c2dc3}
\end{figure}

The results are shown in the scatter plot in Fig.~\ref{plot.monoH_c2dc3}. The yellow (orange) points are those for which there exists a lower $\ETmiss$ cut within the EFT validity region, that allows to observe
a mono-Higgs signature with a significance of least 2 (5) $\Bsigma$ at  the $13$ TeV LHC  with $\unit[3000]{fb^{-1}}$.
The left and right panels distinguish the two cases in which $\at_{2D}$ and $\at_3$ have either opposite or same sign. 
In both cases, in the limit $f_a/\at_{2D}\to \infty$, the $2\Bsigma$ and $5\Bsigma$ sensitivity curves for $f_a/\at_3$ converge towards 
 values close to the optimal ones found in the $\A_3$-only analysis (the discrepancy is due to the different treatment of the $\ETmiss^\mathrm{min}$ cut).
  An analogous behavior is also observed in the orthogonal direction.
 
More interesting is the region where $\at_{2D}$ and $\at_3$ are close in absolute value. In particular, it shows that the contributions to the 
mono-Higgs process stemming from the two operators produce destructive interference when $\at_{2D}$ and $\at_3$ have the same sign: for 
$\at_{2D}\simeq \at_3$ (right panel) the signal is reduced compared to the case in which one of the two operators dominates, and the sensitivity is 
therefore lower in this region of the parameter space. On the other hand, for $\at_{2D}\simeq - \at_3$ (left panel) constructive interference effects 
enhance mono-Higgs production, so that the LHC would be sensitive to larger values of $f_a/\at_i$ than in the one-operator case.
 
As with the previous study, the results obtained for projected mono-Higgs searches in the $(\at_{2D}, \at_3)$ plane can be easily contrasted with the 
bound inferred in Sect.~\ref{Sect:hZax_Br} from the current upper limits on $\Br(h\to {\rm BSM})$. This is depicted as a grey-shaded region 
in Fig.~\ref{plot.monoH_c2dc3} and seen to be more stringent for same sign $\at_{2D}$ and $\at_3$, as no cancellation can then take place   
 in the dominant expression in $\Br(h\to {\rm BSM})$, see Eq.~\eqref{Hinv_bound_kh}.
 
As a result of the combination of the existing bound with the projected reach, it appears that mono-Higgs searches may be useful for probing a relevant 
region of the parameter space, namely that with $\unit[300]{GeV}\lesssim |f_a/\at_3| \lesssim \unit[700]{GeV}$, 
where the lower bound is a direct consequence of requiring the EFT validity. In this region, $|f_a/\at_{2D}|$  may be no smaller than $2$-$\unit[3]{TeV}$, as 
lower values are already excluded by the $h\to {\rm BSM}$ constraint that we derived from present data in Sect.~\ref{Sect:hZax_Br}. 
Overall, we find that although mono-Higgs searches at the LHC are sensitive to the presence of both operators $\A_{2D}$ and $\A_3$, they are not competitive in constraining $f_a/\at_{2D}$ with 
the Br($h\to {\rm BSM}$) bound, neither with the fermionic-induced bound in Eq.~(\ref{boundc2Dfermions}) when that coupling is considered just by itself. On the other hand, they are more sensitive to the presence of $\at_3$ and therefore they may provide valuable, 
complementary, information in the study of the ALP's coupling to the Higgs.

In conclusion, if interpreted in terms of the presence of a light pseudo-Goldstone boson, and barring fine-tunings, the observation of a mono-Higgs 
signature at the LHC represents a smoking gun of non-linearity in the EWSB sector, as couplings such as $aZ(\g)h$ are not to be found in the NLO Lagrangian 
of linear EWSB setups (see~~\ref{FR.AAa}, \ref{FR.Zah}). Within the  effective Lagrangian in Eq.~\eqref{Lchiral}, the observation of this signal at foreseen 
LHC data can only be attributed to the presence of $\A_3$ (or eventually $\A_{10}$), as  $\at_{2D}$ is out of reach in that data set given the range of values allowed 
by the current bounds.

\subsubsection*{A comment on di-Higgs production}
\label{Sect:diHiggs}
The $aZhh$ interaction allows for di-Higgs final state, due to a quark-initiated $h h + \ETmiss$ production via Drell-Yan (see Fig.~\ref{Plots_monoh_dih} (right)). 
This is in contrast to di-Higgs production in the SM, which is exclusively 
gluon-fusion initiated. Moreover, the presence of $\ETmiss$ in the final state could serve as an additional handle to suppress SM backgrounds to the 
di-Higgs process. 
This discussion highlights that $a$-$h$ interactions could constitute a very promising avenue for non-linear ALP phenomenology at the LHC, which 
we intend to explore in the future.

%%%%%%%%%%%%%%%%%%%%%%%%%%%%%%%%%%%%%%%%%%%%%%%%%%
%%%%%%%%%%%%%%%%%%%%%%%%%%%%%%%%%%%%%%%%%%%%%%%%%%
\subsection{Coupling to fermions}
%%%%%%%%%%%%%%%%%%%%%%%%%%%%%%%%%%%%%%%%%%%%%%%%%%
%%%%%%%%%%%%%%%%%%%%%%%%%%%%%%%%%%%%%%%%%%%%%%%%%%
In this paper we have focused on the relation of the ALP with the EWSB sector via bosonic operators, and explored the impact of couplings of the ALP to SM bosons. However, in \mbox{Sects.~\ref{Sect:2pointF_linear} and ~\ref{Sect:basis}} we noticed that bosonic operators 
would lead to ALP-fermion couplings via a field redefinition. 
\begin{figure}[t!]
\begin{center}
\hspace{-1.cm}
\includegraphics[width=.8\textwidth]{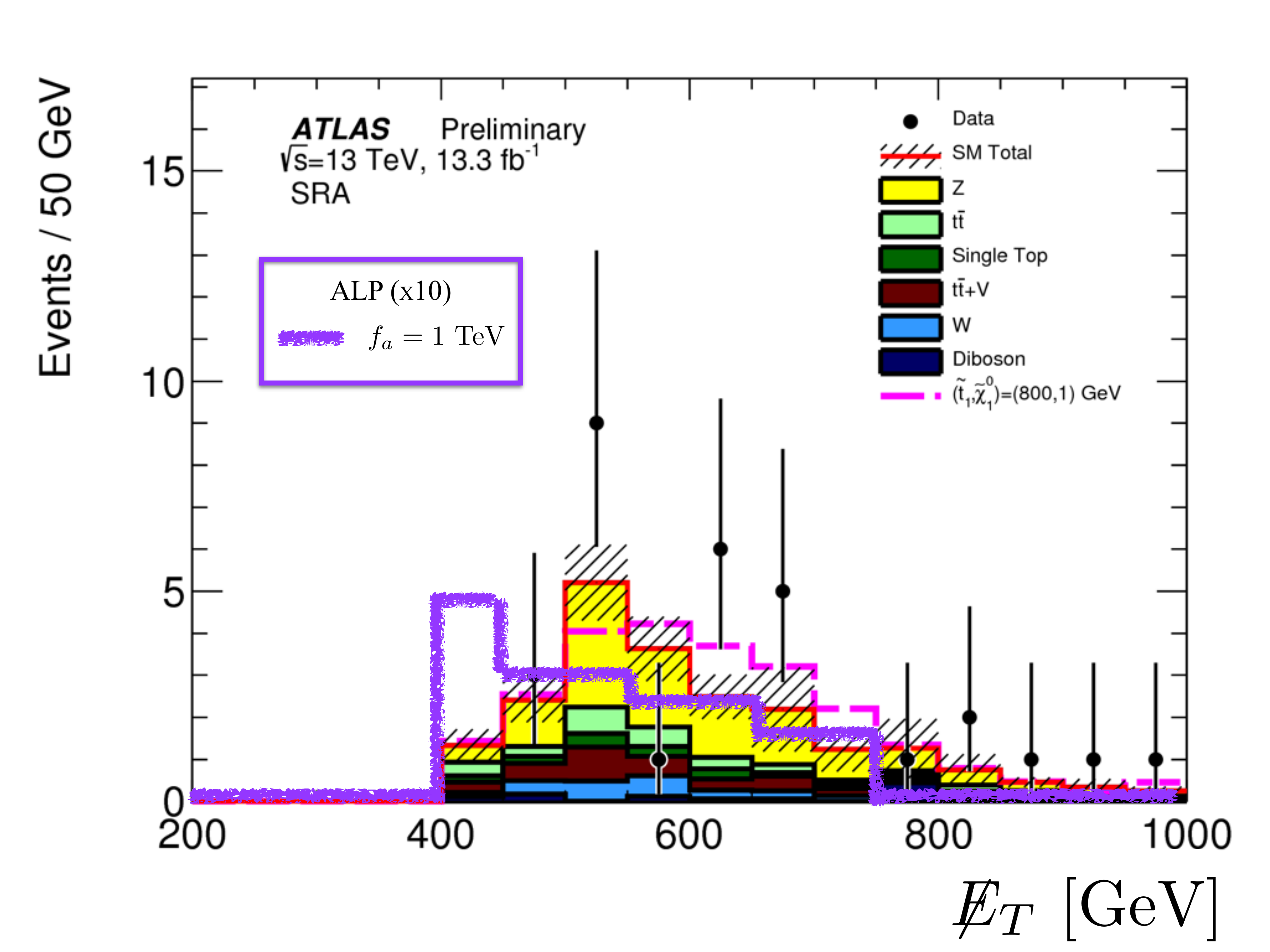}
\caption{\it \small Missing energy distribution for the production of an light ALP in association with $t \bar t$ for 13.3 fb$^{-1}$ of 13 TeV data. The normalization has been chosen with $f_a$= 1 TeV and then multiplied by a factor 10. We show the corresponding simulation of supersymmetric scenarios by ATLAS, as well as their event count.} \label{ttbar_met}
\end{center}
\end{figure}

Although the bounds we obtained in Eqs.~(\ref{boundaphifermions}) and (\ref{boundc2Dfermions}) when considering  operators one at a time are very strong,  it is worth exploring complementary searches at the LHC. 
The structure of these fermionic couplings is very specific, proportional to the Yukawa matrices, see Feynman rules in App.~\ref{App:feynman_rules}. One would then expect the ALP to couple more strongly to third generation quarks,  provided the matrices $X_\psi$ in Eq.~(\ref{general-fermionic-Yukawa}) are generic.  
We then consider the characteristics of the leading ALP production in association with a $t\bar t$ pair at LHC. 

For ALPs stable on LHC scales, this final state is similar to searches for supersymmetric scenarios, where two stops are strongly produced and produce a signature of  $t\bar t$ in association with two neutralinos (Dark Matter candidates). For example, via the LO coupling $c_{2D}$ the production cross section of the final state $t \bar t$+ALP, where the ALP is emitted as final state radiation --see \ref{FR.auu},  is given by
\bea
\sigma (p \, p \to t \, \bar t \, a) [\sqrt{s} = 13 \textrm{ TeV }] = c^2_{2 D} \, \left(\frac{1 \textrm{ TeV}}{f_a}\right)^2 \, (50 \textrm{ fb}).
\eea

In these searches, final states are selected by requiring a number of jets, b-jets with characteristics matching those of top decays. More importantly, a substantial cut on missing energy is required. For example, in a recent study with 13 TeV data by ATLAS~\cite{ATLAS-CONF-2016-077},  the cut on missing energy for the channel of interest ({\bf \tt TT}) (topology of two tops) is 400 GeV . In our scenario, with single-production of a light pseudoscalar via strong production of two tops, the distribution of missing energy is not as hard as in scenarios where heavy stops are pair produced and inject a large boost into the neutralino. This is shown in Fig.~\ref{ttbar_met}, where we compare our results for $f_a$= 1 TeV with the ATLAS data and Monte Carlo simulations for a supersymmetric scenario with 800 GeV stops and a light neutralino.  

This type of analysis opens the way to further phenomenological explorations of the fermionic signals associated to ALP production. This is most relevant and promising in order to tackle the ALP-fermionic couplings identified in App.~\ref{App:FermionicCouplings}, which are part of the 
complete NLO basis of operators --bosonic and fermionic-- involving one ALP and established in this work. See also the phenomenological signals discussed just before Sect.~\ref{Sect:monoZW_correl}.

%%%%%%%%%%%%%%%%%%%%%%%%%%%%%%%%%%%%%%%%%%%%%%%%%% HERE GO THE CONCLUSIONS
%%%%%%%%%%%%%%%%%%%%%%%%%%%%%%%%%%%%%%%%%%%%%%%%%%

\section{Summary and Outlook}
\label{Sect:Conclusions}

In this paper we have developed a systematic approach to describe interactions of an axion or an axion-like particle (ALP) with special attention to the  
sector responsible for electroweak symmetry breaking (EWSB), obtaining the complete --bosonic and fermionic-- NLO Lagrangian in the case that the latter is non-linearly realized. With this theoretical framework in place, we have then studied new collider 
phenomenology associated with ALPs, as well as explored the sensitivity of the LHC in the high-luminosity phase (HL-LHC). Both the approach 
and the phenomenological results in this paper are novel, and will hopefully guide new searches at the LHC and the study of  complementarity 
with other experiments at lower energies. 
 \begin{figure}[t!]\centering
\includegraphics[width=.85 \textwidth]{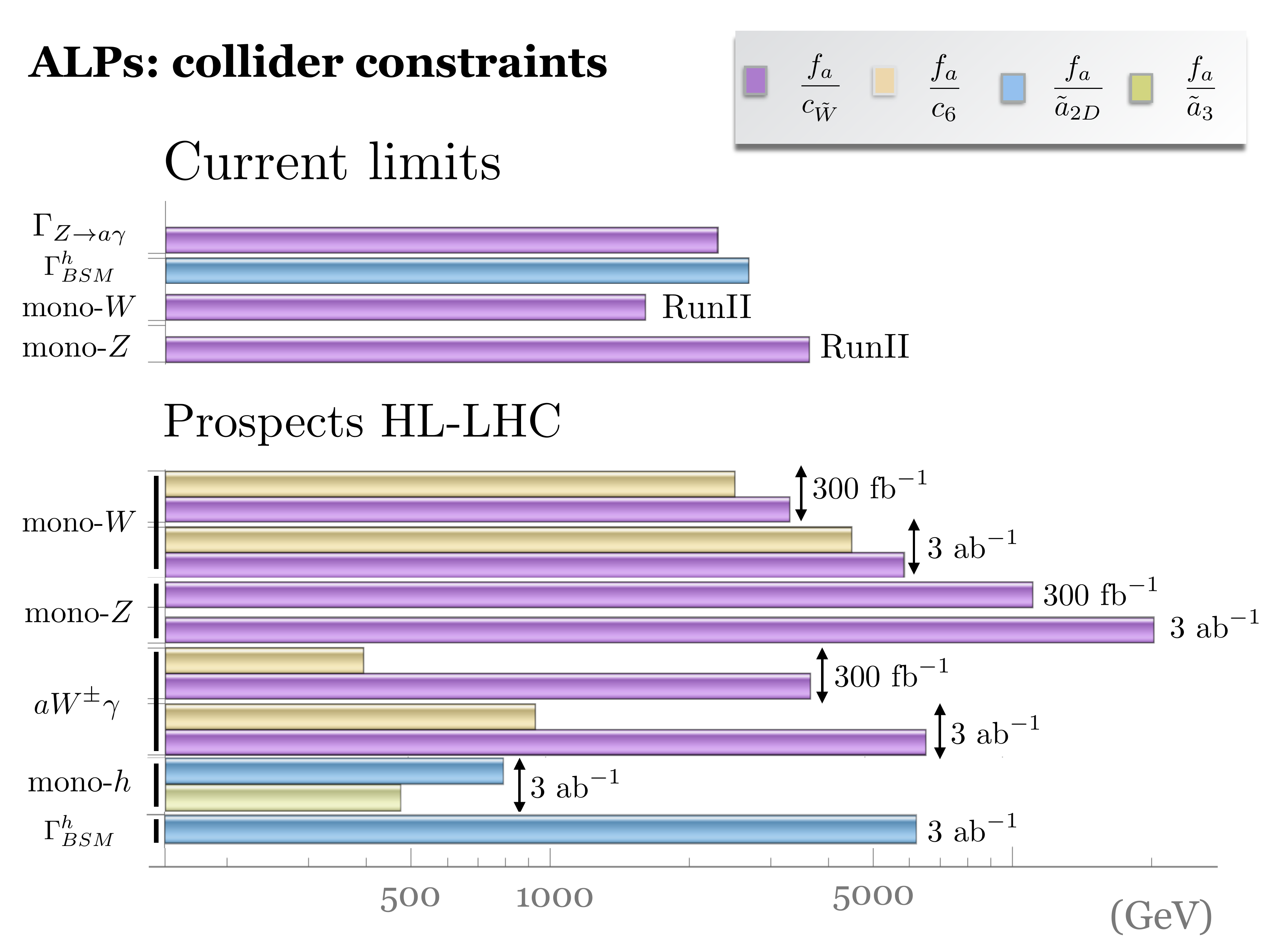}
 \caption{\it \small Summary of the most significant constraints stemming from the studies on tree-level ALP couplings presented in this work, upon the 
 assumption $g_{a\g\g}=0$ or equivalently $c_{\tilde{B}}=-t_\theta^2 c_{\tilde{W}}$. The bars for $\Gamma_{Z\to a\gamma}$, $\Gamma^h_{BSM}$, $\text{mono}-W$ and  $\text{mono}-Z$ correspond to $95\cl$, both in existing constraints and expected reaches for $\sqrt{s}=\unit[13]{TeV}$, inferred in Sects.~\ref{Sect:PhenoAnalysisPresent} and \ref{Sect:PhenoAnalysisFuture} respectively. The bars for $aW^{\pm}\gamma$ and $\text{mono}-h$, instead, indicate the 2\Bsigma projected reach of given searches at the LHC with $\sqrt{s}=\unit[13]{TeV}$, see Sect.~\ref{Sect:PhenoAnalysisFuture}. Systematic uncertainties are taken into account for the present constraints but are neglected in the projected ones.
  }\label{plot:summary}
\end{figure}

\vspace{-2mm}

\subsubsection*{Theoretical developments}

\vspace{-2mm}

 Neglecting ALP masses, we have developed a complete list of bosonic operators under two scenarios, with EWSB linearly and non-linearly realized, 
 valid in all generality  for any value of the axionic scale larger than the electroweak scale (and in the non-linear case also larger than its implicit electroweak BSM scale).  In the linear case, in which the couplings 
 involving an ALP first appear at $d=5$,  special attention has been paid to recalling the subtle effect of the 
 operator $(\Phi^\dag\overleftrightarrow{D}_\mu\Phi)\frac{\de^\mu a}{f_a}$, which induces a contribution to the two-point 
 function involving {\it longitudinal} gauge bosons, and can be removed via a Higgs field redefinition. This redefinition 
 generates new couplings of the ALP to fermions, with the distinctive feature of being proportional to the Yukawa couplings. 
 
 In the non-linear realization, we have employed a systematic approach to classify the new operators order-by-order, and much care 
 has been paid to define the expansion in both its non-linear and ALP sectors.  A complete and non-redundant basis of operators  involving 
 an ALP has been determined, even though  the impact analysis has focused on the bosonic couplings. Several interesting features arise when 
 considering ALPs coupled to a non-linear realization of EWSB, in particular the existence --already at the leading order in the 
 derivative expansion-- of interaction vertices involving the Higgs and gauge bosons with the ALP. This is due to the fact that the two-point function stemming from the operator $\A_{2D}$  cannot be entirely traded 
 by fermionic couplings (in contrast to the linear case above).  Additionally, we find that the non-linear effects induce new Lorentz structures beyond those 
 in the {\it traditional} (linear) ALPs couplings.

 Furthermore, a detailed comparison of the differences and correspondences between 
 the operators in the linear and non-linear setups has been developed, as well as a prospective study on how to  disentangle {\it a priori} 
 both expansions if a signal is found.

For the most part, phenomenological studies on the ALP effective Lagrangians have focused on couplings to photons, gluons and fermions. 
However,  if the ALP couples to photons SM gauge invariance also implies the existence of similar couplings 
to the massive gauge bosons, irrespective of whether the mechanism behind EWSB gives rise to a linear or a non-linear expansion. In this work we propose for the first time signals in accelerator searches with heavy SM bosons in the final state.

In this paper we have obtained new constraints on ALP couplings to SM particles, as well as provided a guide for future searches of 
ALPs and the sensitivity HL-LHC could reach, for ALP scales of $\cal{O}($TeV) or somewhat above.  Special attention has been paid 
to the consistency of the kinematic regions used for each search with the assumption of validity of the ALP expansion in powers of $1/f_a$.

\vspace{-2mm}

\subsubsection*{Current constraints}

\vspace{-2mm}

We started by looking at new constraints on (linear) ALP couplings to all EW gauge bosons.
 In particular, we looked for observables sensitive to the linear combination of $SU(2)_L\times U(1)_Y$ operators 
 $(c_\Bt, \, c_\Wt)$  orthogonal to the coupling to photons, i.e. orthogonal to $g_{a\gamma\gamma} \sim c_\Wt s_\theta^2 + c_\Bt c_\theta^2$. 

To account for the strong constraints on the value of $g_{a\gamma\gamma}$ we then imposed $ c_\Bt \simeq -  t_\theta^2 c_\Wt$ in our analyses, 
effectively reducing the number of parameter by one.  In Fig.~\ref{plot:summary} one can see that LEP constraints on the invisible width of 
the $Z$ boson and LHC searches for final states with one massive boson and missing energy (mono-$Z$ and mono-$W$ channels) provide handles to probe 
the Wilson coefficient $c_\Wt$. We find that mono-$Z$ limits impose at present a constraint $f_a/c_{\tilde W} \gtrsim \unit[4]{TeV} $.  

We also discussed the impact on bounds from rare-decays to mesons and missing energy, and how they provide a complementary approach to accelerator searches.
Besides the stringent constraints existing in the literature on $f_a/c_{\tilde W}$ from the former searches --which strictly speaking only apply when all other operators are set to zero-- a similar new bound on the strength of the linear operator $\O_{a\Phi}$  has been obtained here.

In non-linear realizations of EWSB in particular, many other operators affect LHC physics.  For ALP masses under $3$ GeV, data on rare meson decays allowed to strongly bound $c_{2D}$ if considered by itself.  Furthermore, of particular interest are operators which induce new type of couplings, 
specifically new couplings of the ALP to Higgs particles, e.g. ALP-$Zh$ or ALP-$Zhh$, which are dominantly generated by the non-linear 
operators $\A_{2D}$, $\A_3$ and $\A_{10}$. In the LHC RunI the coupling ALP-$Zh$ can be probed via  non-standard Higgs decays; if the impact of the different operators contributing is considered one at a time, a bound on $f_a/\tilde a_{2D}$ of the order of 3 TeV follows for ALP masses in the range $3$-$34$ GeV.

\vspace{-2mm}

\subsubsection*{Future sensitivity}

\vspace{-2mm}

We then moved on to examine the capability of the HL-LHC to search for ALPs. Apart from improvements on 
current channels (non-standard Higgs decays, mono-$W$ and mono-$Z$), we proposed and evaluated possible new channels at 13 TeV which could dramatically change 
 our understanding of ALPs both in the linear and non-linear realizations. 
 
 Future improvements of mono-$Z$ searches with 3 ab$^{-1}$ of data could bring the collider sensitivity to the linear operator coefficient $f_a/c_{\tilde W}$ to above 
 20 TeV. But the most striking signatures stemming from bosonic operators,  i.e.  mono-Higgs and associated $W^\pm \gamma$ production plus missing energy, would access the non-linear 
 operators mentioned before,  $\A_3$ and $\A_{2D}$, and a new one, $\A_6$.  We also propose the study of different channels, 
 like mono-$W$ in combination with $aW\g$ production, to disentangle the presence of  two different operators.  On the other hand, the very sensitive mono-$Z$ and mono-$W$ signals may play a specially important role in probing fermion-ALP interactions; this will be tackled in a future study.
 
Besides these signals, we proposed to use the searches on stops in on-shell top final states to look for ALPs, whose couplings to 
quarks are derived from couplings to the bosonic sector and are proportional to the fermion mass.    

This study motivates further work on ALP physics beyond the usual framework of couplings to photons and gluons, and more emphasis was 
placed on the effects in the sector responsible for electroweak symmetry. Additionally, we propose to perform dedicated experimental 
analyses in channels like mono-Higgs and new channels involving the ALP and two bosons in the final state, such as $W^\pm \gamma$ and missing energy.

Although in this paper we presented a rather comprehensive analysis of the effective theory for ALPs as well as their phenomenology,  
there are a number of open issues that deserve further study. To name a few: the extension of the collider analysis to higher ALP mass regions (including signals from ALP decays), 
the study of vector-boson fusion channels,  the analysis of ALP-fermion signals to probe the complete NLO basis of operators --bosonic and fermionic-- established in this work, the combination of collider constraints with lower-energy experiments (particularly rare decays of mesons), and the evaluation of modifications 
to the history of the axion in the Early Universe due to the non-linear effects.

%%%
%%%%%%%%%%%%%%%%%%%%%%%%%%  
%%%
\section*{Acknowledgements}

We want to thank Giacomo Polesello for very useful discussions on the LHC sensitivity analysis,  to Javier Redondo for discussions on cosmological/astrophysical bounds on ALPs, and to Brian Shuve for feedback on the impact of rare meson decay searches.  This work has been supported by the Royal Society International Exchanges programme.  
I.B. research was supported by an ESR contract of the EU network FP7 ITN INVISIBLES (Marie Curie Actions, PITN-GA-2011-289442) and by the Villum Fonden. I.B. also acknowledges partial support by the Danish National Research Foundation (DNRF91). I.B., M.B.G., L.M., R.dR. acknowledge partial 
support of the European Union network FP7 ITN INVISIBLES, of CiCYT through the projects FPA2012-31880 and FPA2016-78645, and of the Spanish MINECO's ``Centro de Excelencia Severo Ochoa'' Programme under 
grant SEV-2012-0249. M.B.G. and L.M. acknowledge partial support by a grant from the Simons Foundation and  the Aspen Center for Physics, where part of this work has been developed, 
which is supported by National Science Foundation grant PHY-1066293. L.M. research is supported by the Spanish MINECO through the ``Ram\'on y Cajal'' programme (RYC-2015-17173). K. M. is supported in part by the Belgian Federal Science Policy Office through the Interuniversity Attraction Pole P7/37. The work of K.M. and V.S. is supported by the Science Technology and Facilities Council (STFC) under grant number ST/L000504/1.
J.M.N. has been supported by the People Programme (Marie Curie Actions) of the European Union Seventh Framework Programme (FP7/2007-2013) under REA grant agreement PIEF-GA-2013-625809 and the European Research Council under the  European Unions Horizon 2020 program (ERC Grant Agreement no.648680 DARKHORIZONS).

\pagebreak 
\begin{appendices}
%%%
%%%%%%%%%%%%%%%%%%%%%%%%%% A. 
%%%

\section{Fermionic Chiral ALP Lagrangian and Complete Basis}
\label{App:FermionicCouplings}

In what follows, a complete basis of operators -bosonic plus fermionic- which include an ALP insertion is determined, up to NLO for the chiral EWSB.  
Consider the following set of independent fermionic structures, assuming only one flavour family:
\[
\begin{aligned}
\B^q_1=&\bar Q_L\U Q_R\,\partial_\mu\dfrac{a}{f_a}\partial^\mu\F(h)\,,
&\B^\ell_1=&\bar L_L\U L_R\,\partial_\mu\dfrac{a}{f_a}\partial^\mu\F(h)\,,\\
\B^q_2=&\bar Q_L\T\U Q_R\,\partial_\mu\dfrac{a}{f_a}\partial^\mu\F(h)\,,\\
\B^q_3=&\bar Q_L\V_\mu \U Q_R\,\partial^\mu\dfrac{a}{f_a}\F(h)\,,\\
\B^q_4=&\bar Q_L\left\{\V_\mu,\T\right\} \U Q_R\,\partial^\mu\dfrac{a}{f_a}\F(h)\,,
&\B^\ell_2=&\bar L_L\left\{\V_\mu,\T\right\} \U L_R\,\partial^\mu\dfrac{a}{f_a}\F(h)\,,\\
\B^q_5=&\bar Q_L\left[\V_\mu,\T\right] \U Q_R\,\partial^\mu\dfrac{a}{f_a}\F(h)\,,
&\B^\ell_3=&\bar L_L\left[\V_\mu,\T\right] \U L_R\,\partial^\mu\dfrac{a}{f_a}\F(h)\,,\\
\B^q_6=&\bar Q_L\T\V_\mu\T \U Q_R\,\partial^\mu\dfrac{a}{f_a}\F(h)\,,\\
\B^q_7=&\bar Q_L\sigma^{\mu\nu}\V_\mu\U Q_R\,\partial^\mu\dfrac{a}{f_a}\F(h)\,,\\
\B^q_8=&\bar Q_L\sigma^{\mu\nu}\left\{\V_\mu,\T\right\}\U Q_R\,\partial^\mu\dfrac{a}{f_a}\F(h)\,,
&\B^\ell_4=&\bar L_L\sigma^{\mu\nu}\left\{\V_\mu,\T\right\}\U L_R\,\partial^\mu\dfrac{a}{f_a}\F(h)\,,\\
\B^q_9=&\bar Q_L\sigma^{\mu\nu}\left[\V_\mu,\T\right]\U Q_R\,\partial^\mu\dfrac{a}{f_a}\F(h)\,,
&\B^\ell_5=&\bar L_L\sigma^{\mu\nu}\left[\V_\mu,\T\right]\U L_R\,\partial^\mu\dfrac{a}{f_a}\F(h)\,,\\
\B^q_{10}=&\bar Q_L\sigma^{\mu\nu}\T\V_\mu\T\U Q_R\,\partial^\mu\dfrac{a}{f_a}\F(h)\,,
\end{aligned} \label{fermionicoperators}
\]
Would the neutral components be added to  the $SU(2)_R$ doublet $L_R\equiv (0,E_R)$, the number of leptonic operators above would double. When considering several generations, each of the structures in Eq.~(\ref{fermionicoperators}) encodes all possible independent flavour operators $\B^q_{i,\alpha\beta}$, where greek indices denote flavour.

A complete basis can be constructed combining the set of fermionic operators above with the bosonic Lagrangian in Eq.~\eqref{Complete} while avoiding redundancies. This can be enforced using the EOM which may relate some bosonic and fermionic operators. Given the form of the chiral LO Lagrangian in Eq.~(\ref{Lchiral-LO}), the relevant EOMs read 
\begin{align}
&\begin{aligned}
 i\slashed{D}\psi_L& =  \frac{v}{\sqrt{2}} \U \mathcal{Y}_\psi(h)  \psi_R +iv\sqrt2\frac{a}{f_a}c_{2D}\,\U\mathcal{Y}_\psi(h)\s^3\psi_R\,,\\[2mm]
 i\slashed{D}\psi_R&=  \frac{v}{\sqrt{2}} \mathcal{Y}^\dag_\psi(h) \U^\dag\psi_L  -iv\sqrt2\frac{a}{f_a}c_{2D}\,\s^3 \mathcal{Y}^\dag_\psi(h) \U^\dag\psi_L\,,
\label{EOM_psi}
 \end{aligned}\\
 \nn\\
& (D^\mu\WWd)^a = \sum_{\psi=Q,L}\frac{g}{2}\bar{\psi}_L\s^a \g_\nu \psi_L 
+\dfrac{igv^2}{4}\tr[\V_\nu \s^a] \cF_C(h)\label{EOM_W}\,,\\[2mm]
& \de^\mu \BBd =g\ct\sum_{\parbox{12mm}{\scriptsize$i=L,R$\\
$\psi=Q,L$}} \bar{\psi}_i\mathbf{h}_{\psi_i}\g_\nu \psi_i 
-\frac{ig\ct v^2}{4}\tr[\T\V_\mu]\cF_C(h)\label{EOM_B}\,,\\
 \nn\\
&\begin{aligned}
\square h =& - V'(h)-\frac{v^2}{4}\tr[\V_\mu\V^\mu]  
\cF'_C(h)+\\[2mm]
&-\frac{v}{\sqrt2}\sum_{\psi=Q,L}\left(\bar\psi_L\U\mathcal{Y}'_\psi(h)
\psi_R+\text{h.c.}\right)+v^2 c_T\tr(\TL\VL_\mu)^2\cF'_T(h)+\\
&+ic_{2D}v^2\left[ \tr\left[\T\V_\mu\right] \frac{\de^\mu a}{f_a} \hat{\F}'_{2D}(h)-\sqrt{2}{v}\frac{a}{f_a}\sum_{\psi=Q,L }\left(\bar\psi_L\U\mathcal{Y}'_\psi(h) \psi_R\right)+\text{h.c.}\right]\,,\label{EOM_h} \\
\end{aligned} \\[2mm] 
&\square \frac{a}{f_a}=-ic_{2D}\frac{v^2}{f_a^2}\left[\partial_\mu\left(\tr\left[\T\V_\mu\right]\hat{\cF}_{2D}(h)\right)+\frac{\sqrt2}{v}\sum_{\psi=Q,L}\left(\bar\psi_L\U\mathcal{Y}_\psi(h) \psi_R\right)+\text{h.c.}\right]\,, \label{EOM_a}
\end{align}
where $\mathcal{Y}_\psi(h)$ has been defined in Eq.~\eqref{curly_Y}, $\hat{\F}_{2D}(h)$ is defined as $\F_{2D}(h)$ without its $h$-independent term and the prime on the $\cF_i$ and $\mathcal{Y}_\psi$ functions denotes the first derivative with respect to $h$.  $\mathbf{h}_{\psi_i}$  in Eq.~\eqref{EOM_B} are the hypercharges given in the $2\times2$ matrix 
notation
\begin{equation}
 \begin{aligned}
  \mathbf{h}_{Q_L}&=\diag\left(1/6,1/6\right)\,,\qquad&	\mathbf{h}_{Q_R}&
=\diag\left(2/3,-1/3\right)\,,\\
  \mathbf{h}_{L_L}&= \diag\left(-1/2,-1/2\right)\,,&	\mathbf{h}_{L_R}&
=\diag\left(0,-1\right)\,.
 \end{aligned}
\end{equation} 
A consequence of Eqs.~\eqref{EOM_psi} and~\eqref{EOM_W} is~\cite{Brivio:2013pma,Brivio:2016fzo}
\begin{equation}
 \D_\mu \left(\V^\mu \cF_C\right) = \frac{i}{v^2}D_\mu
\left(\sum_{\psi=Q,L}\bar{\psi}_L\s^j \g^\mu \psi_L\right)\s^j=
 \frac{1}{\sqrt2 v}\sum_{\psi=Q,L}\left(\bar\psi_L\s^j\U\mathcal{Y}_\psi(h) \psi_R
-\bar\psi_R\mathcal{Y}^\dag_\psi(h)\U^\dag\s^j \psi_L\right)\s^j\,,
\end{equation} 
which can be recast as
\begin{equation}\label{DmuVmu}
\tr(\s^j\D_\mu\V^\mu)\cF_C(h) =
\frac{\sqrt2}{v}\sum_{\psi=Q,L}\left(\bar\psi_L\s^j\U\mathcal{Y}_\psi(h) \psi_R
-\bar\psi_R\mathcal{Y}^\dag_\psi(h)\U^\dag\s^j \psi_L\right)-\tr(\s^j\V_\mu)
\de^\mu\cF_C(h)\,,
\end{equation}
and is valid order by order in the $h$ expansion.

Applying the EOMs above, the operators $\A_{8}$, $\A_{11}$, $\A_{13}$, $\A_{17}$ in Eq.~(\ref{bosonic_basis}) can be removed as redundant, because tradable by  flavour-blind structures of the type in Eq.~(\ref{fermionicoperators}). In summary,  the complete basis of LO plus NLO operators of the EWSB chiral expansion  which include an ALP insertion includes a total of 32 independent operators, considering only one fermion generation and disregarding the different coefficients inside the $\cF_i(h)$ functions; the extension to three generations is obvious.

%%%
%%%%%%%%%%%%%%%%%%%%%%%%%% B. 
%%%
\section{Feynman rules for the bosonic basis}
\label{App:feynman_rules}

This section provides a complete list of the Feynman rules for vertices involving an ALP  and 
resulting from  the NLO linear Lagrangian Eq.~(\ref{deltaLbosonic-lin}) and the chiral one 
Eq.~(\ref{Complete}),  up to
four legs. 
The coefficients $\at_i$ and $\bt_i$ have been defined in Eq.~(\ref{atilde}) in terms of the parameters in those Lagrangians; this is extended below for the operators $\A_{15}$ and $\A_{16}$, that contain two 
functions $\F_i(h)$ and $\F_i^\prime (h)$, redefining $c_ia_ia_i^\prime \to \at_i$. The rules are computed:
\begin{itemize}
 \item choosing the momenta to flow inwards in the vertices
 \item in unitary gauge
 \item neglecting flavor effects, {\it i.e.} assuming hermitian and diagonal Yukawa matrices ($\cY_\psi \equiv \cY_\psi^\dag$) and $V_{\rm CKM}\equiv \mathbbm{1}$ (greek indices will indicate flavour).
\end{itemize}
The Feynman rules for the linear case be easily obtained from those for the non-linear Lagrangian, in the limit
\begin{equation}
 c_{1},\dots, c_{17}\to 0\,,\qquad
 c_{\tilde{B}}\to c_{\tilde{B}}\,,\qquad
 c_{\tilde{W}}\to c_{\tilde{W}}\,,\qquad
 c_{\tilde{G}}\to c_{\tilde{G}}\,,\qquad
\end{equation}
A final replacement, $c_{2D}\to -c_\Phi/2$, only applies for the fermionic couplings stemming from the corresponding chiral and linear operators.  In the table of Feynman rules below, the left, center and middle columns show respectively the phenomenological vertex, the amplitude in the chiral case and that in the linear case when non-vanishing, up to NLO.

\fancypagestyle{mylandscape}{%
  \fancyhf{}% Clear header/footer
  \fancyfoot{% Footer
    \makebox[\textwidth][r]{% Right
     \rotatebox{90}{\thepage}}}% Rotate counter-clockwise
  \renewcommand{\headrulewidth}{0pt}% No header rule
  \renewcommand{\footrulewidth}{0pt}% No footer rule
}

\begin{landscape}
\thispagestyle{mylandscape}
\addtolength{\voffset}{1cm}
\renewcommand{\arraystretch}{6}
\footnotesize
\begin{center}
\vspace*{-2cm}
\begin{longtable}{m{8mm}@{\hspace*{5mm}}>{\centering}m{5cm}m{10cm}@{\hspace*{2cm}}m{3cm}}
\multicolumn{2}{m{5.8cm}}{\centering \bf Coupling} & \bf Chiral Lagrangian& \bf Linear Lagrangian\\[-3mm]\hline \\[-3cm]

%%% AAa
\nr\label{FR.AAa}& \parbox{3cm}{\input{Fdiagrams/AAa}} & 
$-\frac{4i}{f_a}p_{A1\a}p_{A2\b}\e^{\mu\nu\a\b}\left(\ct^2c_{\tilde{B}} + \st^2c_{\tilde{W}}\right)$& 
$-\frac{4i}{f_a}p_{A1\a}p_{A2\b}\e^{\mu\nu\a\b}\left(\ct^2c_{\tilde{B}} + \st^2c_{\tilde{W}}\right)$\\

%%% ZZa
\nr\label{FR.ZZa}& \parbox{3cm}{\input{Fdiagrams/ZZa}} & 
$-\frac{4i}{f_a}p_{Z1\a}p_{Z2\b}\e^{\mu\nu\a\b}\left(\st^2 c_{\tilde{B}} + \ct^2 c_{\tilde{W}}+\frac{g}{16\pi}(c_2+2c_7-2t_\theta c_1)\right)$&  
$-\frac{4i}{f_a}p_{Z1\a}p_{Z2\b}\e^{\mu\nu\a\b}\left(\st^2 c_{\tilde{B}} + \ct^2 c_{\tilde{W}}\right)$\\
  
%%%ZAa
\nr\label{FR.ZAa}& \parbox{3cm}{\input{Fdiagrams/ZAa}} & 
$\frac{i}{f_a}p_{Z\a}p_{A\b}\e^{\mu\nu\a\b}\left(2\sdt(c_{\tilde{B}}-c_{\tilde{W}})-\frac{g}{8\pi}(2c_1+t_\theta(c_2+2c_7))\right)$&
$\frac{2i\sdt}{f_a}p_{Z\a}p_{A\b}\e^{\mu\nu\a\b}\left(c_{\tilde{B}}-c_{\tilde{W}}\right)$\\

%%% WWa
\nr\label{FR.WWa}& \parbox{3cm}{\input{Fdiagrams/WWa}} & 
$\begin{array}{l}
\frac{g}{4\pi f_a}\left[c_6g^{\mu\nu}(p_+^2-p_-^2)+\left(\frac{g}{4\pi}c_8-c_6\right)\left(p_+^\mu p_+^\nu -p_-^\mu p_-^\nu\right)\right]+\\[-1.5cm]
-\frac{4i}{f_a}\left(c_{\tilde{W}}+\frac{g}{16\pi}c_2\right)p_{+\a}p_{-\b}\e^{\mu\nu\a\b}
\end{array}$&  
$-\frac{4i}{f_a}c_{\tilde{W}}p_{+\a}p_{-\b}\e^{\mu\nu\a\b}$\\

%%% GGa
\nr\label{FR.GGa}& \parbox{3cm}{\input{Fdiagrams/GGa}} & 
$-\frac{4i}{f_a} c_{\tilde{G}}\,p_{G1\a}p_{G2\b}\,\e^{\mu\nu\a\b}$&
$-\frac{4i}{f_a} c_{\tilde{G}}\,p_{G1\a}p_{G2\b}\,\e^{\mu\nu\a\b}$\\

%%% Aah
\nr\label{FR.Aah}& \parbox{3cm}{\input{Fdiagrams/Aah}} & 
$\frac{1}{2\pi v f_a}(\at_3\ct+\at_{10}\st)\left(p_A^\mu p_a\cdot p_A-p_A^2 p_a^\mu\right)$& \\

%%% Zah
\nr\label{FR.Zah}& \parbox[c][1cm][c]{3cm}{\input{Fdiagrams/Zha}} &
$\begin{array}{r} 
\frac{1}{4\pi^2 \sdt v f_a}\Big[
e\, p_h^\mu (p_a^2  \at_{11}+ p_a\cdot p_h \at_{14}) + e\, p_a^\mu (p_h^2 \at_{13} + p_a\cdot p_h \at_{12}) \\[-1.5cm]
+2\pi\sdt(\at_3\st-\at_{10}\ct)(p_Z^\mu p_h\cdot p_Z-p_h^\mu p_Z^2)\\[-1.5cm]
-e\, p_a^\mu (16\pi^2 v^2\at_{2D}-\at_{17}p_a^2)
\Big]\end{array}$ & \\ 

%%% ahhZ
\nr\label{FR.Zahh}& \parbox[c][1cm][c]{3cm}{\input{Fdiagrams/Zhha}} & $ \begin{array}{r}
\frac{1}{4\pi^2 \sdt v^2 f_a}\Big[
e\, (p_{h1}+p_{h2})^\mu (p_a^2  \bt_{11}+ p_a\cdot (p_{h1}+p_{h2}) \bt_{14}) \\[-1.5cm]
+ e\, p_a^\mu ((p_{h1}+p_{h2})^2 \bt_{13} + p_a\cdot (p_{h1}+p_{h2}) \bt_{12}) \\[-1.5cm]
+2e\at_{16} (p_{h1}^\mu p_a\cdot p_{h2}+p_{h2}^\mu p_a\cdot p_{h1})\\[-1.5cm]
+4e \at_{15} p_a^\mu p_{h1}\cdot p_{h2}
-e\, p_a^\mu (16\pi^2 v^2\bt_{2D}-\bt_{17}p_a^2)
\\[-1.5cm]
+2\pi\sdt(\bt_3\st-\bt_{10}\ct)( p_Z^2 p_a^\mu-p_Z^\mu p_a\cdot p_Z  )
\Big]\end{array}$& \\ 

%%% Aahh
\nr\label{FR.Aahh}& \parbox[c][1cm][c]{3cm}{\input{Fdiagrams/Aahh}} & 
$\frac{1}{2\pi v^2 f_a}(\bt_3\ct+\bt_{10}\st)\left(p_A^\mu p_a\cdot p_A-p_A^2 p_a^\mu\right)$&\\
 
%%% WWZa
\nr\label{FR.WWZa}& \parbox[c][1cm][c]{3cm}{\input{Fdiagrams/WWZa}} & $\begin{array}{l}
\frac{g^2}{4\pi\ct f_a}\left[2\left(\frac{g}{8\pi}c_5+ \ct^2 c_6\right)g^{\mu\nu}p_a^\rho+
\left(\frac{g}{8\pi}c_4-\ct^2 c_6-\frac{e\st}{4\pi}c_8\right)(g^{\mu\rho}p_a^\nu+g^{\nu\rho}p_a^\mu)\right]+\\[-1.5cm]
-\frac{4ig\ct}{ f_a}\left(c_{\tilde{W}}+c_2\frac{g(1+2\ct^2)}{32\pi\ct^2}+\frac{g}{16\pi\ct^2}c_7\right)\e^{\mu\nu\rho\a}p_{a\a}
\end{array}$&
$-\frac{4ig\ct}{ f_a}c_{\tilde{W}}\e^{\mu\nu\rho\a}p_{a\a}$\\

%%% WWAa
\nr\label{FR.WWAa}& \parbox[c][1cm][c]{3cm}{\input{Fdiagrams/WWAa}} & $
\begin{array}{l}
\frac{ge}{4\pi f_a}\left[\left(\frac{g}{4\pi}c_8-c_6\right)\left(g^{\mu\rho}p_a^\nu+g^{\nu\rho}p_a^\mu\right)
+2c_6 g^{\mu\nu}p_a^\rho\right]+ \\[-1.5cm]
-\frac{4ig}{f_a}\left(c_{\tilde{W}}+\frac{g}{16\pi}c_2\right)\e^{\mu\nu\rho\a}p_{a\a}
\end{array}$& 
$-\frac{4ig}{f_a}c_{\tilde{W}}\e^{\mu\nu\rho\a}p_{a\a}$\\

%%% ZZZa
\nr\label{FR.ZZZa}& \parbox[c][1cm][c]{3cm}{\input{Fdiagrams/ZZZa}} & 
$\frac{g^3}{16\pi^2 \ct^3 f_a}(c_4+c_5+2c_9)\left(g^{\mu\nu}p_a^\rho+g^{\mu\rho}p_a^\nu+g^{\nu\rho}p_a^\mu\right) $&\\

%%% ahZZ
\nr\label{FR.ahZZ}& \parbox[c][1cm][c]{3cm}{\input{Fdiagrams/ahZZ}} &
$-\frac{ig}{4\pi v f_a}p_{a\a}(p_{Z1-}p_{Z2})_\b \e^{\mu\nu\a\b}\left(\at_2+2\at_7-2 t_\theta\at_1\right)$
\\

%%% ahAZ
\nr\label{FR.ahAZ}& \parbox[c][1cm][c]{3cm}{\input{Fdiagrams/ahAZ}} & 
$-\frac{ig}{4\pi vf_a}p_{a\a}p_{A\b}\e^{\mu\nu\a\b}\left(2\at_1+t_\theta(\at_2+2\at_7)\right)$
\\

%%% ahWW
\nr\label{FR.ahWW}& \parbox[c][1cm][c]{3cm}{\input{Fdiagrams/ahWW}} & 
$\begin{array}{l}
  \frac{g}{2\pi v f_a}\left[\at_6g^{\mu\nu}p_a\cdot(p_+-p_-)
  +\left(\frac{g}{4\pi}\at_8-\at_6\right)\left(p_a^\mu p_+^\nu -p_-^\mu p_a^\nu\right)\right]+\\[-1.5cm]
-\frac{ig}{4\pi v f_a}\at_2p_{a\a}(p_+-p_-)_\b\e^{\mu\nu\a\b}
+\frac{g}{2\pi vf_a}\left(\at_{10}-\frac{g}{4\pi}\at_8\right)(p_a^\mu p_h^\nu-p_a^\nu p_h^\mu)
 \end{array}$& \\
 
%%% GGGa
\nr\label{FR.GGGa}& \parbox[c][1cm][c]{3cm}{\input{Fdiagrams/GGGa}} & 
$\frac{4g_s}{f_a}c_{\tilde{G}}f^{abc}\e^{\mu\nu\rho\a}p_{a\a} $&   $\frac{4g_s}{f_a}c_{\tilde{G}}f^{abc}\e^{\mu\nu\rho\a}p_{a\a} $\\

%%% auu
\nr\label{FR.auu}&  \parbox{2.5cm}{\input{Fdiagrams/auu}} & 
$\frac{2\sqrt2 v}{ f_a}\left(c_{2D}-\frac{c_{17}}{16\pi^2}\frac{p_a^2}{v^2}\right)(\cY_U)_{\a}\delta_{\a\b}\, \g_5 $ & 
$-\frac{\sqrt{2}v}{ f_a}c_{a\Phi}(Y_U)_\a \delta_{\a\b}\, \g_5 $ \\
 
%%% add
\nr\label{FR.add}&  \parbox{2.5cm}{\input{Fdiagrams/add}} &
$-\frac{2\sqrt2 v}{ f_a}\left(c_{2D}-\frac{c_{17}}{16\pi^2}\frac{p_a^2}{v^2}\right)\left(Y_D\right)_\a\delta_{\a\b}\, \g_5 $ &
$\frac{\sqrt2v}{ f_a}c_{a\Phi}(Y_D)_\a\delta_{\a\b}\, \g_5 $ \\

%%% aee
\nr\label{FR.aee}&  \parbox{2.5cm}{\input{Fdiagrams/aee}} & 
$-\frac{2\sqrt2 v}{ f_a}\left(c_{2D}-\frac{c_{17}}{16\pi^2}\frac{p_a^2}{v^2}\right)(Y_E)_\a\delta_{\a\b}\, \g_5$ &
$\frac{\sqrt2v}{f_a}c_{a\Phi}(Y_E)_\a\delta_{\a\b}\, \g_5$ \\

%%% auuh
\nr\label{FR.auuh}&  \parbox{2.5cm}{\input{Fdiagrams/auuh}} & 
$\frac{4\sqrt2}{ f_a}a_U\left(c_{2D}-\frac{c_{17}}{16\pi^2}\frac{p_a^2}{v^2}\right)(Y_U)_\a\delta_{\a\b}\, \g_5 $ & 
$-\frac{\sqrt{2}}{ f_a}c_{a\Phi}(Y_U)_\a \delta_{\a\b}\, \g_5 $\\
 
%%% addh
\nr\label{FR.addh}&  \parbox{2.5cm}{\input{Fdiagrams/addh}} &
$-\frac{4\sqrt2 }{ f_a}a_D\left(c_{2D}-\frac{c_{17}}{16\pi^2}\frac{p_a^2}{v^2}\right)(Y_D)_\a\delta_{\a\b}\, \g_5 $ & 
$\frac{\sqrt2}{ f_a}c_{a\Phi}(Y_D)_\a\delta_{\a\b}\, \g_5 $ \\

%%% aeeh
\nr\label{FR.aeeh}&  \parbox{2.5cm}{\input{Fdiagrams/aeeh}} & 
$-\frac{4\sqrt2 }{ f_a}a_E\left(c_{2D}-\frac{c_{17}}{16\pi^2}\frac{p_a^2}{v^2}\right)(Y_E)_\a\delta_{\a\b}\, \g_5$ & 
$\frac{\sqrt2}{f_a}c_{a\Phi}(Y_E)_\a\delta_{\a\b}\, \g_5$ \\
\end{longtable}
\end{center}
\end{landscape}

% \clearpage

\newpage

%%%
\section{Linear siblings}
\label{App:linear_siblings}
The interaction vertices  described by the chiral operators in Sect.~\ref{Sect:basis} can also be described in the context of  linearly realized EWSB, through linear operators in which the Higgs resonance is embedded within the SM Higgs doublet. In this section, the connection between the two expansions is shown.  Operators up to NNLO of the linear expansion have to be taken into account in order to encompass all the interaction vertices appearing in the chiral framework up to NLO. The chiral couplings involving an ALP discussed in this work can be grouped as those
\paragraph{Connected to $d=5$ operators in the linear expansion}
\[
\hspace{-4.1cm}\begin{array}{rl}
\A_{2D}\longrightarrow& \frac{-i}{2}(\f^\dagger\overleftrightarrow{D}^\m \f)  \frac{\de_\m a}{f_a}   \\[3mm]
\A_{\tilde{B}}\longrightarrow &-\BBd\tilde{B}^{\mu\nu}\dfrac{a}{f_a}\\
\A_{\tilde{W}}\longrightarrow&-\WWd^a\tilde{W}^{a\mu\nu}\dfrac{a}{f_a}\\
\A_{\tilde{G}} \longrightarrow&-G^a_{\mu\nu}\tilde{G}^{a\mu\nu}\dfrac{a}{f_a}\\

\end{array}
\]

\paragraph{Connected to $d=7$ operators in the linear expansion}

\[
\hspace{-0.85cm}\begin{array}{rl}
\A_1 &\longrightarrow  -\frac{2i}{(4\pi) v^2}^\prime\tilde{B}_{\m\n}(\f^\dagger\overleftrightarrow{D}_\m \f) \de_\n\frac{ a}{f_a} \\[3mm]
\A_2 &\longrightarrow  -\frac{i}{(4\pi) v^2}(D_\m \fda \Wt^{\m\n}\f-\fda \Wt^{\m\n}D_\m \f)\frac{\de^\nu a}{f_a} \\[3mm]
\A_3 &\longrightarrow \frac{-2}{(4\pi) v^2}\BBd \frac{\de^\m a}{f_a} D^(\fda\f) \\[3mm]
\A_4,\,\A_8 &\longrightarrow \frac{4i}{(4\pi)^2v^2} (D^\m \fda D_\m D_\n \f - D_\m D_\n \fda D^\m \f) \frac{\de^\n a}{f_a} \\[3mm]
\A_5 &\longrightarrow \frac{4i}{(4\pi)^2v^2} (D^\n \fda \Box \f - \Box \fda D^\m \f) \frac{\de_\n a}{f_a} \\[3mm]
\A_6 &\longrightarrow -\frac{4}{(4\pi)i v^2} (\fda W_{\m\n} D^\m \f + D^\m \fda W_{\m\n}\f)\frac{\de^\n a}{f_a} \\[3mm]
\A_{10} &\longrightarrow \frac{4}{(4\pi) v^2} (\fda W_{\m\n} D^\m \f + D^\m \fda W_{\m\n}\f)\frac{\de^\n a}{f_a} \\[3mm]
\A_{11} &\longrightarrow -\frac{2i}{(4\pi)^2 v^2} (\fda \Box \f-\f \Box \fda)\frac{\Box a}{f_a} \\[3mm]
\A_{12} &\longrightarrow -\frac{2i}{(4\pi)^2 v^2} (\fda \olra{D_\m D_\n}\f) \frac{\de^\m \de^\n a}{f_a} \\[3mm]
\A_{15},\,\A_{16} &\longrightarrow -\frac{8i}{(4\pi)^2 v^2} (D^\mu \fda D_\m D_\n \f - D_\m D_\n \fda D^\m \f) \frac{\de^\n a}{f_a} \\[3mm]
\A_{17}&\longrightarrow 2\frac{2i}{(4\pi)^2 v^2} (\fda D_\m \f) \frac{\de^\m \Box a}{f_a}
\end{array}
\]

\paragraph{Connected to $d=9$ operators in the linear expansion}
\[
\hspace{-1.9cm}\begin{array}{rl}
\A_7 &\longrightarrow \frac{8i}{(4\pi)^2 v^4}(\fda \Wt_{\m\n}\f)(\fda \olra{\D^\m} \f)\frac{\de^\n a}{f_a} \\[3mm]
\A_{13} &\longrightarrow -\frac{4i}{(4\pi)^2v^4} (\fda \olra{D}_\m \f ) \Box[\fda \f ]\frac{\de^\m a}{f_a} \\[3mm]
\A_{14} &\longrightarrow-\frac{4i}{(4\pi)^2v^4} (\fda \olra{D}_\m \f ) \de^\m\de^\n[\fda \f ]\frac{\de_\n a}{f_a} \\[3mm]

\end{array}
\]

\paragraph{Connected to $d=11$ operators in the linear expansion}
\[
\hspace{-1.9cm}\begin{array}{rl}
\A_9 \longrightarrow -\frac{i}{2\pi v^6}(\fda D_\m\f)(\fda D^\m\f)(\fda D_\n\f)\frac{\de^\n a}{f_a} \\[3mm]
\end{array}
\]
This shows that operators of the linear expansion up to $d=11$ ``collapse'' into NLO or LO operators of the chiral one.
Note that the leading corrections of the non-linear bosonic set encompass $1\text{ (2 derivatives)}+20\text{ (4 derivatives)}=21$ couplings while the 
linear $d=5$ level has only 4.

%%%%%%%%%%%%%%%%%%%%%%%%%% C. 
%%%
\section{Effects of fields redefinitions}
\label{App:field_redef}

The field redefinitions performed to remove the $a$-$Z$ two-point function stemming from the operator $\O_{a\Phi}$ in the linear EFT, Eq.~(\ref{OaPhi}), and from its sibling $\A_{2D}$  in the chiral EFT, Eq.~(\ref{Eq:A2D}), can be generalized. The effects of generic redefinitions of the GB matrix $\U$ and of fermionic fields in both the linear and chiral cases will be discussed and compared next.

\subsection{Chiral EFT}

In the chiral EFT case, the most general redefinition of the GB matrix $\U$ and of fermionic fields can be schematically written as:
\begin{subequations}
\label{general.redef}
\begin{align}
\U &\to \U \,\exp\left\{i x_U\, \frac{a}{f_a}\, \s^3\right\}\,,\\
\psi_L &\to \exp\left\{i x_{\psi L}\, \frac{a}{f_a}\right\}\, \psi_L\,,\hspace*{3,2cm} \psi=\{Q,\,L\}\,,\\
Q_R &\to \exp\left\{i \bigg((x_{u R}+x_{d R})\frac{\mathbf{1}}{2}+(x_{u R}-x_{d R})\frac{\s^3}{2}\bigg)\,\frac{a}{f_a}\right\}\,Q_R\,,\\
L_R &\to \exp\left\{i x_{e R}\frac{(\mathbf{1}-\s^3)}{2}\,\frac{a}{f_a}\right\}\,L_R\,.
\end{align}
\end{subequations}
Although the parameters $x_{\psi L,R}$ are generically $3\times3$ hermitian matrices in flavor space, they will be taken to be flavor universal, $x_{\psi L,R}\equiv x_{\psi L,R}\,\unity$. Moreover, without loss of generality, all the arbitrary $x_i$ parameters are taken to be real, and $f=v$ will be assumed in oder to simplify the notation.

Applying the redefinitions in Eq.~\eqref{general.redef} on the leading order Lagrangian $\LL_{\text{HEFT}}^\text{LO}$ in Eq.~\eqref{Eq:SMLO} leads to additional terms:
\beq
\LL^{\text{LO}}_{\text{HEFT}}
\to\LL_{\text{HEFT}}^\text{LO} +\Delta\LL_{\text{HEFT}}^\text{LO}\,,
\eeq
with
\begin{equation}\label{L0_redefined}
\begin{split}
 \Delta\LL_{\text{HEFT}}^\text{LO}= &
 -\frac{iv^2}{2}x_U\tr(\T\V_\mu)\frac{\de^\mu a}{f_a}\F_C(h)+\\
 & -\frac{v}{2\sqrt2}\frac{ia}{f_a}\bigg[\bar{Q}_{L}
 \mathcal{Y}_Q(h) \U\s^3Q_{R}\,(2x_U+x_{u R}-x_{d R})+\bar{Q}_{L}\mathcal{Y}_Q(h)\U Q_{R}\,(x_{u R}+x_{d R}-2x_{Q L}) +\\
 &\qquad\qquad+\bar{L}_{L}\mathcal{Y}_L(h)\U\s^3L_{R}\, (2x_U-x_{e R})+ \bar{L}_{L}\mathcal{Y}_L(h)\U L_{R} \,(x_{e R}-2x_{LL}) +\hc\bigg]+\\   
 &+\frac{\de_\mu a}{f_a}\bigg[(\bar{Q}_{L}\g^\mu Q_{L})\left(x_{Q L}-x_{u R}-x_{d R}\right)+(\bar{L}_{L}\g^\mu L_{L})\left(x_{L L}-x_{e R}\right)\bigg]+\\
   &-\dfrac{\alpha_1}{8\pi}\,\BBd\tilde{B}^{\mu\nu}\frac{a}{f_a}\sum\left(\frac{1}{3}x_{QL}-\frac{8}{3}x_{uR}-\frac{2}{3}x_{dR}+x_{LL}-2x_{eR}\right)+\\
 &-\dfrac{\alpha_2}{8\pi}\, \WWd^a\tilde{W}^{a\mu\nu}\frac{a}{f_a}\sum\left(3x_{QL}+x_{LL}\right)+\\
 &-\dfrac{\alpha_3}{8\pi}\, G^a_{\mu\nu}\tilde{G}^{a\mu\nu}\frac{a}{f_a}\sum \left(2x_{QL}-x_{uR}-x_{dR}\right)\,,
\end{split}
\end{equation}
where $\alpha_i\equiv g_i^2/4\pi$. The contributions in the last three lines, proportional to $aX_{\mu\nu}\tilde{X}^{\mu\nu}$, arise from the anomaly triangle and the sum runs over the three fermion generations. This is consistent with the result shown in Ref.~\cite{Salvio:2013iaa}.

The $a$-$Z$ two-point function stemming from the operator $\A_{2D}$ can be completely removed by choosing $x_U=2c_{2D}$: this corresponds to the procedure described in Sect.~\ref{Sect:2pointF}. In addition, it remains the freedom to choose the six fermionic transformations so as to remove two fermionic terms among $(\bar{\psi}_L \g_\mu\psi_L)\de^\mu a$, $(\bar{\psi}_L\g_\mu\s^3\psi_L)\de^\mu a$ and $i a(\bar{\psi}_L\U\psi_R)$. For example, requiring that
\begin{equation}
\begin{gathered}
x_{uR}-x_{dR} = -2x_U=- x_{eR} = -2x_{LL}=-4c_{2D}\\  
x_{uR}+x_{dR} - 2x_{QL} =0
\end{gathered} 
\end{equation} 
it results 
\begin{equation}
\begin{split}
 \Delta\LL_{\text{HEFT}}^\text{LO} = &
 -iv^2c_{2D}\tr(\T\V_\mu)\de^\mu\frac{a}{f_a}\F_C(h)
-2\frac{\de_\mu a}{f_a} c_{2D}\left(
\bar{Q}_{L}\g^\mu\s^3Q_{L}  +\bar{L}_{L}\g^\mu\s^3 L_{L}\right)+\\
 &+\dfrac{\alpha_2}{8\pi}\,\WWd^a\tilde{W}^{a\mu\nu}\frac{a}{f_a}\sum\left(3x_{QL}+2c_{2D}\right)+\dfrac{\alpha_1}{8\pi}\,\BBd\tilde{B}^{\mu\nu}\frac{a}{f_a}\sum\left(-\frac{3}{2}x_{QL} -c_{2D} \right)\,.
\end{split}
\end{equation}
The parameter $x_{QL}$ is still free and can be set to zero: this corresponds to recasting the impact of the $a$-$Z$ two-point function into a redefinition of the coupling $c_{\tilde W}$ plus the insertion of the fermionic term $(\de_\mu a)(\bar{\psi}\g^\mu\g_5\s^3\psi)$. This result is equivalent to that reported in Eq.~\eqref{Apsi_2D_da}.

\subsection{Linear EFT}
It is useful to reformulate the discussion of the previous paragraph for the linear EFT, in order to point out a few worthy differences. The most general field redefinition for this case is  
\begin{subequations}\label{general.redef.linear}
\begin{align}
\Phi &\to \exp\{i x_\Phi\, a/f_a\}\,\Phi\,,\\
\psi_L &\to \exp\{i x_{\psi L}\, a/f_a\}\, \psi_L\,,\\
\psi_R &\to \exp\{i x_{\psi R}\, a/f_a\}\, \psi_R\,,
\end{align}
\end{subequations}
for $\psi_L=\{Q_L,L_L\}$, $\psi_R=\{u_R,d_R,e_R\}$. As above, the fermion redefinitions generically act as $3\times3$ hermitian matrices in flavor space, while $x_\Phi\in\mathbbm{R}$ is chosen $x_\Phi\in\mathbbm{R}$. The action of these redefinitions on the LO linear Lagrangian, $\LL_{\text{SM}}$, Eq.~(\ref{LSM_lin}), is 
\begin{equation}\label{LSM_transformed}
\begin{split}
\LL_{\text{SM}}\to \LL_{\text{SM}}&
+\frac{ia}{f_a}\sum_{\psi= Q,\,L}\left[\bar{\psi}_{L}{\bf\Phi} \left(x_\Phi \sigma^3 {\bf Y}_\psi+x_{\psi L}{\bf Y}_\psi-{\bf Y}_\psi x_{\psi R}\right) \psi_{R}+\hc\right]\,+\\
&+\sum_{\psi=u,d,e,\nu}\frac{\de_\mu a}{2f_a}(\bar{\psi}_{\a}\g^\mu\g_5\psi_{\b})(x_{\psi L}-x_{\psi R})_{\a\b}+\\
&-ix_\Phi (\Phi^\dag\overleftrightarrow{D}_\mu\Phi)\frac{\de^\mu a}{f_a}+\\
&-\dfrac{\alpha_1}{8\pi}\,\BBd\tilde{B}^{\mu\nu}\frac{a}{f_a}\sum\left(\frac{1}{3}x_{QL}-\frac{8}{3}x_{uR}-\frac{2}{3}x_{dR}+x_{LL}-2x_{eR}\right)+\\
&-\dfrac{\alpha_2}{8\pi}\, \WWd^a\tilde{W}^{a\mu\nu}\frac{a}{f_a}\sum\left(3x_{QL}+x_{LL}\right)+\\
&-\dfrac{\alpha_3}{8\pi}\, G^a_{\mu\nu}\tilde{G}^{a\mu\nu}\frac{a}{f_a}\sum \left(2x_{QL}-x_{uR}-x_{dR}\right)\,,
\end{split}
\end{equation}
where $\Phi^\dag\overleftrightarrow{D}_\mu\Phi\equiv \Phi^\dag D_\mu\Phi-(D_\mu\Phi)^\dag \Phi$, and the last three lines are identical to those for non-linear case, Eq.~(\ref{L0_redefined}).

The parameter $x_\Phi$ can be conveniently chosen so as to remove the $a$-$Z$ two-point function contained in the operator $\O_{a\Phi}=(\Phi^\dag\overleftrightarrow{D}_\mu\Phi)\de^\mu a$: this is similar to what happened in the chiral case choosing conveniently the parameter $x_U$. Moreover, it is also possible to choose in this linear case  only one of the two axion-fermion operators (either the Yukawa-like or the vector-axial structure) by tuning the fermion field redefinitions, as described for the chiral case. For instance, focusing on the $a\bar{d}d$ vertex, it is possible to retain the structure $ia(\bar{Q}_L\Phi d_R)$ choosing $x_{QL}=x_{dR}$; alternatively,  the coupling $\de_\mu a(\bar{d}\g^\mu\g_5 d)$ can be selected  setting $x_\Phi Y_D-x_{QL}^TY_D+Y_Dx_{dR}\equiv0$. 

The major difference of the impact of the field redefinitions  on the linear and chiral EFTs resides instead in the Higgs couplings: while in the linear case the operator $\O_{a\Phi}$ is completely removed from the Lagrangian, including its couplings containing Higgs legs, this is not the case in the chiral case where only the pure $a$-$Z$ two-point coupling is redefined away, as illustrated in Sect.~\ref{Sect:2pointF}, but in general not those involving the ALP, gauge bosons and Higgs legs. This follows from the fact that Higgs couplings and pure-gauge interactions are correlated in the linear case, while they are independent in the chiral one. The presence of couplings with the structure $(Z_\mu \de^\mu a) h^n,\, n\geq 1$, among the dominant deviations from the SM expectations, is a \underline{smoking gun of non-linearity}, as such vertices appear in the linear EFT case only at NNLO (operators with $d\ge 7$, see Sect.~\ref{linvsnonlin}).

\end{appendices}

%
%\bibliography{biblio}{}
%\bibliographystyle{BiblioStyle}

\providecommand{\href}[2]{#2}\begingroup\raggedright\endgroup

\end{document}

%% file: Fdiagrams/apsipsi.tex
\begin{fmffile}{Fdiagrams/apsipsicoupl}
\begin{fmfgraph*}(50,50)
\fmfleft{i1,i2,i3}
\fmfright{o3,o2,o1}
  \fmf{dashes}{i2,v1}
  \fmf{fermion}{o1,v1}
  \fmf{fermion}{v1,o3}
  
\fmfv{lab=$a$}{i2}
\fmfv{lab=$\bar{\psi}_\a$,label.angle=-5}{o1}
\fmfv{lab=$\psi_\a$,label.angle=5}{o3}
\end{fmfgraph*}
\end{fmffile}

%% file: Fdiagrams/apsipsih.tex
\begin{fmffile}{Fdiagrams/apsipsihcoupl}
\begin{fmfgraph*}(50,50)
\fmfleft{i1,i2,i3}
\fmfright{o3,o2,o1}
  \fmf{dashes}{i1,v1}
   \fmf{dashes}{i3,v1}
  \fmf{fermion}{o1,v1}
  \fmf{fermion}{v1,o3}
  
\fmfv{lab=$h$,label.angle=180}{i1}
\fmfv{lab=$a$,label.angle=180}{i3}
\fmfv{lab=$\bar{\psi}_\a$,label.angle=-5}{o1}
\fmfv{lab=$\psi_\a$,label.angle=5}{o3}
\end{fmfgraph*}
\end{fmffile}

%% file: Fdiagrams/Zha.tex
\begin{fmffile}{Fdiagrams/zhacoupl}
\begin{fmfgraph*}(50,50)

\fmfleft{i1}
\fmfright{o2,o1}
  \fmf{dashes}{i1,v1}
  \fmf{dashes}{v1,o1}
  \fmf{boson}{v1,o2}
  
\fmfv{lab=$h$,l.angle=0}{o1}
\fmfv{lab=$Z_\mu$,l.angle=0}{o2}
\fmfv{lab=$a$,l.angle=180}{i1}
\end{fmfgraph*}
\end{fmffile}

%% file: Fdiagrams/Zhhanarrow.tex
\begin{fmffile}{Fdiagrams/zhhanarrowcoupl}
\begin{fmfgraph*}(70,50)

\fmfleft{i1,i2}
\fmfright{o2,o1}
  \fmf{dashes}{i2,v1}
  \fmf{dashes}{i1,v1}
  \fmf{boson}{v1,o1}
  \fmf{dashes}{v1,o2}
  
\fmfv{lab=$Z_\mu$,l.angle=0}{o1}
\fmfv{lab=$h$,l.angle=0}{o2}
\fmfv{lab=$h$,l.angle=180}{i1}
\fmfv{lab=$a$,l.angle=180}{i2}
\end{fmfgraph*}
\end{fmffile}

%% file: Fdiagrams/monoZW_nonlinear.tex
\begin{fmffile}{Fdiagrams/monoZW_nonlinear}
\fmfset{arrow_len}{2.5mm}
\parbox[c][2.5cm]{4.1cm}{\centering
\begin{fmfgraph*}(100,50)	% qqbar -> w -> wa
\fmfleft{i1,i2}
\fmfright{o2,o1}
  \fmf{fermion,tension=1.2}{i2,v1,i1}
  \fmf{boson,lab=$W^\pm$,l.side=left,tension=1.2}{v1,v2}
  \fmf{boson}{v2,o1}
  \fmf{dashes}{v2,o2}  
   
\fmfv{lab=$W^\pm$,l.angle=0}{o1}
\fmfv{lab=$a$,l.angle=0}{o2}
\fmfv{lab=$q$,l.angle=180}{i2}
\fmfv{lab=$\bar{q}^\prime$,l.angle=180}{i1}
\fmfv{d.shape=circle,d.size=3,foreground=red,lab={\parbox{2.8cm}{\centering\color{red}\tiny$c_{\tilde{W}}+c_2+c_6+c_8$}},l.angle=0}{v2}
\end{fmfgraph*}
}
~
\parbox[c][2.5cm]{4.1cm}{\centering
\begin{fmfgraph*}(120,50)	% qqbar -> A-> Za
\fmfleft{i1,i2}
\fmfright{o2,o1}
  \fmf{fermion,tension=1.2}{i2,v1,i1}
  \fmf{boson,lab=$\g$,l.side=left,tension=1.2}{v1,v2}
  \fmf{boson}{v2,o1}
  \fmf{dashes}{v2,o2}  
   
\fmfv{lab=$Z$,l.angle=0}{o1}
\fmfv{lab=$a$,l.angle=0}{o2}
\fmfv{lab=$q$,l.angle=180}{i2}
\fmfv{lab=$\bar{q}$,l.angle=180}{i1}
\fmfv{d.shape=circle,d.size=3,foreground=red,lab={\parbox{1.8cm}{\centering\color{red}\tiny$c_{\tilde{B}}+c_{\tilde{W}}+$\\$c_1+c_2+c_7$}},l.angle=0}{v2}
\end{fmfgraph*}
}
~
\parbox[c][2.5cm]{4.1cm}{\centering
\begin{fmfgraph*}(100,50)	% qqbar -> z -> Za
\fmfleft{i1,i2}
\fmfright{o2,o1}
  \fmf{fermion,tension=1.2}{i2,v1,i1}
  \fmf{boson,lab=$Z$,l.side=left,tension=1.2}{v1,v2}
  \fmf{boson}{v2,o1}
  \fmf{dashes}{v2,o2}  
   
\fmfv{lab=$Z$,l.angle=0}{o1}
\fmfv{lab=$a$,l.angle=0}{o2}
\fmfv{lab=$q$,l.angle=180}{i2}
\fmfv{lab=$\bar{q}$,l.angle=180}{i1}
\fmfv{d.shape=circle,d.size=3,foreground=red,lab={\parbox{1.8cm}{\centering\color{red}\tiny$c_{\tilde{B}}+c_{\tilde{W}}+$\\$c_1+c_2+c_7$}},l.angle=0}{v2}
\end{fmfgraph*}
}
~
\end{fmffile}

%% file: Fdiagrams/AP_VBF_diagrams.tex
\begin{fmffile}{Fdiagrams/AP_VBF_diag}
\fmfset{arrow_len}{2.5mm}
%\parbox[c][3cm]{4.7cm}{\centering
%\begin{fmfgraph*}(100,40)
%\fmfleft{i1,i2,i3}
%\fmfright{o3,o2,o1}
%  \fmf{fermion}{i3,v1}
%  \fmf{fermion}{v1,i1}
%  \fmf{boson,tension=2,lab=$W^\pm$}{v1,v2}
%  \fmf{boson}{v2,o3}
%  \fmf{dashes}{v2,o1}
%\fmfv{lab=$W^\pm$,l.angle=0}{o3}
%\fmfv{lab=$a$}{o1}
%\fmfv{lab=$q$}{i3}
%\fmfv{lab=$\bar{q}^\prime$}{i1}
%\fmfv{d.shape=circle,d.size=3,foreground=red,lab={\parbox{1.8cm}{\centering\color{red}\tiny$c_{\tilde{W}}+c_2+c_6+c_8$}},l.angle=0,l.dist=10}{v2}
%\end{fmfgraph*} 
%}
% ~
\centering{
\parbox[c]{1cm}{\flushright (i)}
\parbox[c][3cm]{5cm}{\centering
\begin{fmfgraph*}(100,50)
\fmfleft{i1,i2,i3}
\fmfright{o3,o2,o1}
  \fmf{fermion,tension=2}{i3,v1}
  \fmf{fermion,tension=2}{v1,i1}
  \fmf{boson,tension=2,l.side=left,lab={\color{red}\tiny$\begin{array} c c_{\tilde{W}}+c_2 \\ +c_6+c_8\end{array}$}}{v1,v2}
  \fmf{boson}{v2,o3}
  \fmf{dashes}{v2,o1}
  \fmf{boson}{v2,o2}
  
\fmfv{lab=$W^\pm$,l.angle=0}{o3}
\fmfv{lab=$a$}{o1}
\fmfv{lab=$\gamma$}{o2}
\fmfv{lab=$q$}{i3}
\fmfv{lab=$\bar{q}^\prime$}{i1}
\fmfv{d.shape=circle,d.size=3,foreground=red,lab=$W^\pm$,l.angle=-120}{v2}
\end{fmfgraph*}
}
~~~~~
\parbox[c]{1cm}{\flushright (ii)}
\parbox[c][3cm]{5cm}{\centering
\begin{fmfgraph*}(100,50)
\fmfleft{i1,i2,i3}
\fmfright{o3,o2,o1}
  \fmf{fermion,tension=2}{i3,v1}
  \fmf{fermion,tension=2}{v1,i1}
  \fmf{boson,tension=2,lab=$W^\pm$}{v1,v2}
  \fmf{boson}{v2,o3}
  \fmf{boson,label=$Z/\gamma$,l.side=left,l.dist=5}{v2,v3}
  \fmf{dashes}{v3,o1}
  \fmf{boson}{v3,o2}
\fmfv{lab=$W^\pm$,l.angle=0}{o3}
\fmfv{lab=$a$}{o1}
\fmfv{lab=$\gamma$}{o2}
\fmfv{lab=$q$}{i3}
\fmfv{lab=$\bar{q}^\prime$}{i1}
\fmfv{d.shape=circle,d.size=3,foreground=red,
lab=\parbox{3cm}{\centering\color{red}\tiny$\begin{array} c c_{\tilde{W}}+c_\Bt +\\+ c_1 + c_2 + c_7 \end{array}$},l.angle=30}{v3}
\end{fmfgraph*}
}
}\\
\centering{
\parbox[c]{1cm}{\flushright (iii)}
\parbox[c][3cm]{5cm}{\centering
\begin{fmfgraph*}(100,50)

\fmfleft{i1,i2,i3}
\fmfright{o1,o2,o3}
  \fmf{fermion,tension=2}{i3,v1,o3}
  \fmf{fermion,tension=2}{i1,v3,o1}
  \fmf{boson,tension=2}{v1,v2,v3}
  \fmf{dashes}{v2,o2}

\fmfv{lab=$a$}{o2}
\fmfv{lab=$j$}{o1,o3}
\fmfv{lab=$q$}{i3}
\fmfv{lab=$\bar{q}^\prime$}{i1}
\fmfv{lab=$W^\pm$,l.angle=-120}{v1}
\fmfv{lab=$W^\mp$,l.angle=120}{v3}
\fmfv{d.shape=circle,d.size=3,foreground=red,lab={\parbox{1.8cm}{\centering\color{red}\tiny$c_{\tilde{W}}+c_2+c_6+c_8$}},l.angle=-50}{v2}
\end{fmfgraph*} 
}
~~~~~
\parbox[c]{1cm}{\flushright (iv)}
\parbox[c][3cm]{5cm}{\centering
\begin{fmfgraph*}(100,50)

\fmfleft{i1,i2,i3}
\fmfright{o1,o2,o3,o4}
  \fmf{fermion,tension=2}{i3,v1,o4}
  \fmf{fermion,tension=2}{i1,v3,o1}
  \fmf{boson,tension=2}{v1,v2,v3}
  \fmf{dashes}{v2,o3}
  \fmf{boson}{v2,o2}

\fmfv{lab=$a$}{o3}
\fmfv{lab=$\gamma$}{o2}
\fmfv{lab=$j$}{o1,o4}
\fmfv{lab=$q$}{i3}
\fmfv{lab=$\bar{q}^\prime$}{i1}
\fmfv{lab=$W^\pm$,l.angle=-120}{v1}
\fmfv{lab=$W^\mp$,l.angle=120}{v3}
\fmfv{d.shape=circle,d.size=3,foreground=red,lab={\parbox{1.8cm}{\centering\color{red}\tiny$c_{\tilde{W}}+c_2+c_6+c_8$}},l.angle=0,l.dist=20}{v2}
\end{fmfgraph*} 
}
}
%\parbox[c][3cm]{4cm}{\centering
%\begin{fmfgraph*}(80,60)
%
%\fmfleft{i3,i2,i1}
%\fmfright{o3,o2,o1}
%\fmf{fermion,tension=2}{i3,v3,v1,i1}
%\fmf{boson,tension=1.5}{v1,v4,o2}
%\fmf{dashes}{v4,o1}
%\fmf{boson}{v3,o3}
%\fmfv{lab=$a$}{o1}
%\fmfv{lab=$\gamma$}{o2}
%\fmfv{lab=$q$}{i3}
%\fmfv{lab=$\bar{q}^\prime$}{i1}
%\fmfv{lab=$W^\pm$}{o3}
%\end{fmfgraph*} 
%}

\end{fmffile}

%% file: Fdiagrams/mono_di_Higgs.tex
\begin{fmffile}{Fdiagrams/mono_di_Higgs}
\fmfset{arrowlen}{2mm}
\parbox{7cm}{\centering
\begin{fmfgraph*}(160,60)
\fmfleft{i3,i2,i1}
\fmfright{o3,o2,o1}
  \fmf{fermion}{i1,v1}
  \fmf{fermion}{v1,i3}
  \fmf{boson,lab=$Z/\gamma$,l.side=left}{v1,v2}
  \fmf{dashes}{v2,o3}
  \fmf{dashes}{v2,o1}
  
\fmfv{lab=$h$,l.angle=0}{o3}
\fmfv{lab=$a$,l.angle=0}{o1}
\fmfv{lab=$q$,l.angle=180}{i1}
\fmfv{lab=$\bar{q}$,l.angle=180}{i3}
\fmfv{d.shape=circle,d.size=3,foreground=red,lab={\parbox{2cm}{\centering\color{red}\tiny$\at_{2D}+\at_3+\quad (\at_{10}+\at_{11}+\at_{12}+\at_{13}+\at_{14}+\at_{17})$}},l.angle=0,l.dist=15}{v2}
\end{fmfgraph*}
}
~
\parbox{7cm}{\centering
\begin{fmfgraph*}(160,60)
\fmfleft{i3,i2,i1}
\fmfright{o3,o2,o1}
  \fmf{fermion}{i1,v1}
  \fmf{fermion}{v1,i3}
  \fmf{boson,lab=$Z/\gamma$,l.side=left}{v1,v2}
  \fmf{dashes,tension=.7}{v2,o3}
  \fmf{dashes,tension=.7}{v2,o2}
  \fmf{dashes,tension=.7}{v2,o1}
  
\fmfv{lab=$h$,l.angle=0}{o3}
\fmfv{lab=$h$,l.angle=0}{o2}
\fmfv{lab=$a$,l.angle=0}{o1}
\fmfv{lab=$q$,l.angle=180}{i1}
\fmfv{lab=$\bar{q}$,l.angle=180}{i3}
\fmfv{d.shape=circle,d.size=3,foreground=red,lab={\parbox{1.8cm}{\centering\color{red}\tiny$\bt_3+\bt_{10}+ \bt_{11}+\bt_{12}+\bt_{13}+\bt_{14}+\bt_{16}+\bt_{17}$}},l.angle=-120,l.dist=10}{v2}
\end{fmfgraph*}
}
\end{fmffile}

%% file: Fdiagrams/AAa.tex
\begin{fmffile}{Fdiagrams/AAacoupl}
\begin{fmfgraph*}(50,50)

\fmfleft{i1,i2,i3}
\fmfright{o3,o2,o1}
  \fmf{dashes}{i2,v1}
  \fmf{boson,label=$p_{A1}$,l.side=left}{v1,o1}
  \fmf{boson,label=$p_{A2}$,l.side=right}{v1,o3}
  
\fmfv{lab=$A_\mu$,l.angle=0}{o1}
\fmfv{lab=$A_\nu$,l.angle=0}{o3}
\fmfv{lab=$a$}{i2}
\end{fmfgraph*}
\end{fmffile}

%% file: Fdiagrams/ZZa.tex
\begin{fmffile}{Fdiagrams/zzacoupl}
\begin{fmfgraph*}(50,50)

\fmfleft{i1,i2,i3}
\fmfright{o3,o2,o1}
  \fmf{dashes}{i2,v1}
  \fmf{boson,label=$p_{Z1}$,l.side=left}{v1,o1}
  \fmf{boson,label=$p_{Z2}$,l.side=right}{v1,o3}
  
\fmfv{lab=$Z_\mu$,l.angle=0}{o1}
\fmfv{lab=$Z_\nu$,l.angle=0}{o3}
\fmfv{lab=$a$}{i2}
\end{fmfgraph*}
\end{fmffile}

%% file: Fdiagrams/ZAa.tex
\begin{fmffile}{Fdiagrams/ZAacoupl}
\begin{fmfgraph*}(50,50)

\fmfleft{i1,i2,i3}
\fmfright{o3,o2,o1}
  \fmf{dashes}{i2,v1}
  \fmf{boson}{v1,o1}
  \fmf{boson }{v1,o3}
  
\fmfv{lab=$Z_\mu$,l.angle=0}{o1}
\fmfv{lab=$A_\nu$,l.angle=0}{o3}
\fmfv{lab=$a$}{i2}
\end{fmfgraph*}
\end{fmffile}

%% file: Fdiagrams/WWa.tex
\begin{fmffile}{Fdiagrams/wwacoupl}
\begin{fmfgraph*}(50,50)

\fmfleft{i1,i2,i3}
\fmfright{o3,o2,o1}
  \fmf{dashes}{i2,v1}
  \fmf{boson}{v1,o1}
  \fmf{boson}{v1,o3}
  
\fmfv{lab=$W^+_\mu$,l.angle=0}{o1}
\fmfv{lab=$W^-_\nu$,l.angle=0}{o3}
\fmfv{lab=$a$}{i2}
\end{fmfgraph*}
\end{fmffile}

%% file: Fdiagrams/GGa.tex
\begin{fmffile}{Fdiagrams/GGacoupl}
\begin{fmfgraph*}(50,50)

\fmfleft{i1,i2,i3}
\fmfright{o3,o2,o1}
  \fmf{dashes}{i2,v1}
  \fmf{gluon,label=$p_{G1}$,l.side=left}{v1,o1}
  \fmf{gluon,label=$p_{G2}$,l.side=right}{v1,o3}
  
\fmfv{lab=$G_\mu$,l.angle=0}{o1}
\fmfv{lab=$G_\nu$,l.angle=0}{o3}
\fmfv{lab=$a$}{i2}
\end{fmfgraph*}
\end{fmffile}

%% file: Fdiagrams/Aah.tex
\begin{fmffile}{Fdiagrams/Aahcoupl}
\begin{fmfgraph*}(50,50)

\fmfleft{i1}
\fmfright{o2,o1}
  \fmf{dashes}{i1,v1}
  \fmf{dashes}{v1,o1}
  \fmf{boson}{v1,o2}
  
\fmfv{lab=$h$,l.angle=0}{o1}
\fmfv{lab=$A_\mu$,l.angle=0}{o2}
\fmfv{lab=$a$,l.angle=180}{i1}
\end{fmfgraph*}
\end{fmffile}

%% file: Fdiagrams/Zhha.tex
\begin{fmffile}{Fdiagrams/zhhacoupl}
\begin{fmfgraph*}(90,50)

\fmfleft{i1,i2}
\fmfright{o2,o1}
  \fmf{dashes}{i2,v1}
  \fmf{dashes}{i1,v1}
  \fmf{boson}{v1,o1}
  \fmf{dashes}{v1,o2}
  
\fmfv{lab=$Z_\mu$,l.angle=0}{o1}
\fmfv{lab=$h$,l.angle=0}{o2}
\fmfv{lab=$h$,l.angle=180}{i1}
\fmfv{lab=$a$,l.angle=180}{i2}
\end{fmfgraph*}
\end{fmffile}

%% file: Fdiagrams/Aahh.tex
\begin{fmffile}{Fdiagrams/Aahhcoupl}
\begin{fmfgraph*}(90,50)

\fmfleft{i1,i2}
\fmfright{o2,o1}
  \fmf{dashes}{i2,v1}
  \fmf{dashes}{i1,v1}
  \fmf{boson}{v1,o1}
  \fmf{dashes}{v1,o2}
  
\fmfv{lab=$A_\mu$,l.angle=0}{o1}
\fmfv{lab=$h$,l.angle=0}{o2}
\fmfv{lab=$h$,l.angle=180}{i1}
\fmfv{lab=$a$,l.angle=180}{i2}
\end{fmfgraph*}
\end{fmffile}

%% file: Fdiagrams/WWZa.tex
\begin{fmffile}{Fdiagrams/wwzacoupl}
\begin{fmfgraph*}(90,50)

\fmfleft{i1,i2}
\fmfright{o2,o1}
  \fmf{dashes}{i2,v1}
  \fmf{boson}{i1,v1}
  \fmf{boson}{v1,o1}
  \fmf{boson}{v1,o2}
  
\fmfv{lab=$W^+_\mu$,l.angle=0}{o1}
\fmfv{lab=$W^-_\nu$,l.angle=0}{o2}
\fmfv{lab=$Z_\rho$,l.angle=180}{i1}
\fmfv{lab=$a$,l.angle=180}{i2}
\end{fmfgraph*}
\end{fmffile}

%% file: Fdiagrams/WWAa.tex
\begin{fmffile}{Fdiagrams/wwAacoupl}
\begin{fmfgraph*}(90,50)

\fmfleft{i1,i2}
\fmfright{o2,o1}
  \fmf{dashes}{i2,v1}
  \fmf{boson}{i1,v1}
  \fmf{boson}{v1,o1}
  \fmf{boson}{v1,o2}
  
\fmfv{lab=$W^+_\mu$,l.angle=0}{o1}
\fmfv{lab=$W^-_\nu$,l.angle=0}{o2}
\fmfv{lab=$A_\rho$,l.angle=180}{i1}
\fmfv{lab=$a$,l.angle=180}{i2}
\end{fmfgraph*}
\end{fmffile}

%% file: Fdiagrams/ZZZa.tex
\begin{fmffile}{Fdiagrams/zzzacoupl}
\begin{fmfgraph*}(90,50)

\fmfleft{i1,i2}
\fmfright{o2,o1}
  \fmf{dashes}{i2,v1}
  \fmf{boson}{i1,v1}
  \fmf{boson}{v1,o1}
  \fmf{boson}{v1,o2}
  
\fmfv{lab=$Z_\mu$,l.angle=0}{o1}
\fmfv{lab=$Z_\nu$,l.angle=0}{o2}
\fmfv{lab=$Z_\rho$,l.angle=180}{i1}
\fmfv{lab=$a$,l.angle=180}{i2}
\end{fmfgraph*}
\end{fmffile}

%% file: Fdiagrams/ahZZ.tex
\begin{fmffile}{Fdiagrams/ahZZcoupl}
\begin{fmfgraph*}(90,50)

\fmfleft{i1,i2}
\fmfright{o2,o1}
  \fmf{boson}{i1,v1}
  \fmf{boson}{i2,v1}
  \fmf{dashes}{v1,o1}
  \fmf{dashes}{v1,o2}

\fmfv{lab=$Z_\nu$,l.angle=180}{i1}  
\fmfv{lab=$Z_\mu$,l.angle=180}{i2}
\fmfv{lab=$a$,l.angle=0}{o1}
\fmfv{lab=$h$,l.angle=0}{o2}
\end{fmfgraph*}
\end{fmffile}

%% file: Fdiagrams/ahAZ.tex
\begin{fmffile}{Fdiagrams/ahAZcoupl}
\begin{fmfgraph*}(90,50)

\fmfleft{i1,i2}
\fmfright{o2,o1}
  \fmf{boson}{i1,v1}
  \fmf{boson}{i2,v1}
  \fmf{dashes}{v1,o1}
  \fmf{dashes}{v1,o2}

\fmfv{lab=$Z_\nu$,l.angle=180}{i1}  
\fmfv{lab=$A_\mu$,l.angle=180}{i2}
\fmfv{lab=$a$,l.angle=0}{o1}
\fmfv{lab=$h$,l.angle=0}{o2}
\end{fmfgraph*}
\end{fmffile}

%% file: Fdiagrams/ahWW.tex
\begin{fmffile}{Fdiagrams/ahWWcoupl}
\begin{fmfgraph*}(90,50)

\fmfleft{i1,i2}
\fmfright{o2,o1}
  \fmf{boson}{i1,v1}
  \fmf{boson}{i2,v1}
  \fmf{dashes}{v1,o1}
  \fmf{dashes}{v1,o2}

\fmfv{lab=$W^+_\mu$,l.angle=180}{i1}  
\fmfv{lab=$W^-_\nu$,l.angle=180}{i2}
\fmfv{lab=$a$,l.angle=0}{o1}
\fmfv{lab=$h$,l.angle=0}{o2}
\end{fmfgraph*}
\end{fmffile}

%% file: Fdiagrams/GGGa.tex
\begin{fmffile}{Fdiagrams/gggacoupl}
\begin{fmfgraph*}(90,50)

\fmfleft{i1,i2}
\fmfright{o2,o1}
  \fmf{dashes}{i2,v1}
  \fmf{gluon}{i1,v1}
  \fmf{gluon}{v1,o1}
  \fmf{gluon}{v1,o2}
  
\fmfv{lab=$G^a_\mu$,l.angle=0}{o1}
\fmfv{lab=$G^b_\nu$,l.angle=0}{o2}
\fmfv{lab=$G^c_\rho$,l.angle=180}{i1}
\fmfv{lab=$a$,l.angle=180}{i2}
\end{fmfgraph*}
\end{fmffile}

%% file: Fdiagrams/auu.tex
\begin{fmffile}{Fdiagrams/auucoupl}
\begin{fmfgraph*}(50,50)

\fmfleft{i1,i2,i3}
\fmfright{o3,o2,o1}
  \fmf{dashes}{i2,v1}
  \fmf{fermion,label=$U_\beta$}{o1,v1}
  \fmf{fermion,label=$U_\alpha$}{v1,o3}
  
\fmfv{lab=$a$}{i2}
\end{fmfgraph*}
\end{fmffile}

%% file: Fdiagrams/add.tex
\begin{fmffile}{Fdiagrams/addcoupl}
\begin{fmfgraph*}(50,50)

\fmfleft{i1,i2,i3}
\fmfright{o3,o2,o1}
  \fmf{dashes}{i2,v1}
  \fmf{fermion,label=$D_\beta$}{o1,v1}
  \fmf{fermion,label=$D_\alpha$}{v1,o3}
  
\fmfv{lab=$a$}{i2}
\end{fmfgraph*}
\end{fmffile}

%% file: Fdiagrams/aee.tex
\begin{fmffile}{Fdiagrams/aeecoupl}
\begin{fmfgraph*}(50,50)

\fmfleft{i1,i2,i3}
\fmfright{o3,o2,o1}
  \fmf{dashes}{i2,v1}
  \fmf{fermion,label=$E_\beta$}{o1,v1}
  \fmf{fermion,label=$E_\alpha$}{v1,o3}
  
\fmfv{lab=$a$}{i2}
\end{fmfgraph*}
\end{fmffile}

%% file: Fdiagrams/auuh.tex
\begin{fmffile}{Fdiagrams/auuhcoupl}
\begin{fmfgraph*}(90,50)

\fmfleft{i1,i2}
\fmfright{o2,o1}
  \fmf{dashes}{i2,v1,i1}
  \fmf{fermion}{o1,v1}
  \fmf{fermion}{v1,o2}
  
\fmfv{lab=$a$,l.angle=180}{i2}
\fmfv{lab=$h$,l.angle=180}{i1}
\fmfv{lab=$U_\beta$,l.angle=0}{o1}
\fmfv{lab=$U_\alpha$,l.angle=0}{o2}
\end{fmfgraph*}
\end{fmffile}

%% file: Fdiagrams/addh.tex
\begin{fmffile}{Fdiagrams/addhcoupl}
\begin{fmfgraph*}(90,50)

\fmfleft{i1,i2}
\fmfright{o2,o1}
  \fmf{dashes}{i2,v1,i1}
  \fmf{fermion}{o1,v1}
  \fmf{fermion}{v1,o2}
  
\fmfv{lab=$a$,l.angle=180}{i2}
\fmfv{lab=$h$,l.angle=180}{i1}
\fmfv{lab=$D_\beta$,l.angle=0}{o1}
\fmfv{lab=$D_\alpha$,l.angle=0}{o2}
\end{fmfgraph*}
\end{fmffile}

%% file: Fdiagrams/aeeh.tex
\begin{fmffile}{Fdiagrams/aeehcoupl}
\begin{fmfgraph*}(90,50)

\fmfleft{i1,i2}
\fmfright{o2,o1}
  \fmf{dashes}{i2,v1,i1}
  \fmf{fermion}{o1,v1}
  \fmf{fermion}{v1,o2}
  
\fmfv{lab=$a$,l.angle=180}{i2}
\fmfv{lab=$h$,l.angle=180}{i1}
\fmfv{lab=$E_\beta$,l.angle=0}{o1}
\fmfv{lab=$E_\alpha$,l.angle=0}{o2}
\end{fmfgraph*}
\end{fmffile}